%% file: main.tex
\documentclass[acmsmall]{acmart}

\AtBeginDocument{%
  \providecommand\BibTeX{{%
    \normalfont B\kern-0.5em{\scshape i\kern-0.25em b}\kern-0.8em\TeX}}}

\setcopyright{acmcopyright}
\copyrightyear{2018}
\acmYear{2018}
\acmDOI{10.1145/1122445.1122456}

\acmJournal{TSAS}
\acmVolume{1}
\acmNumber{1}
\acmArticle{1}
\acmMonth{1}

\usepackage{graphicx}
\usepackage{subfig}
\usepackage[ruled]{algorithm2e} 
\usepackage{multirow}
\usepackage{xcolor}

\usepackage[normalem]{ulem}
\usepackage{threeparttable} 
\usepackage{amsmath} 
\usepackage{url}
\usepackage{tcolorbox} 
\usepackage{adjustbox} 
\usepackage{tabularx}
\usepackage{colortbl} 

\begin{document}

\title[TDEFSI]{TDEFSI: Theory Guided Deep Learning Based Epidemic Forecasting with Synthetic Information}

\author{Lijing Wang}
\orcid{0000-0002-0836-9190}
\affiliation{%
  \institution{Network Systems Science and Advanced Computing Division, Biocomplexity Institute and Initiative \& Department of Computer Science, University of Virginia}
\city{Charlottesville}\state{VA}\postcode {22904}
}
\email{lw8bn@virginia.edu}
\author{Jiangzhuo Chen}
\affiliation{%
  \institution{Network Systems Science and Advanced Computing Division, Biocomplexity Institute \& Initiative, University of Virginia}
\city{Charlottesville}\state{VA}\postcode {22904}
}
\email{chenj@virginia.edu}
\author{Madhav Marathe}
\affiliation{%
 \institution{Network Systems Science and Advanced Computing Division, Biocomplexity Institute and Initiative \& Department of Computer Science, University of Virginia}
\city{Charlottesville}\state{VA}\postcode {22904}
}
\email{marathe@virginia.edu}
%
\renewcommand{\shortauthors}{Wang and Chen and Marathe}

%
\begin{abstract}
\input{sections/abstract.tex}
\end{abstract}

%
%



\maketitle

\input{sections/introduction.tex}

\input{sections/related.tex}
\input{sections/problem.tex}
\input{sections/tdefsi.tex}

\input{sections/experiment.tex}
\input{sections/conclusion.tex}

\begin{acks}
The authors would like to thank members of the Network Systems Science and Advanced Computing (NSSAC) Division for interesting discussion and suggestions related to epidemic science and machine learning. This work has been partially supported by Defense Threat Reduction Agency (DTRA) Grant HDTRA1-17-D-0023, National Institutes of Health (NIH) Grant 1R01GM109718, NSF BIG DATA Grant IIS-1633028, NSF DIBBS Grant ACI-1443054.
\end{acks}

\bibliographystyle{ACM-Reference-Format}
\bibliography{main}

\appendix
\input{sections/appendix.tex}

\end{document}

%% file: sections/abstract.tex
Influenza-like illness (ILI) places a heavy social and economic burden on our society. Traditionally, ILI surveillance data is updated weekly and provided at a spatially coarse resolution. Producing timely and reliable high-resolution spatiotemporal forecasts for ILI is crucial for local preparedness and optimal interventions. 

\smallskip
\noindent
We present TDEFSI{\footnote{The preliminary version of this work \cite{wang2019defsi} was presented at the \emph{Thirty-First Innovative Applications of Artificial Intelligence Conference (IAAI 2019}).}} (\textit{T}heory Guided \textit{D}eep Learning Based \textit{E}pidemic \textit{F}orecasting with \textit{S}ynthetic \textit{I}nformation), an epidemic forecasting framework that integrates the strengths of deep neural networks and high-resolution simulations of epidemic processes over networks. TDEFSI yields accurate high-resolution spatiotemporal forecasts using low-resolution time series data.

\smallskip
\noindent
During the training phase, TDEFSI uses high-resolution simulations of epidemics that explicitly model spatial and social heterogeneity
inherent in urban regions as one component of training data. We train a two-branch recurrent neural network model to take both within-season and between-season low-resolution observations as features, and output high-resolution detailed forecasts. The resulting forecasts are not just driven by observed data but also capture the intricate social, demographic and geographic attributes of specific urban regions and mathematical theories of disease propagation over networks. 

We focus on forecasting the incidence of ILI and evaluate TDEFSI's performance using synthetic and real-world testing datasets at the state and county levels in the USA. The results show that, at the state level, our method achieves comparable/better performance than several state-of-the-art methods. At the county level, TDEFSI outperforms the other methods. The proposed method can be applied to other infectious diseases as well. 

%% file: sections/introduction.tex
\section{Introduction}\label{sec:introduction}
Influenza-like illness (ILI) poses a serious threat to global public health. Worldwide, annually, seasonal influenza causes three to five million cases of severe illness and 290,000 to 650,000 deaths~\cite{whoseasonalinfluenza}. 
Since 2010 in the USA, seasonal influenza has resulted in 10-50 million cases annually, 140,000 to 960,000 hospitalizations, between 12,000 and 79,000 deaths, and is responsible for approximately $\$$87.1 billion in economic losses~\cite{cdcburden,molinari2007}.   
Producing timely, well-informed, and reliable forecasts for ILI of an ongoing flu epidemic is crucial for preparedness and optimal intervention~\cite{doms2018assessing}.
Traditionally, ILI surveillance data from the Centers for Disease Control and Prevention (CDC) has been used as reference data to predict future ILI incidence.
The surveillance data is updated weekly but often delayed by one to four weeks and is provided at a  HHS region (i.e. the ten regions defined by the United States Department of Health $\&$ Human Services) level and recently at the state level.
Considering the heterogeneity between different subregions, accurate predictions with a finer resolution, e.g. at county  or city level in the USA, are crucial for local public health decision making, optimal mitigation resource allocation among subregions, and household or individual level preventive actions informed by neighboring prevalence~\cite{yang2016forecasting}.
Given spatially coarse-grained surveillance data, it is challenging to forecast at a finer spatial level. 

In this paper we use \textbf{\textit{flat-resolution}} forecasting to denote the prediction of ILI incidence with the same resolution as the surveillance data and \textbf{\textit{high-resolution}} forecasting to denote the prediction with a higher geographical resolution than provided in surveillance data.
We focus on state level ILI surveillance and state (flat-resolution) or county level (high-resolution) ILI forecasts. We use the term \emph{deep neural networks (DNN)} to denote multi-layer neural networks with multiple inputs and outputs. 

\subsection{Our contributions}
We propose a novel epidemic forecasting framework, called \textbf{T}heory Guided \textbf{D}eep Learning Based \textbf{E}pidemic \textbf{F}orecasting with \textbf{S}ynthetic \textbf{I}nformation (\textbf{TDEFSI}). 

\smallskip\emph{Overall approach.} \
TDEFSI produces accurate weekly high-resolution ILI forecasts from flat-resolution observations. This is achieved by using a two-branch neural network model for ILI forecasting. It combines within-season observations (observed data points of the previous weeks that characterize the ongoing epidemic) and between-season historical observations (observed data points from similar weeks of the past seasons that characterize general trends around the current week). 
It can generate probabilisitic forecasts by using Monte Carlo Dropout technique~\cite{gal2016dropout}.

A key contribution of the paper is to use theory generated synthetic data to train the neural network. This is necessitated by the fact that disease surveillance data is sparse. Furthermore, the data is noisy and incomplete. We overcome the limitations by training
TDEFSI using data generated by high performance computing based simulations of well accepted causal processes that capture epidemic dynamics. These simulations are based on decades of work and have been extensively validated. The simulations allow us to: 
($i$) use a realistic representation of the underlying social contact network that captures the multi-scale spatial, temporal and social interactions, as well as the inherent heterogeneity of social networks (individual demographic attributes, heavy tailed nature of social contacts, etc.), leading to \emph{forecasts that are context specific and capture the unique properties of a given urban region}; 
($ii$) produce multi-resolution forecasts even though observational data might only be available at an aggregate level, leading to \emph{an ability to forecast disease incidence at a county or a city level as well as forecasts for desired demographic groups}; and
($iii$) capture the underlying causal processes and mathematical theories leading to {\em explainable and generalizable AI}
-- the combination of theory and data driven machine learning is an important and emerging approach to scientific problems that are data sparse. 

\smallskip
\emph{Key findings.}
Extensive experiments were carried out using both real-world as well as synthetic datasets for testing.
($i$) In experiments on synthetic testing data, we evaluate TDEFSI performance with different hyperparameter settings and find that the best look-back window size is 52 weeks, the same as the period of influenza seasons, for both state level and county level forcasting.
($ii$) In experiments on two states of the USA using their real ILI incidence data as ground truth, we compare TDEFSI and its variants with several state-of-the-art forecasting methods, among which four methods can only make state level predictions directly and one method can make both state level and county level predictions directly. 
The results indicate that in most cases TDEFSI methods achieves comparable/better performance than the comparison methods at the state level. For high-resolution forecasting at the county level, TDEFSI significantly outperforms the comparison methods. Between the variants of TDEFSI, we find that the between-season branch of our neural network model improves the forecasting accuracy.
($iii$) We also find that the two physical constraints in our TDEFSI model, which address spatial consistency and non-negative consistency respectively, contribute to the improvement on the forecasting performance.
($iv$) Through a comparison between TDEFSI models trained with datasets generated by no-intervention simulations and those by intervention-aware simulations, we find that in our TDEFSI framework realistic settings in the causal model behind the neural network do improve the generalizability of the trained forecasting model.
($v$) In general, TDEFSI is able to capture the heterogeneity in epidemic dynamics among counties in a state and the spatial spread of the disease across the counties.


To the best of our knowledge, TDEFSI is the first to use a realistic causal high resolution model to train a deep neural network for epidemic forecasting. The basic approach is general and points to the potential utility of the approach to study other problems in social and ecological sciences.  Unlike physical systems, encoding system level constraints is often possible only via simulations; the theories are largely local rules of interactions. In this sense, training the neural network using simulations provides a natural way to place constraints on the concept class that the neural network effectively learns.

A natural question that arises is: \emph{why does one need to use a neural network when simulations are available?} There are multiple reasons to do this: ($i$) computational efficiency (ability to rapidly produce forecasts, ($ii$) generalizability (often simulation parameters might end up overfitting to the data), and ($iii$) ability to incorporate additional data sources. In this sense, \emph{DL+simulations} appears to be a promising approach for forecasting rather than using either of them individually. See the next section for further discussion. 

%% file: sections/related.tex
\input{tables/rel-summary.tex}

\section{Related work}
\label{sec:related}

\subsection{Epidemic forecasting}
\label{subsec:related-epi}
Forecasting the spatial and temporal evolution of infectious disease epidemics has been an area of active research over the past couple of decades~\cite{longini1986predicting,loytonen1996forecasting,viboud2003prediction,goldstein2011predicting,shaman2012,nsoesie2013forecasting,nsoesie2013simulation,shaman2013realtime,biggerstaff2016results,viboud2018rapidd,pei2018forecasting,biggerstaff2018results,mcgowan2019collaborative,reich2019}.
We briefly review related work in epidemic forecasting and deep learning pertinent to our problem; see ~\cite{alessa2018,chretien2014influenza,nsoesie2014systematic} for more details. 
We discuss four ILI forecasting methods: causal methods, statistical methods, artificial neural network methods, and hybrid methods. See Figure~\ref{fig:summaryofrelatedwork} for a brief summary.

\noindent{\bf Causal methods} In epidemiology, within-host progression models for ILI include: susceptible-infectious-recovered (SIR), susceptible-exposed-infectious-recovered (SEIR), susceptible-infectious-recovered-susceptible (SIRS), and their extensions~\cite{bailey1975mathematical,kuznetsov1994bifurcation}. 
Forecasting methods employing these models are called causal methods (or mechanistic methods) because they are based on the causal mechanisms of infectious diseases.
In these methods the underlying epidemic model can be either a compartmental model (CM)~\cite{flahault2006,lee2012,lunelli2009} or an agent-based model (ABM)~\cite{parker2011,chao2010}.
In a compartmental model, a population is divided into compartments (e.g. S, E, I, R). A differential equation system characterizes the change of the sizes of each compartment due to disease propagation and progression. 
To get county level epidemics in a compartmental model, one needs to create compartments in each county, where county population sizes and between county travel data become crucial.
In an agent-based model, disease spreads among heterogeneous agents through an unstructured network. 
Dynamics with individual behavior change exhibit significant impact on epidemic and dynamic forecast models ~\cite{eksin2019systematic}, which can be implemented using a high-performance computing model~\cite{bisset2009}. 
The individual level details in an agent-based model can be easily aggregated to obtain epidemic data of any resolution, e.g. number of newly infected people in a county in a specific week. 
Many forecasting methods have been developed based on either CM or ABM~\cite{hua2018,tuite2010,shaman2012,nsoesie2013simulation,yang2014comparison,yang2015inference,zhao2015simnest,morita2018influenza}.
Shaman et al.~\cite{shaman2012} developed a framework for initializing real-time forecasts of seasonal influenza outbreaks, using a data assimilation technique commonly applied in numerical weather prediction.
Tuite et al.~\cite{tuite2010} used an SIR CM to estimate parameters and morbidity in pandemic H1N1. 
Yang et al.~\cite{yang2014comparison} applied various filter methods to model and forecast influenza activity using an SIRS CM. 
In~\cite{nsoesie2013simulation}, the authors proposed a simulation optimization approach based on the SEIR ABM for epidemic forecasting. 
Hua et al.~\cite{hua2018} and Zhao et al.~\cite{zhao2015simnest} infer the parameters of the SEIR ABM from social media data for ILI forecasting.
{\em Limitations: Causal methods are generally computationally expensive as they require the parameter estimation over a high dimensional space. As a result the use of such methods for real-time forecasting is challenging.} 

\noindent{\bf Statistical methods}
Statistical methods employ statistical and time series based methodologies to learn patterns in historical epidemic data and leverage those patterns for forecasting~\cite{brooks2018nonmechanistic,kandula2019near}.
Popular statistical methods for ILI forecasting include e.g. generalized linear models (GLM), autoregressive integrated moving average (ARIMA), and generalized autoregressive moving average (GARMA)~\cite{bardak2015prediction,benjamin2003generalized,dugas2013}. 
Wang et al.~\cite{wang2015dynamic} proposed a dynamic Poisson autoregressive model with exogenous input variables (DPARX) for flu forecasting. 
Yang et al.~\cite{yang2015accurate} proposed ARGO, an autoregressive-based influenza tracking model for nowcasting incorporating CDC ILI data and Google search data. The extensive work based on ARGO is discussed in~\cite{yang2017}.
{\em Limitations: Statistical methods are fast. But they crucially depend on the availability of training data and as such can only produce flat-resolution forecasts. High-resolution forecasts must be calculated by multiplying the flat-resolution forecasts with high-resolution population proportions. The trained models could not capture the heterogeneous dynamics between high-resolution regions. Furthermore, since they are purely data driven, they do not capture the underlying causal mechanisms. As a result epidemic dynamics affected by behavioral adaptations are usually hard to capture. }


\noindent{\bf Artificial neural network methods}
Artificial neural networks (ANN) have gained increased prominence in epidemic forecasting due to their self-learning ability without prior knowledge.
Xu et al.~\cite{xu2017forecasting} firstly introduced feed-forward neural network (FNN) into surveillance of infectious diseases and investigated its predictive utility using CDC ILI data, Google search data, and meteorological data.
Recurrent neural network (RNN) has been demonstrated to be able to capture dynamic temporal behavior of a time sequence.
In~\cite{volkova2017forecasting} Volkova et al. built an LSTM model for short-term ILI forecasting using CDC ILI and Twitter data. 
Venna et al.~\cite{venna2019novel} proposed an LSTM based method that integrates the impacts of climatic factors and geographical proximity to achieve better forecasting performance.
Wu et al.~\cite{wu2018deep} constructed a deep learning structure combining RNN and convolutional neural network to fuse information from different sources.
Deng et al.~\cite{deng2019graph} recently designed a cross-location attention based graph neural network for learning time series embeddings and location aware attentions.
{\em Limitations: Just like statistical methods, ANN based forecasting methods are data driven and have similar limitations. In addition, the model performance usually depends on the availability of a very large training dataset. Another well known limitation of ANN methods is their ability to explain the resulting forecasts.} 

\noindent{\bf Hybrid methods}
Hybrid methods combine data driven and causal methods. They are attractive as they can borrow the best from both worlds~\cite{kandula2018evaluation}. 
The authors in~\cite{osthus2019dynamic} proposed a dynamic Bayesian model for influenza forecasting which combines the machine learning approach and a compartmental model to explicitly account for systematic deviations between mechanistic models and the observed data.
Such methods have shown promise as evidenced in recent papers on the study of physical and biological systems~\cite{faghmous2014,fischer2006,hautier2010,kawale2013,khandelwal2015,khandelwal2017,karpatne2017,wong2009,xu2015} -- see~\cite{karpatne2017} for a discussion on this subject.


\noindent{\bf TDEFSI method}
Our method combines the deep neural networks and high-reslution epidemic simulations to enable accurate weekly high-resolution ILI forecasts from flat-resolution observations.
Compared with causal methods, TDEFSI avoids searching optimal disease model parameters over a high dimensional space because it does not need to identify any specific causal models for the forecasting. Compared with data driven methods (statistical and neural network methods), TDEFSI explicitly models spatial and social heterogeneity in a region from the training data. It can capture the heterogeneous dynamics between high-resolution regions, as well as underlying causal processes and mathematical theories. In addition, the large volume of synthetic training data helps TDEFSI to overcome  the risk of overfitting due to sparse observation data.

\subsection{Data augmentation for time series}
\label{subsec:related-aug}
Data augmentation in deep neural networks is the process of generating artificial data in order to reduce overfitting. It has been shown to improve deep neural network's generalization capabilities in many tasks especially in computer vision tasks such as image or video recognition~\cite{schluter2015exploring}. Various augmentation techniques have been applied to specific problems, including affine transformation of the original images~\cite{vasconcelos2017increasing,rizk2019effectiveness,wong2016understanding} and unsupervised generation of new data using Generative Adversarial Nets (GANs)~\cite{perez2017effectiveness,gurumurthy2017deligan,marchesi2017megapixel,zhu2017unpaired} or variational autoencoder (VAE) models~\cite{rizk2019effectiveness}, etc. However, the techniques for image augmentation do not generalize well to time series. The main reason is that image augmentation is not expected to change the class of an image, while for time series data, one cannot confirm the effect of such transformations on the nature of a time series. In what follows we introduce related work on time series data augmentation.

\noindent{\bf Data augmentation for time series classification}
For time series classification (TSC) problems, one of the most popular methods is the slicing window technique, originally introduced for deep CNNs in~\cite{cui2016multi}. The method was inspired by the image cropping technique for computer vision tasks~\cite{zhang2016understanding}. In~\cite{kvamme2018predicting}, it was adopted to improve the CNNs' mortgage delinquency prediction using customer's historical transactional data. The authors in~\cite{krell2018data} used it to improve the Support Vector Machines accuracy for classifying electroencephalographic time series. The authors in~\cite{um2017data} proposed a novel data augmentation method (including window slicing, permutating, rotating, time-warping, scaling, magnitude-wrapping, jiterring, cropping) specific to wearable sensor collected time series data.
Le Guennec et al.~\cite{le2016data} extended the slicing window technique with a warping window that generates synthetic time series by warping the data through time. It extracts multiple small-size windows from a single window and lengthens/shortens a part of the window data, respectively. The methods are reported to reduce classification error on several types of time series data.
Forestier et al.~\cite{forestier2017generating} proposed to average a set of time series as a new synthetic series. It relies on an extension of Dynamic Time Warping (DTW) Barycentric Averaging (DBA).

\noindent{\bf Data augmentation for time series regression}
Unlike data augmentations for TSC, data augmentation
for time series regression (TSR) has not been well investigated yet to the best of our knowledge. 
Bergmeir et al.~\cite{bergmeir2016bagging} presented a method
using Box-Coxfor  transformation followed by an STL decomposition to separate the time series into trend, seasonal part, and remainder. The remainder was then bootstrapped using a moving
block bootstrap, and a new series was assembled using this bootstrapped remainder.

All above methods for TSC or TSR apply techniques directly on observed time sequences, which generate synthetic data at the same resolution as the original data. In our problem, we try to forecast at a higher resolution when there is no or very sparse high-resolution observations.

\noindent{\bf TDEFSI method}
We generate synthetic high-resolution data using high performance computing based simulations of well accepted causal processes that capture epdemic dynamics. Different from data augmentation techniques introduced above, we synthesize high-resolution data which is not available or quite sparse in the real world. 

%% file: tables/rel-summary.tex
\begin{figure}[!t]
    \centering
    \begin{tcolorbox}[standard jigsaw,opacityback=0]
    \begin{itemize}
        \item \textbf{Epidemic forecasting}
            \begin{itemize}
            \item \textbf{Causal methods}
            \begin{itemize}
                \item[] \textit{Pros:} employ mathematical models of disease transmission; make multi-fidelity predictions. The models can often capture human decision making and thus provide a path for counterfactual forecasts.
                \item[] \textit{Cons:} generally computationally expensive as they require parameter estimation over a high dimensional space. For networked model, obtaining the needed data to build realistic social networks can be challenging.
            \end{itemize}
            \item \textbf{Statistical methods}
            \begin{itemize}
                \item[] \textit{Pros:} learn patterns from the historical time series; easy to implement the model; fast to train and forecast.
                \item[] \textit{Cons:} usually assume a simple relationship between the inputs and outputs; unable to make heterogeneous high-resolution forecasting.
            \end{itemize}
            \item \textbf{Artificial neural network (ANN) methods}
            \begin{itemize}
                \item[] \textit{Pros:} learn patterns from the historical data; capture non-linear relationship between the inputs and outputs.
                \item[] \textit{Cons:} model performance depends on the availability of large amount of training data; unable to make heterogeneous high-resolution forecasting; lack of explainability; overfitting is a concern due to the small size of the training dataset.
            \end{itemize}
            \item \textbf{Hybrid methods}
            \begin{itemize}
                \item[] \textit{Pros:} combine data driven (statistical and ANN methods) and causal methods; integrate strengths of both methods.
                \item[] \textit{Cons:} have not been explored until now in epidemic forecasting domain.
            \end{itemize}
            \item \textbf{TDEFSI method}
            \begin{itemize}
                \item[] \textit{Pros:} combine deep neural networks and high-resolution epidemic simulations; avoid overfitting using large volume of training data from causal models; enable heterogeneous high-resolution forecasting; yield accurate high-resolution spatiotemporal forecasts using low-resolution time-series data.
            \end{itemize}
        \end{itemize}
        \item \textbf{Data augmentation for time series}
            \begin{itemize}
            \item \textbf{Data augmentation for TSC}
            \begin{itemize}
                \item[] \textit{Pros:} generate artificial time series data to reduce classification error using techniques, such as slicing window, warping window, permutating, scaling, cropping, VAEs, GANs, etc.
                \item[] \textit{Cons:} difficult to apply to time series regression problems. 
            \end{itemize}
            \item \textbf{Data augmentation for TSR}
            \begin{itemize}
                \item[] \textit{Pros:} generate new time series using transformation or decomposition techniques.
                \item[] \textit{Cons:} not well investigated in epidemic forecasting domain; difficult to generate high-resolution time series.
            \end{itemize}
            \item \textbf{TDEFSI method}
            \begin{itemize}
                \item[] \textit{Pros:} synthesize large volume of high-resolution time series from simulations of causal processes based on mathematical epidemiology theory.
                \item[] \textit{Cons:} challenging to minimize the difference between synthetic data and real data.
            \end{itemize}
        \end{itemize}
    \end{itemize}
    \end{tcolorbox}
    \caption{Brief summary of existing ILI forecasting methods and data augmentation techniques.}
    \label{fig:summaryofrelatedwork}
\end{figure}

%% file: sections/problem.tex
\section{Problem Setup}\label{sec:problem}
Given an observed time series of weekly ILI incidence for a specific region, we focus on predicting ILI incidence for both the region and its subregions in short-term.
Without loss of generality, in this paper we consider making predictions for a state of the USA and all counties in the state, using observations only from CDC state level ILI incidence data~\cite{cdcfluview}. In this setting, state level forecasting is flat-resolution, while county level forecasting is high-resolution. 
The proposed framework is not limited to this setting and can be generalized for subregion forecasting in any region, e.g. state level forecasting in a country where only national level surveillance data is available. 
Our proposed method is different from traditional ILI incidence forecasting methods in that the model is trained on synthetic ILI incidence data but forecasts by taking ILI surveillance data as inputs.

Let $\mathbf{y}=\langle y_1,y_2,\cdots,y_T,\cdots \rangle$ denote the sequence of weekly state level ILI incidence, where $y_i \in \mathbb{R}$.
Let $\mathbf{y}^C=\langle y^C_1,y^C_2,\cdots,y^C_T,\cdots \rangle$ denote the sequence of weekly ILI incidence for a particular county $C$ within the state.
Assume that there are $K$ counties $\mathcal{D}=\{C_1, C_2, \cdots, C_{K}\}$ in the state. Let $\mathbf{y}^{\mathcal{D}}_t=\{y^C_t | C \in \mathcal{D}\}$ denote ILI incidence of all counties in the state at week $t$.
Suppose we are given only state level ILI incidence up to week $T$.
The problem is defined as predicting both state level and county level incidence at week $t$, where $t=T+1$, denoted as $\mathbf{z}_t=(y_t, \mathbf{y}^{\mathcal{D}}_t), \mathbf{z}_t \in \mathbb{R}^{K+1}$, given $\langle y_1,y_2,\cdots,y_T \rangle$.

In our problem, when training the deep neural network models, we consider three types of physical consistency requirements based on epidemiologic domain knowledge.
They are \textbf{temporal consistency}, \textbf{spatial consistency}, and \textbf{non-negative consistency}.
($i$) Temporal consistency: the ILI diseases transmit via person to person contacts. The number of infected cases at the current time point depends on the number of infected cases at the previous time points. In addition, infected persons' incubation periods and infectious periods vary due to the heterogeneity among individuals.
In our work, we use the long short term memory (LSTM) network~\cite{hochreiter1997} to capture the temporal dependencies among variables.
($ii$) Spatial consistency: the high-resolution ILI incidence should be consistent with the flat-resolution ILI incidence. In our problem, this consistency is represented as $y_t=\sum \limits_{C \in \mathcal{D}} y^C_t$, i.e., the state incidence equals the sum of ILI incidence at the county level. 
($iii$) Non-negative consistency: the number of infected cases at time $t$ is either zero or a positive value, denoted as $y_t, y^C_t \geq 0$.


\begin{figure}[!t]
\centering
\includegraphics[width=\textwidth]{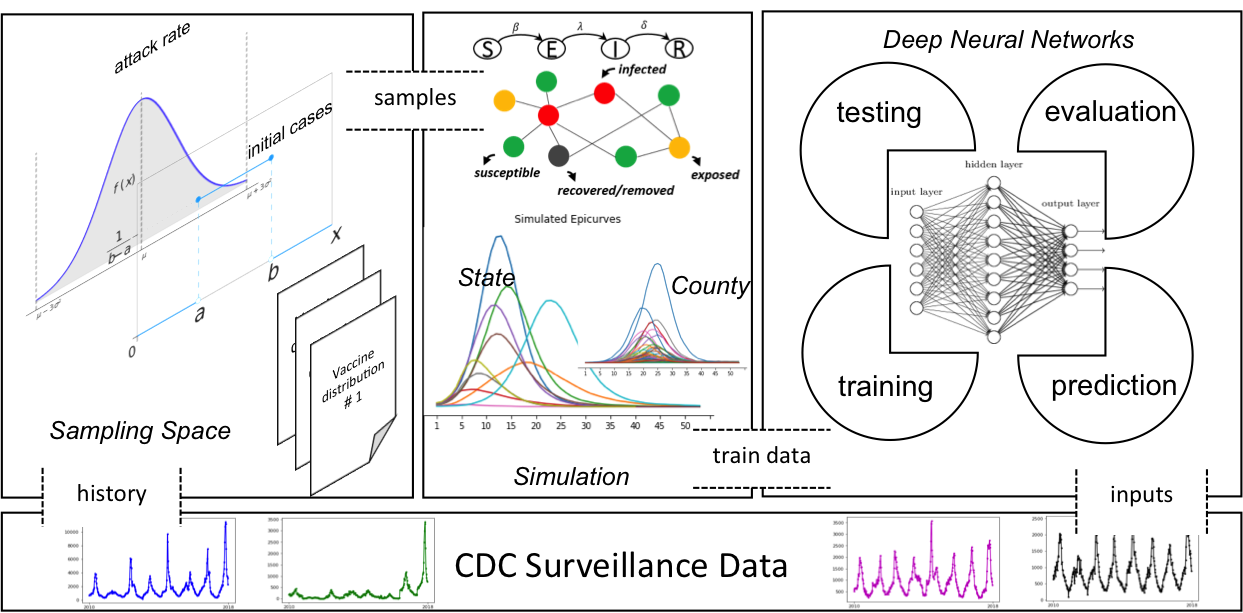}
 \caption{TDEFSI framework. In this framework, a region-specific disease parameter space for a disease model is constructed based on historical surveillance data. Synthetic training data consisting of both state level and county level weekly ILI incidence curves is generated by simulations parameterized by samples from the parameter space. An LSTM based deep neural network model is trained on the synthetic data. The trained model produces forecasts by taking surveillance data as the input.}
  \label{fig:framework}
\end{figure}

%% file: sections/tdefsi.tex
\section{TDEFSI}\label{sec:defsi}
\subsection{Framework}\label{subsec:framework}
The TDEFSI framework consists of three major components (shown in Figure~\ref{fig:framework}): ($i$) \textit{Disease model parameter space construction}: given a state and an existing disease model, we estimate a marginal distribution for each model parameter based on the surveillance data of the state and its neighbors; ($ii$) \textit{Synthetic training data generation}: we generate a synthetic training dataset at both flat-resolution and high-resolution scales for that state by running simulations parameterized from the parameter space; ($iii$) \textit{Deep neural network training and forecasting}: we design a two-branch deep neural network model trained on the synthetic training dataset and use surveillance data as its inputs for forecasting. 
We will elaborate on the details in the following subsections.

\subsection{SEIR-based Epidemic Simulation}\label{subsec:seir}
We simulate the spread of the disease in a synthetic population via its social contact network. In this work we use the synthetic social contact network of each state in the USA (a brief description of the methodology used for constructing the synthetic population and the social network can be found in Appendix~\ref{sec:appendix}).
The SEIR disease model is widely used for ILI diseases~\cite{kuznetsov1994bifurcation}.
Each person is in one of the following four health states at any time: susceptible (S), exposed (E), infectious (I), recovered or removed (R). A person $v$ is in the susceptible state until he becomes exposed. If $v$ becomes exposed, he remains so for $p_E(v)$ days, called the incubation period, during which he is not infectious. Then he becomes infectious and remains so for $p_I(v)$ days, called the infectious period. 
Both $p_E(v)$ and $p_I(v)$ are sampled from corresponding distributions, as shown in Algorithm \ref{method:calibrate}, e.g. $p_E(v)\sim\{1:0.3, 2:0.5, 3:0.2\}$ means that an exposed person will remain so for 1 day with probability 0.3, 2 days with probability 0.5, and 3 days with probability 0.2, similar to $P_I(v)$.
Finally he becomes removed (or recovered) and remains so permanently. 
While the SEIR model characterizes within-host disease progression, between-host disease propagation is modeled by transmissions from person to person with a probability parameter $\tau$, through either complete mixing or heterogeneous connections between people.
With our contact network model, the disease spreads in a population in the following way. It can only be transmitted from an infectious node to a susceptible node. On any day, if node $u$ is infectious and $v$ is susceptible, disease transmission from $u$ to $v$ occurs with probability $p(\tau,w(u, v))$, where $w(u, v)$ represents the contact duration between node $u$ and node $v$. The disease propagates probabilistically along the edges of the contact network.

Various simulators are developed to model human mobility, disease spread, and public health intervention. 
They include compartment-based patch models~\cite{flahault2006,lee2012,lunelli2009}, as well as agent-based models such as EpiFast~\cite{bisset2009}, GSAM~\cite{parker2011}, and FluTE~\cite{chao2010}. Any of these simulators can be used in TDEFSI to generate synthetic training data.
In this work, we adopt an agent-based simulator EpiFast~\cite{bisset2009}.
The outputs are individual infections with their days of being infected in a simulated season. They can be aggregated to any temporal and spatial scale, such as daily (weekly) state (county) level ILI incidence.
Vaccine intervention $I_V$ can be implemented in EpiFast simulations, by specifying the quantity of vaccines applied to the population in each week.
Next we describe how to estimate a distribution on the parameter space $\mathcal{P}(p_E,p_I,\tau,N_I,I_V)$ from CDC historical data, where $N_{I}$ denotes the initial number of infections.
In our simulations, $N_I$ of the population are infectious while all the rest are susceptible at the beginning of the simulation.

\subsection{Disease Model Parameter Space}\label{subsec:paramspace}
Of the parameters, $(p_E,p_I)$ can be taken from literature~\cite{achla2011b}. 
We assume that each of $(\tau,N_I,I_V)$ follows a distribution that can be estimated from historical data.
For clarity, we define an epidemiological week in a calendar year as \textbf{ew}, and a seasonal week in a flu season as \textbf{sw}, where $ew(40)$ is $sw(1)$.
The historical time series of CDC surveillance data (refers to historical training data) used to construct parameter space is split into seasons at $ew(40)$ of each year. That is, each flu season starts from $ew(40)$ of a calendar year and ends in $ew(39)$ of the next year. Note that this applies to the USA, but {\bf sw} may be specified differently for other countries. 

We want to highlight that the number of clinically attended cases and the reported or tested cases are lower than the actual number of cases in the population. Additionally, reporting rates can vary between regions. To address the gap between ILINet case count and population case count, we scale the former with a scaling factor, called surveillance ratio. The ratio is different among different states. See more details of the surveillance ratio in Appendix~\ref{subsec:hosp}.

Firstly, we collect observations of each parameter value as follows: 
\begin{itemize}
\item \textbf{Initial Case Number ($N_{I}$)}: 
We collect the ILI incidence of $sw(1)$ of each season for the target state and its neighboring states (i.e. geographically contiguous states).
\item \textbf{Vaccine Intervention ($I_V$)}: We collect vaccination schedules of the past influenza seasons in the USA~\cite{vacschedule}. 
Each schedule consists of timing and percentage coverage of vaccine application throughout the season. Vaccine efficacy (reduction of disease transmission probability) and compliance rate (probability that a person will take the vaccine) are set according to a survey used in~\cite{wang2018framework}, which is conducted by Gfk.com, under the National Institutes of Health grant no. 1R01GM109718. This survey collects data on demographics of the respondents and their preventive health behaviors during a hypothetical influenza outbreak. 
We assume that each person follows a common compliance rate and the state level vaccine schedule is the same as the nationwide schedule.
\item \textbf{Transmissibility ($\tau$)}:
First we compute the overall attack rate (i.e. the fraction of population getting infected in the season) of each historical season for the target state and its neighboring states.
Then for each attack rate $ar$, say of season $s$ and state $r$, we calibrate a transmissibility value as the solution to $\min_{\tau}|AR(EpiFast(\tau, P_E, P_I, N_I, I_V))-ar|$, where $p_E$ and $p_I$ are sampled for each person from the distributions shown in Table~\ref{tab:distr}; $N_I$ is the initial case number of season $s$ and state $r$; $I_V$ is the vaccination schedule for season $s$; $EpiFast(\cdot)$ is a simulation run on the population of state $j$ with the parameters $(\tau, P_E, P_I, N_I, I_V)$; and $AR(\cdot)$ computes attack rate from the output of $EpiFast(\cdot)$.
Details of this process are shown in Algorithm~\ref{method:calibrate}.
\end{itemize}

\begin{algorithm}[t]
\SetAlgoNoLine
\KwIn{Simulator PS, CDC historical data $histCDC$, and synthetic social contact networks $Network$.}
\KwOut{Calibrated $\tau^*$.}
$p_E \sim \{1:0.3, 2:0.5, 3:0.2\}$~\cite{achla2011b,wang2018framework}\;
$p_I \sim \{3:0.3, 4:0.4, 5:0.2, 6:0.1\}$~\cite{achla2011b,wang2018framework}\;
$I_V = \emptyset$\;
$regions$ = \text{\{state and its adjacent neighbors\}}\;
$seasons$ = \text{\{available seasons of $histCDC$\}}\;
$\tau^* = \emptyset$\;
\For{$r$ in $regions$}{ 
    \For{$s$ in $seasons$}{
        $totalili_{(r,s)} = TOTAL(histCDC_{(r,s)})$ \;
		$ar_{(r,s)}$ =  $\frac{totalili_{(r,s)}}{population_{(r)}}$\;
        $\tau^*_{(r,s)}$ = $\min_{\tau}|AR(EpiFast(\tau, P_E, P_I, I_V, N_{I(r,s)}, Network_{(r,s)}))-ar_{(r,s)}|$\;
        $\tau^* = \tau^* \cup \tau^*_{(r,s)}$}
        }
\caption{Calibrating disease model parameter $\tau$}
\label{method:calibrate}
\end{algorithm}

Secondly, for $\tau$ and $N_I$, we fit the collected samples to several distributions including normal, uniform. 
Then we run KS-test to choose a best well fit distribution (refer to Appendix~\ref{sec:appendix} for more details). 
For $I_V$, we assume the six vaccination schedules follow a discrete uniform distribution.  
In this way, a region-specific parameter space $\mathcal{P}$ is constructed. 

We first implement our TDEFSI framework without considering interventions in the simulations. Then we add $I_V$ to $\mathcal{P}$ to generate more realistic synthetic training data. This will improve the forecasting performance of TDEFSI.
We will discuss the impact of including $I_V$ on the forecasting performance of TDEFSI in Section~\ref{subsec:interv}.

\subsection{Training Dataset from Simulations}\label{subsec:sim}
For each simulation run, a specific parameter setting is sampled from $\mathcal{P}$, and the simulator is called to generate daily individual health states. 
These individual health states are aggregated to get state and county level weekly incidences, called \textit{synthetic epicurves}. 
Week 1 in the synthetic epicurve corresponds to $sw(1)$ of a flu season.
Large volumes of high-resolution synthetic data are generated by repeating the sampling and simulating process. 
Let us denote all simulated epicurves by $\mathit{\Omega}=\{(\mathbf{y}_{(i)},\mathbf{y}^{\mathcal{D}}_{(i)})\in \mathbb{R}^{\ell \times (K+1)}|i=1,2,\cdots,r\}$, where $\ell$ is the length of an epicurve (number of weeks), $K$ is the number of counties in the state, and $r$ is the total number of simulation runs. Algorithm~\ref{method:sim} describes the generating process. 

Compared with CDC surveillance data, the training dataset $\mathit{\Omega}$ is prominent in two aspects: ($i$) it includes high-resolution spatial dependencies between subregions; ($ii$) the large volume of synthetic training data reduces the possibility of overfitting when training a deep neural network model. Thus the trained model has better generalization ability.

\begin{algorithm}[t]
\SetAlgoNoLine
\KwIn{Simulator PS, and Parameter space $\mathcal{P}$.}
\KwOut{Simulated epicurves $\mathit{\Omega}=\{(\mathbf{y}_{(i)},\mathbf{y}^{\mathcal{D}}_{(i)})|i=1,2,\cdots,r\}$.}
$\mathit{\Omega} = \emptyset$\;
\For{$i = 1$ to $r$}{ 
		$P$ = Sample($\mathcal{P}$)\;
        $(\mathbf{y}_{(i)},\mathbf{y}^{\mathcal{D}}_{(i)})$ = PS($P$)\;
        $\mathit{\Omega} = \mathit{\Omega} \cup (\mathbf{y}_{(i)},\mathbf{y}^{\mathcal{D}}_{(i)})$}
\caption{Generating Training Dataset for TDEFSI}
\label{method:sim}
\end{algorithm}

\begin{figure}[t]
\centering
\includegraphics[width=0.6\columnwidth]{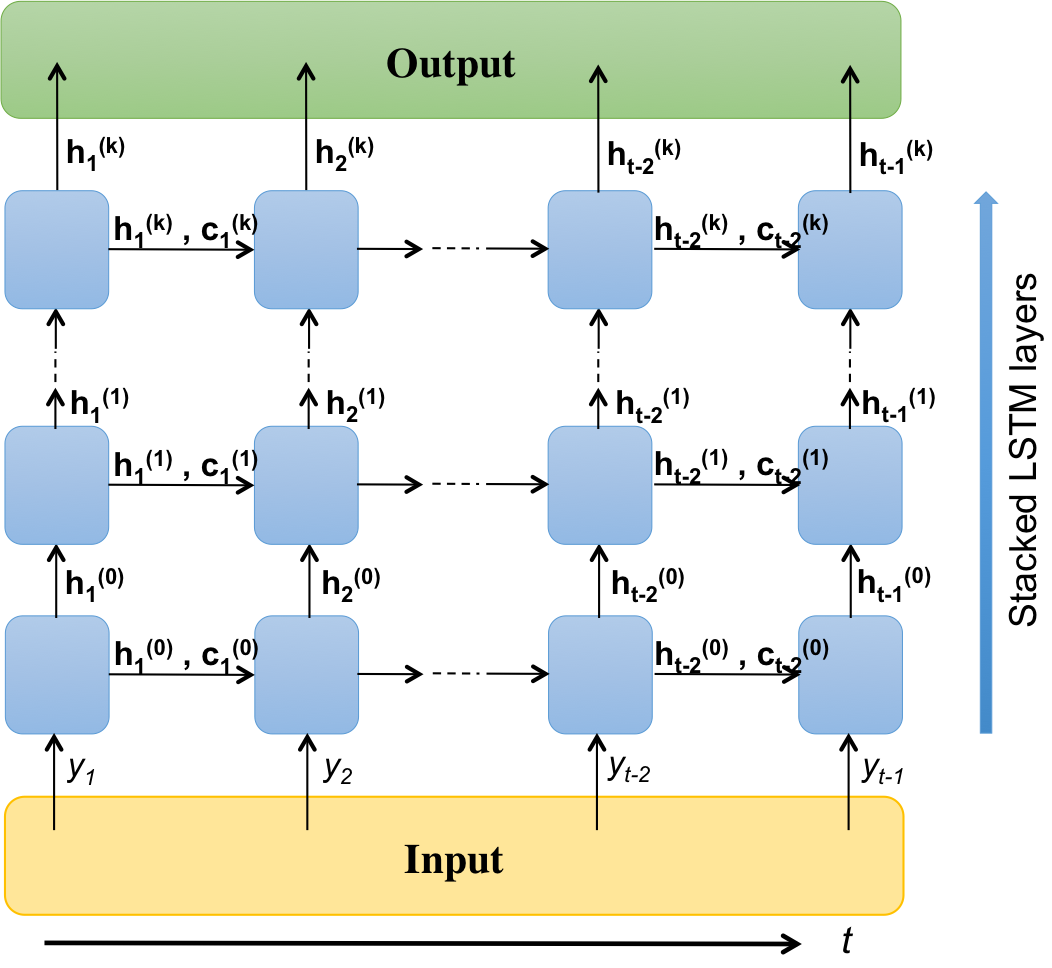}
  \caption{Unrolled k-stacked LSTM layers. Each LSTM layer consists of a sequence of cells. The number of cells depends on the number of input time points. In this figure, the input is a time series of $y_1,...,y_{t-1}$, the output comprises all the cell outputs $\mathbf{h}^{(k)}$ from the last layer $k$ ("last" depth-wise, not time-wise). Each LSTM layer consists of $t-1$ cells. 
  In the first LSTM layer, a cell will work as described in~\ref{equ:cell}, e.g. cell $2$ takes $y_{1}$, cell state $\mathbf{c}_{1}^{(0)}$ and cell output $\mathbf{h}_{1}^{(0)}$ from the previous cell $1$ as inputs, then outputs $(\mathbf{c}_{2}^{(0)}, \mathbf{h}_{2}^{(0)})$ so you could feed them into next cell and feed $\mathbf{h}_{2}^{(0)}$ into next layer. 
  The first LSTM layer take $y_1,...,y_{t-1}$ as the input, the second layer take $\mathbf{h}_1^{(0)},...,\mathbf{h}_{t-1}^{(0)}$ as the input, and rest of the layers behave in the same manner. }
  \label{fig:lstm}
\end{figure}

\subsection{TDEFSI: A Deep Neural Network Model}\label{subsec:defsi-nn}
The Long Short Term Memory (LSTM) network~\cite{hochreiter1997} is adopted in our neural network architecture to capture the inherent temporal dependency in the weekly incidence data. 
Figure~\ref{fig:lstm} shows unrolled k-stacked LSTM layers. 
Each LSTM layer consists of a sequence of cells. The number of cells depends on the number of input time points. In this figure, the input is a time series of $y_1,...,y_{t-1}$, the output comprises all the cell outputs $\mathbf{h}^{(k)}$ from the last layer $k$ ("last" depth-wise, not time-wise). Each LSTM layer consists of $t-1$ cells.
In the first LSTM layer (layer 0), a cell will work as described in~\ref{equ:cell}, e.g. cell $2$ takes $y_{1}$, cell state $\mathbf{c}_{1}^{(0)}$ and cell output $\mathbf{h}_{1}^{(0)}$ from the previous cell $1$ as inputs, then outputs $(\mathbf{c}_{2}^{(0)}, \mathbf{h}_{2}^{(0)})$ so you could feed them into the next cell and feed $\mathbf{h}_{2}^{(0)}$ into the next layer (layer 1). 
The first LSTM layer takes $y_1,...,y_{t-1}$ as the input, the second layer takes $\mathbf{h}_1^{(0)},...,\mathbf{h}_{t-1}^{(0)}$ as the input, and the rest of the layers behave in the same manner.

Let $H^{(i)}, 0 \leq i \leq k$ be the dimension of the hidden state in layer $i$. For the first layer, assume the input of the current cell is $y_{t-1}$. Then the computation within the cell is described mathematically as:

\begin{equation}
\label{equ:cell}
\begin{aligned}
&\mathbf{i}_{t-1}^{(0)} = \sigma(\mathbf{W}_i^{(0)} \cdot y_{t-1} + \mathbf{U}_i^{(0)} \cdot \mathbf{h}_{t-2}^{(0)} + \mathbf{b}_i^{(0)}) \in \mathbb{R}^{H^{(0)}}\\
&\mathbf{f}_{t-1}^{(0)} = \sigma(\mathbf{W}_f^{(0)} \cdot y_{t-1} + \mathbf{U}_f^{(0)} \cdot \mathbf{h}_{t-2}^{(0)} + \mathbf{b}_f^{(0)}) \in \mathbb{R}^{H^{(0)}}\\
&\mathbf{o}_{t-1}^{(0)} = \sigma(\mathbf{W}_o^{(0)} \cdot y_{t-1} + \mathbf{U}_o^{(0)} \cdot \mathbf{h}_{t-2}^{(0)} + \mathbf{b}_o^{(0)}) \in \mathbb{R}^{H^{(0)}}\\
&\widetilde{\mathbf{C}}_{t-1}^{(0)} = tanh(\mathbf{W}_C^{(0)} \cdot y_{t-1} + \mathbf{U}_C^{(0)} \cdot \mathbf{h}_{t-2}^{(0)} + \mathbf{b}_C^{(0)}) \in \mathbb{R}^{H^{(0)}}\\
&\mathbf{C}_{t-1}^{(0)} = \mathbf{f}_{t-1}^{(0)} \circ \mathbf{C}_{t-2}^{(0)} + \mathbf{i}_{t-1}^{(0)} \circ \widetilde{\mathbf{C}}_{t-1}^{(0)} \in \mathbb{R}^{H^{(0)}}\\
&\mathbf{h}_{t-1}^{(0)} = \mathbf{o}_{t-1}^{(0)} \circ \mathbf{C}_{t-1}^{(0)} \in \mathbb{R}^{H^{(0)}}
\end{aligned}
\end{equation}
where $\sigma$ and $tanh$ are sigmoid and tanh activation functions. $\mathbf{W} \in \mathbb{R}^{H^{(0)}}, \mathbf{U} \in \mathbb{R}^{H^{(0)} \times H^{(0)}}$, and $\mathbf{b} \in \mathbb{R}^{H^{(0)}}$ are learned weights and bias.
 $\mathbf{C}_{t-2}^{(0)}, \mathbf{h}_{t-2}^{(0)}$ are the cell state and output of the previous cell. Operator $\circ$ denotes element wise product (Hadamard product). The cell computation is similar in the layer $i$, but with $y_{t-1}$ being replaced by $\mathbf{h}_{t-1}^{(i-1)} \in \mathbb{R}^{H^{(i-1)}}$, and $\mathbf{W} \in \mathbb{R}^{H^{(i)} \times H^{(i-1)}}$.

\begin{figure}[t]
\centering
\includegraphics[width=0.6\textwidth]{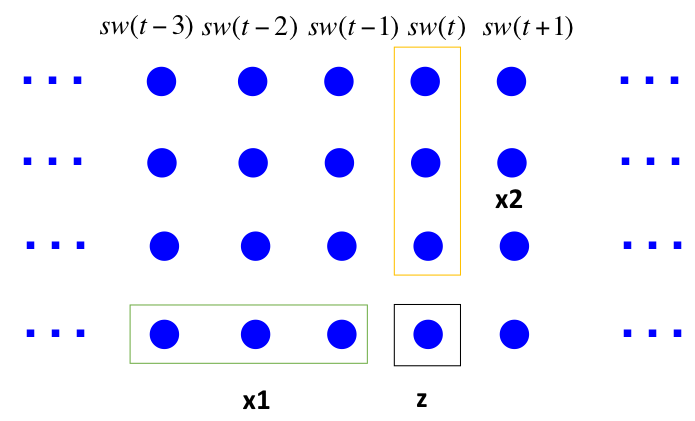}
  \caption{Within-season and between-season observations as the input for the TDEFSI neural network model. In this graph, there are four flu seasons (rows). Nodes in each row denote weekly ILI incidence in each season, which are ordered by $sw$. For a target week $sw(t)$ (black square), the model observes two kinds of information: ($i$) within-season observations $\mathbf{x1}$ - the ILI incidence from the previous weeks back from week $sw(t)$ (green rectangular); ($ii$) between-season observations $\mathbf{x2}$ - the historical ILI incidence from similar weeks of the past seasons (yellow rectangular). $\mathbf{z}$ is the target week of ILI forecasting. $\mathbf{x1}$ and $\mathbf{x2}$ are state level ILI, while $\mathbf{z}$ includes state and county level ILI.}
  \label{fig:inputs}
\end{figure}

In traditional time series models, ILI incidences of the previous few weeks are used as the observations for the prediction of the current week.
In TDEFSI, we use two kinds of observations: 
($i$) \textbf{\textit{Within-season observations}}, denoted as $\mathbf{x1}=\langle y_{t-a},\cdots,y_{t-1}\rangle$, are ILI incidence from previous $a$ weeks which are back from time step $t$. 
($ii$) \textbf{\textit{Between-season observations}}, denoted as $\mathbf{x2}=\langle y_{t-\ell*b},\cdots,y_{t-\ell*1}\rangle$, are ILI incidences of the same $sw$ from the past $b$ seasons. They are used as the surrogate information to improve forecasting performance.
As shown in Figure~\ref{fig:inputs}, for example, there are 4 seasons ordered by $sw$. The within-season observations are ILI incidence of previous $a=3$ weeks in current season. The between-season observations are ILI incidence of the same $sw(t)$ from the past $b=3$ seasons.

\begin{figure}[t]
\centering
\includegraphics[width=0.6\textwidth]{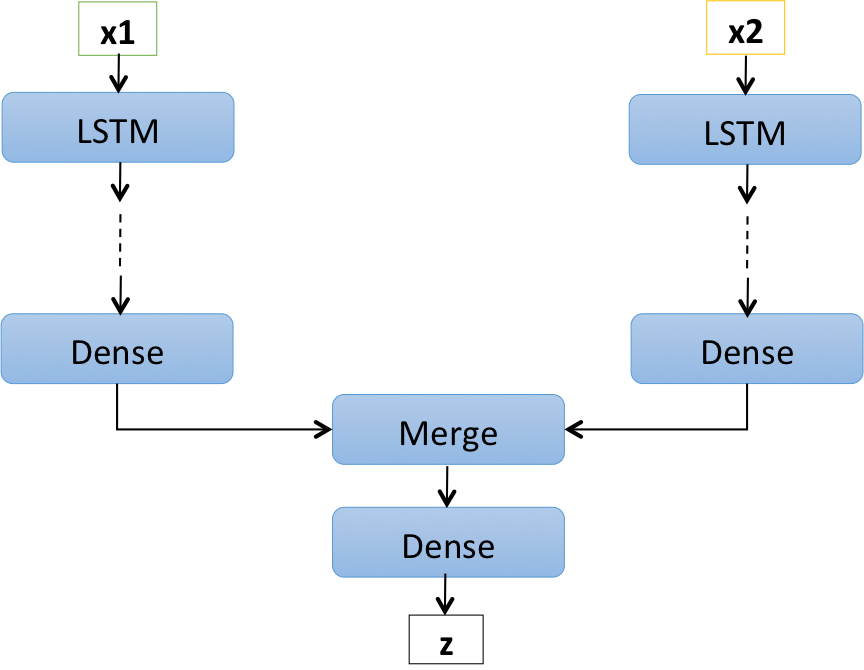}
  \caption{TDEFSI neural network architecture. This architecture consists of two branches. The left branch consists of stacked LSTM layers that encodes state level within-season observations $\mathbf{x1}$, and the right branch consists of stacked LSTM layers that encodes state level  between-season observations $\mathbf{x2}$. A merge layer is added to combine two branches and the output $\mathbf{z}$ is the state and county level predictions.}
  \label{fig:structure}
\end{figure}

In TDEFSI model, we design a two-branch LSTM based deep neural network model to capture temporal dynamics of within-season and between-season observations. 
As shown in Figure~\ref{fig:structure}, the left branch consists of stacked LSTM layers that encode within-season observations $\mathbf{x1}=\langle y_{t-a},\cdots,y_{t-1}\rangle$. The right branch is also LSTM based and encodes between-season observations $\mathbf{x2}=\langle y_{t-\ell*b},\cdots,y_{t-\ell*1}\rangle$. A merge layer is added to combine the outputs of two branches. The final output is $\hat{\mathbf{z}}_t$ which consists of state level and county level predictions (as defined in Section~\ref{sec:problem}).

In the left branch, the output of the Dense layer is:
\begin{equation}
\label{equ:leftbranch}
\mathbf{O}_l = \psi_l(\mathbf{w}_l \cdot \mathbf{h}_{t-1}^{(k_l)} + \mathbf{b}_l) \in \mathbb{R}^{H}
\end{equation}
where $k_l$ is the number of LSTM layers in the left branch, $H$ is the dimension of output of the left branch, $\mathbf{w}_l \in \mathbb{R}^{H \times H^{(k_l)}}$ and $\mathbf{b}_l \in \mathbb{R}^{H}$, $\psi_l$ is the activation function.

Similarly, the output of the Dense layer in the right branch is:
\begin{equation}
\label{equ:rightbranch}
\mathbf{O}_r = \psi_r(\mathbf{w}_r \cdot \mathbf{h}_{t-1}^{(k_r)} + \mathbf{b}_r) \in \mathbb{R}^{H}
\end{equation}
where $k_r$ is the number of LSTM layers in the right branch, $H$ is the dimension of output of the right branch, $\mathbf{w}_r \in \mathbb{R}^{H \times H^{(k_r)}}$ and $\mathbf{b}_r \in \mathbb{R}^{H}$, $\psi_r$ is the activation function.

The merge layer combines the output from two branches by addition, denoted as:
\begin{equation}
\label{equ:mergebranch}
\hat{\mathbf{z}}_t = \psi (\mathbf{w}[\mathbf{O}_l \oplus \mathbf{O}_r] + \mathbf{b}) \in \mathbb{R}^{K+1}
\end{equation}
where $\mathbf{w} \in \mathbb{R}^{(K+1) \times H}$, $\mathbf{b} \in \mathbb{R}^{K+1}$, $\psi$ is the activation function, and $\oplus$ denotes the element-wise addition.

This LSTM based deep neural network model is able to connect historical ILI incidence information to the current prediction. It also allows long-term dependency learning without suffering the gradient vanishing problem.
The number of LSTM layers is a hyperparameter that we tuned by grid searching. 

We are interested in a predictor $f$, which predicts the current week's state level and county level incidence $\mathbf{z}_t$ based on the previous $a$ weeks of within-season state level ILI incidence $\mathbf{x1}$ and the previous $b$ seasons of between-season state level ILI incidence $\mathbf{x2}$:
\begin{equation}
\label{equ:predictor}
\hat{\mathbf{z}}_t = f([\mathbf{x1},\mathbf{x2}]_t, \theta)
\end{equation}
where $\theta$ denotes parameters of the predictor, $\hat{\mathbf{z}}_t$ denotes the prediction of $\mathbf{z}_t$.
Note that {\bf the output of $f$ is always one week ahead forecast} in our model.

The optimization objective is:
\begin{equation}
\label{equ:defsi}
\begin{aligned}
&\min_{\theta} \mathcal{L}(\theta) = \sum\limits_{t} \lVert \mathbf{z}_t - f([\mathbf{x1},\mathbf{x2}]_t,\theta)\rVert^2_2 + \mu\phi(\hat{\mathbf{z}}_t) + \lambda\delta(\hat{\mathbf{z}}_t), 
\end{aligned}
\end{equation}
where $\phi(\hat{\mathbf{z}}_t)$ is an activity regularizer added to the outputs for spatial consistency constraint $\hat{y}_t=\sum \limits_{C \in \mathcal{D}} \hat{y}^C_t$:
\begin{equation}
\label{equ:regularizer:phi}
\phi(\hat{\mathbf{z}}_t) = \left|\hat{y}_t - \sum \limits_{C \in \mathcal{D}} \hat{y}^C_t \right|,
\end{equation}
and $\delta(\hat{\mathbf{z}}_t)$ is an activity regularizer added to the outputs for non-negative consistency constraint $\hat{y}_t, \hat{y}^C_t \geq 0$:
\begin{equation}
\label{equ:regularizer:delta}
\delta(\hat{\mathbf{z}}_t) = \left|\frac{1}{K+1}\sum \max(-\hat{\mathbf{z}}_t, \mathbf{0})\right|,
\end{equation}
$\mu, \lambda$ are two pre-specified hyperparameters, $\min(\hat{\mathbf{z}}_t, \mathbf{0})$ returns element-wise minimum value, $K$ is the number of counties in the state, $\delta(\hat{\mathbf{z}}_t)$ returns the absolute mean of element-wise minimum values. 
The Adam optimization algorithm~\cite{kingma2014adam} is used to learn $\theta$.
How the activity regularizers affect the model performance will be discussed in Section~\ref{subsec:consistency}.

\textbf{Variants of TDEFSI}
The two-branch neural network architecture has multiple variants:
($i$) \textbf{\textit{TDEFSI}}: Two-branch neural network as shown in Figure~\ref{fig:structure}.
($ii$) \textbf{\textit{TDEFSI-LONLY}}: Only the left branch is used to take within-season observations.
($iii$) \textbf{\textit{TDEFSI-RDENSE}}: The left branch comprises of stacked LSTM layers, while the right branch only uses Dense layers, which means that the model does not care about the temporal relationship between between-season data points.
We will discuss the results of different variants in Section~\ref{sec:experiment}.

\textbf{Training and forecasting}
In the training process, we use synthetic training data $\mathit{\Omega}$ to train the TDEFSI models. The historical surveillance data is only used for constructing the disease model parameter space $\mathcal{P}$. 
In the predicting step, the trained model takes state level surveillance as input and makes one week ahead forecasts at both state and county levels. 
TDEFSI models are trained once before the target flu season starts, then can be used for forecasting throughout the season.

\textbf{Multi-step forecasting}
In practical situations, we are interested in making predictions for several weeks ahead using iterative method. 
In TDEFSI, 
the left branch of the model appends the most recent state level prediction to the input for predicting the target of the next week, and the right branch uses the state level ILI incidences from the past seasons with $sw$ equal to the next week number.

%% file: sections/experiment.tex
\input{tables/exp-setting-summary.tex}
\input{tables/exp-result-summary.tex}

\section{Experiments} \label{sec:experiment}
In this section, we will describe datasets, comparison methods, experiment setup, and evaluation metrics. A brief summary of TDEFSI settings is shown in Figure~\ref{fig:summaryofsetting}. And we present results of performance analysis on both simulated testing data and real ILI testing data, and conduct sensitivity analysis on physical consistency constraints and vaccination-based interventions. We also use a case study to demonstrate the capability of TDEFSI model to provide uncertainty in predictions. A brief summary of the experiment results is shown in Figure~\ref{fig:summaryofresult}. In all experiments the models are trained and tested for each state independent of other states.

\input{sections/subsections/exp-dataset.tex}
\input{sections/subsections/exp-baselines.tex}
\input{sections/subsections/exp-setup.tex}

\input{sections/subsections/exp-metrics.tex}

\input{sections/subsections/exp-EDA.tex}
\input{sections/subsections/exp-performance-simulated.tex}
\input{sections/subsections/exp-performance-real-new.tex}
\input{sections/subsections/exp-consistency.tex}
\input{sections/subsections/exp-vaccine-analysis.tex}
\input{sections/subsections/exp-uncertainty.tex}

%% file: tables/exp-setting-summary.tex
\begin{figure}[!t]
    \centering
    \begin{tcolorbox}[standard jigsaw,opacityback=0]
    \begin{itemize}
        \item \textbf{Real Dataset}\\
        Weekly CDC state level ILINet-reported case counts for all states in the USA (2010-2018) (total 397 data points per state)
        \begin{itemize}
        \item \textit{real-training:} the beginning $80\%$ of season 2010-2011 to 2015-2016 (251 data points per state)
        \item \textit{real-validating:} the last $20\%$ of season 2010-2011 to 2015-2016 (63 data points per state)
        \item \textit{real-testing:} season 2016-2017 to 2017-2018 (83 data points per state)
        \end{itemize}
        Weekly county level ILI Lab tested flu positive counts for NJ (2016-2018)
        \begin{itemize}
        \item \textit{County level real-evaluating:} 64 data points per county of NJ
        \end{itemize}
        \item \textbf{Simulated Dataset}\\
        VA: 1000 epicurves in vaccine-case and 1000 epicurves in base-case.\\
        NJ: 1000 epicurves in vaccine-case and 1000 epicurves in base-case.
        \begin{itemize}
        \item \textit{sim-training:} $80\%$ of 1000 epicurves 
        \item \textit{sim-validating:} $15\%$ of 1000 epicurves
        \item \textit{sim-testing:} $5\%$ of 1000 epicurves
        \end{itemize}
        \item \textbf{Disease Model Simulator}\\
        SEIR agent-based model -- EpiFast
        \item \textbf{Disease Model Parameter Space}\\
        $\mathcal{P}(p_E,p_I,\tau,N_I,I_V)$, the learned distribution for $\mathcal{P}$ is shown in Table~\ref{tab:distr}.
        \item \textbf{TDEFSI Neural Network Models}\\
        The architecture (e.g. the number of layers or hidden units) for TDEFSI and its variants is described in Section~\ref{subsec:setup}. The input dimension $a = 52$ and $b=5$, spatial and non-negative coefficients are set with $(\mu,\lambda)_{VA} = (0.1,0.1)$, $(\mu,\lambda)_{NJ} = (1,0.01)$. TDEFSI models are trained with vaccine-case sim-training dataset. We choose the final model by grid searching using sim-validating dataset.  Adam optimizer with all default values are used. In the training process, the best models are selected by early stopping when the validation accuracy does not increase for 50 consecutive epochs, and the maximum epoch number is 300.
        \item \textbf{Prediction Target}\\
        ILINet-report case counts forecasting with $horizon =$ $\{1,$ $2,$ $3,$ $4,$ $5\}$.
    \end{itemize}
    \end{tcolorbox}
    \caption{Brief summary of TDEFSI settings.}
    \label{fig:summaryofsetting}
\end{figure}

%% file: tables/exp-result-summary.tex
\begin{figure}[!t]
    \centering
    \begin{tcolorbox}[standard jigsaw,opacityback=0]
        \begin{itemize}
            \item \textbf{Exploratory Analysis of Spatial Dynamics of NJ County Level Dataset}\\
            The spreading process shows spatial heterogeneity over the counties, which is related with the population size and the commute flow.
            \item \textbf{TDEFSI Performance on a Simulated Testing Dataset}\\
            The best models at the state and county levels are the models with $a = 52$.
            \item \textbf{TDEFSI Performance on a Real Seasonal ILI Testing Dataset}
            \begin{itemize}
                \item \textit{\textbf{Performance of Flat-resolution Forecasting.}}\\
                TDEFSI and its variants achieve comparable or better performance than baselines evaluted on real-testing dataset. 
                \item \textit{\textbf{Performance of High-resolution Forecasting.}}\\
                TDEFSI and its variants outperform baselines evaluated on county level real-evaluating dataset. 
                \item \textit{\textbf{Overall performance.}}\\
                TDEFSI and its variants achieve better forecasting performance. Among the three proposed models, TDEFSI and TDEFSI-RDENSE outperform TDEFSI-LONLY.
            \end{itemize}
            \item \textbf{Physical Consistency Constraints Analysis}\\
            The spatial and non-negative consistency constraints with proper $\lambda$ and $\mu$ help improve the forecasting performance. However, the optimal $\lambda$ and $\mu$ vary between different regions. 
            \item \textbf{Vaccination-based Interventions}\\
            The model learned using simulations that incorporate vaccinations yields better generalizability  to unseen surveillance data. 
            \item \textbf{Prediction Uncertainty Estimation}\\
            The proposed model provides a natural way to estimate the prediction uncertainty using MC Dropout technique.
        \end{itemize}
    \end{tcolorbox}
    \caption{Brief summary of our experimental analysis.}
    \label{fig:summaryofresult}
\end{figure}

%% file: sections/subsections/exp-dataset.tex
\subsection{Datasets} \label{subsec:dataset}
\subsubsection{Real dataset} \label{subsubsec:realdataset}
\textbf{CDC ILI incidence}~\cite{cdcfluview}: The CDC surveillance data used in the experiments is the weekly ILI incidence at state level from 2010 $ew(40)$ to 2018 $ew(18)$. Note that it may be revised continuously until the end of a flu season. We use the finalized data in this paper.
\textbf{ILI Lab tested flu positive counts of New Jersey}~\cite{nj2018}: To evaluate the county level forecasting performance, we collect state level and county level ILI Lab tested flu positive counts of season 2016-2017 and 2017-2018 in NJ. The data is available from $ew(40)$ to the next year's $ew(20)$. We use it as the ground truth when evaluating county level forecasting.
\textbf{Google data}~\cite{gcorr,ght}: The Google correlate terms (keyword: influenza) of each state are queried; we choose the top 100 terms. Then the Google Health Trends of each correlated term for each state is collected and aggregated weekly from 2010 $ew(40)$ to 2018 $ew(18)$. 
\textbf{Weather data}~\cite{cdo}: We download daily weather data (including max temperature, min temperature, precipitation) from Climate Data Online (CDO) for each state and compute weekly data as the average of daily data from 2010 $ew(40)$ to 2018 $ew(18)$. Google data and weather data are used as surrogate information in comparison methods (described in Section~\ref{subsec:baselines}).

We divide the data into: \textit{real-training:} the beginning $80\%$ of season 2010-2011 to season 2015-2016 (251 data points per state). \textit{real-validating:} the last $20\%$ of season 2010-2011 to season 2015-2016 (63 data points per state). \textit{real-testing:} season 2016-2017 to season 2017-2018 (83 data points per state). \textit{County level real-evaluating:} county level ILI lab tested flu positive counts for NJ (64 data points per county of NJ).
For TDEFSI models, we use the training dataset to learn disease parameter space, while for baselines, we use training dataset to train the model directly and use validating dataset to validate and choose the final models. Testing and county level evaluating datasets are used for all methods to evaluate their performance. And the final result of each method is the average value of 10 trials.

\subsubsection{Simulated dataset} \label{subsubsec:syndataset}
For each state, we generate $1000$ simulated curves of weekly ILI incidence at both state level and county level. Of each curve, the first week $sw(1)$ corresponds to epi-week 40 $ew(40)$ of real seasonal curves. We divide the data into: \textit{sim-training:} $80\%$ of 1000 simulated curves. \textit{sim-validating:} $15\%$ of 1000 simulated curves. \textit{sim-testing:} $5\%$ of 1000 simulated curves.
The synthetic data is only used for training and validating of TDEFSI models. No baselines are applied for synthetic data. 

%% file: sections/subsections/exp-baselines.tex
\subsection{Methods used for our comparative analysis} \label{subsec:baselines}
Our method is compared with 5 state-of-the-art ANN methods, statistical methods, and causal methods.
They are:
\begin{itemize}
\item
\textit{LSTM} (CDC data)~\cite{hochreiter1997} and \textit{AdapLSTM} (CDC + weather data)~\cite{venna2019novel} representing artificial neural network methods; 
\item
\textit{SARIMA} (CDC Data)~\cite{benjamin2003generalized} and \textit{ARGO} (CDC + Google data)~\cite{yang2015accurate} representing statistical methods; and
\item
\textit{EpiFast}~\cite{beckman2014} representing causal models. 
\end{itemize}
AdapLSTM, LSTM, ARGO, and SARIMA can make flat-resolution forecasting directly from the model, then flat-resolution forecasts can be turned into high-resolution forecasts by multiplying by county level population proportions. EpiFast is applied for both flat-resolution and high-resolution forecasting directly. 

%% file: sections/subsections/exp-setup.tex
\subsection{Experiment Setup}\label{subsec:setup}
In this section, we describe the experiment settings, including simulation setting and TDEFSI model setting. Note that we conduct the experiments on two states of the USA i.e. VA and NJ. State level forecasting performance will be evaluated on both VA and NJ, while county level forecasting performance is evaluated on NJ only due to the limitation on the availability of high-resolution observations. 

\input{tables/exp-hyper.tex}

\textbf{Disease model settings for generating simulated training data.} The simulation parameter settings are listed in Section~\ref{subsec:fit} Table~\ref{tab:distr}. The length of a simulated epicurve is set to $\ell=52$, and the total runs of simulations is $r=1000$. We adopt EpiFast as the simulator, PS='EpiFast'. More details on parameter space learning are described in Section~\ref{subsec:fit}.

\textbf{TDEFSI model settings.} We set up the architectures for TDEFSI and its variants as follows:
\begin{itemize} 
\item
\textbf{\textit{TDEFSI}}: The left branch consists of two stacked LSTM layers, one dense layer; the right branch consists of one LSTM layer, one dense layer. $k_l=2$, $k_r=1$, $H^{(k_l)}=H^{(k_r)}=128$, $H=256$, $\psi_l,\psi_r, \psi$ are linear functions.
\item
\textbf{\textit{TDEFSI-LONLY}}: The left branch consists of two stacked LSTM layers, one dense layer and no right branch. $k_l=2$, $H^{(k_l)}=128$, $H=256$, $\psi_l, \psi$ are linear functions.
\item 
\textbf{\textit{TDEFSI-RDENSE}}: The left branch consists of two stacked LSTM layers, one dense layer; the right branch consists of one dense layer. $k_l=2$, $k_r=0$, $H^{(k_l)}=H^{(k_r)}=128$, $H=256$, $\psi_l,\psi_r, \psi$ are linear functions.
\end{itemize}
For all TDEFSI models, we set $a=52$, $b=5$, $(\mu,\lambda)_{VA} = (0.1,0.1)$, $(\mu,\lambda)_{NJ} = (1,0.01)$. We use Adam optimizer with all default values. We choose the final model using grid searching with sim-validating dataset.
The grid searching space is about 500 models, including $a(10,20,30,40,50)$, $b(5)$, $\mu(0,0.001,0.01,0.1,1)$, $\lambda(0,0.001,0.01,0.1,1)$, $k_l(1,2)$, $H(128,256)$.
In the training process, the best models are selected by early stopping when the validation accuracy does not increase for 50 consecutive epochs, and the maximum epoch number is 300. Unless explicitly noted, in our experiments, these hyperparameters are set with the values described above. 
The settings of comparison methods are elaborated in Appendix~\ref{subsec:baselinesettings}.

Our experiments are conducted on two testing datasets: 
($i$) synthetic testing dataset and ($ii$) real seasonal ILI dataset.

\textbf{Experimental setup for testing on simulated dataset.}
We make predictions for ten weeks ahead, i.e. $horizon =$ $\{1,$ $2,$ $3,$ $4,$ $5,$ $6,$ $7,$ $8,$ $9,$ $10\}$. Only TDEFSI is tested and analyzed using sim-testing dataset. No comparison methods are applied since there is no surrogate information corresponding to the simulated seasons. 

\textbf{Experimental setup for testing on real seasonal ILI dataset.}
In these experiments, we evaluate TDEFSI models and all comparison methods. The experiments are performed on two states: Virginia (VA) and New Jersey (NJ). The county level evaluation is conducted on NJ counties. 
For TDEFSI and its variants, the real-training dataset is used to estimate disease parameter space, while for all baselines, real-training and real-validating are used for training directly. 
The county level real-evaluating dataset is only used for evaluation of the performance of county level predictions. 
At each time step in the testing season, each model makes predictions up to five weeks ahead, i.e. $horizon = \{1,2,3,4,5\}$.

%% file: tables/exp-hyper.tex
\begin{table}
  \caption{Hyperparameters of TDEFSI Model and Their Values for Sensitivity Analysis. }
  \label{tab:hyper}
  \begin{tabular}{lll}
    \toprule
    Parameters&Description&Values\\
    \midrule
    $a$ & length of within-season observations & 10, 20, 30, 40, 52\\
    $b$ & length of between-season observations & 5\\
    $\lambda$ & coefficient of spatial regularizer & 0, 0.001, 0.01, 0.1, 1, 10, 100\\
    $\mu$ & coefficient of non-negative regularizer & 0, 0.001, 0.01, 0.1, 1, 10, 100\\
  \bottomrule
\end{tabular}
\begin{tablenotes}
  \small
  \item There are many hyperparameters in TDEFSI models, such as input dimension $a, b$, consistency coefficiency $\mu, \lambda$, number of hidden layers $k_r, k_l$, number of hidden units $H^{(k_l)}, H^{(k_r)}, H$, learning rate, training epoch, and so on. In our experiments, we choose the final model by using grid searching on the hyperparameters using sim-validating dataset. In the training process, the best models are selected by early stopping when the validation accuracy does not increase for 50 consecutive epochs, and the maximum epoch number is 300.   
\end{tablenotes}
\end{table}

%% file: sections/subsections/exp-metrics.tex
\subsection{Performance Metrics}
The metrics used to evaluate the forecasting performance are: \textit{root mean squared error (RMSE)}, \textit{mean absolute percentage error (MAPE)}, \textit{Pearson correlation (PCORR)}.

\begin{itemize}
\item \textbf{Root mean squared error ($\mathbf{RMSE}$)}: 
\begin{equation}
\label{equ:rmse}
RMSE = \sqrt{\frac{1}{n}\sum_{i=1}^{n}(y_i - \hat{y_i})^2}
\end{equation}

\item \textbf{Mean absolute percentage error ($\mathbf{MAPE}$)}:
\begin{equation}
\label{equ:mape}
MAPE = (\frac{1}{n}\sum_{i=1}^{n}|\frac{y_i - \hat{y_i}}{y_i+1}|)*100
\end{equation}
where the denominator is smoothed by 1 to avoid zero values.

\item \textbf{Pearson correlation ($\mathbf{PCORR}$)}: 
\begin{equation}
\label{equ:pcorr}
PCORR = \frac{cov(\mathbf{y},\mathbf{\hat{y}})}{\sigma_{\mathbf{y}}\sigma_{\mathbf{\hat{y}}}}
\end{equation}
where $cov(\mathbf{y},\mathbf{\hat{y}})$ is the covariance of $\mathbf{y}$ and $\mathbf{\hat{y}}$, and $\sigma$ is the standard deviation.
\end{itemize}

Among these metrics, RMSE and MAPE evaluate ILI incidence prediction accuracy, PCORR evaluates linear correlation between the true curve and the predicted curve.

%% file: sections/subsections/exp-EDA.tex
\begin{figure}[t]
\centering
\subfloat[Population of NJ counties, 2010]{\label{fig:njpop}\includegraphics[width=2.5in]{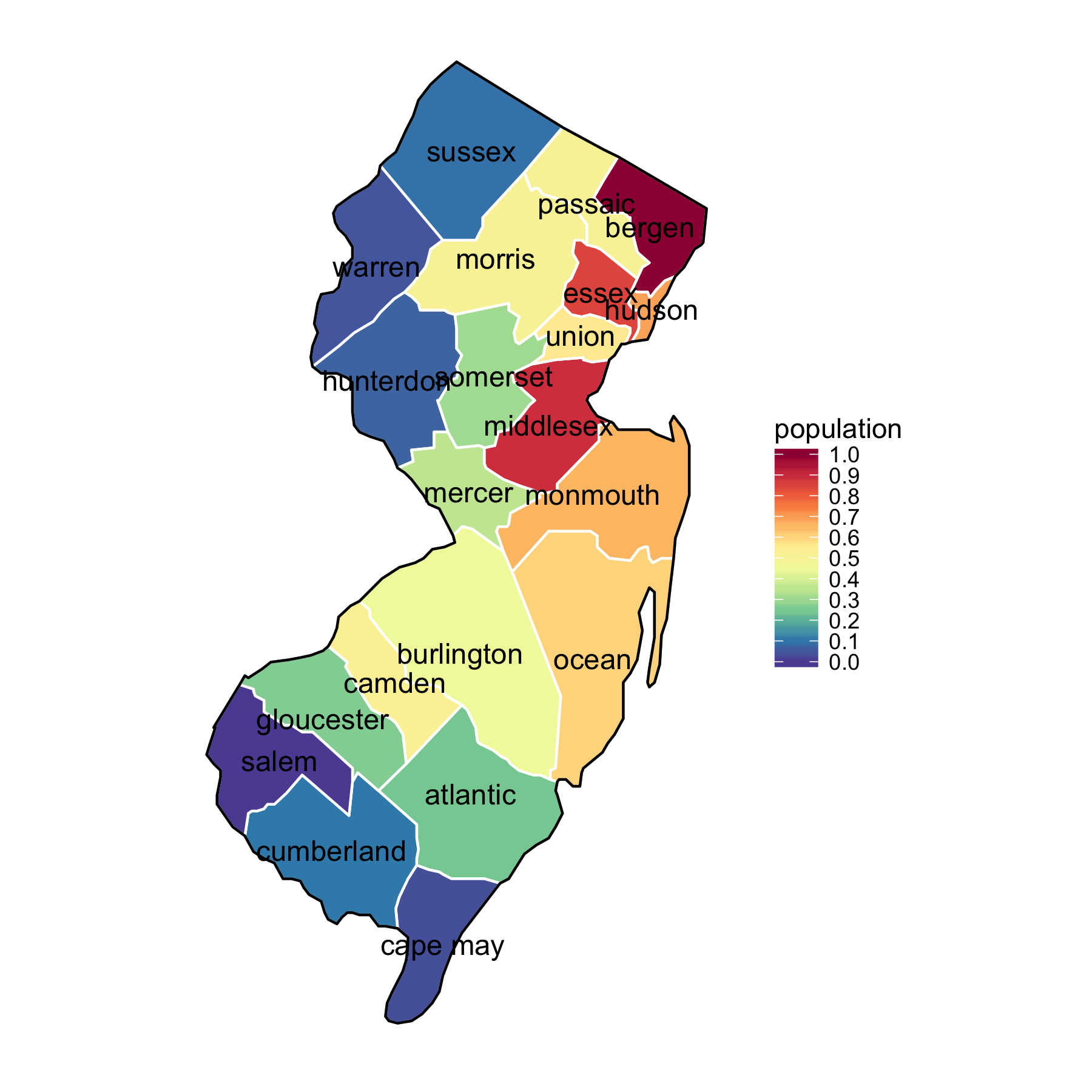}}
\hfil
\subfloat[Population density of NJ counties, 2010]{\label{fig:njden}\includegraphics[width=2.5in]{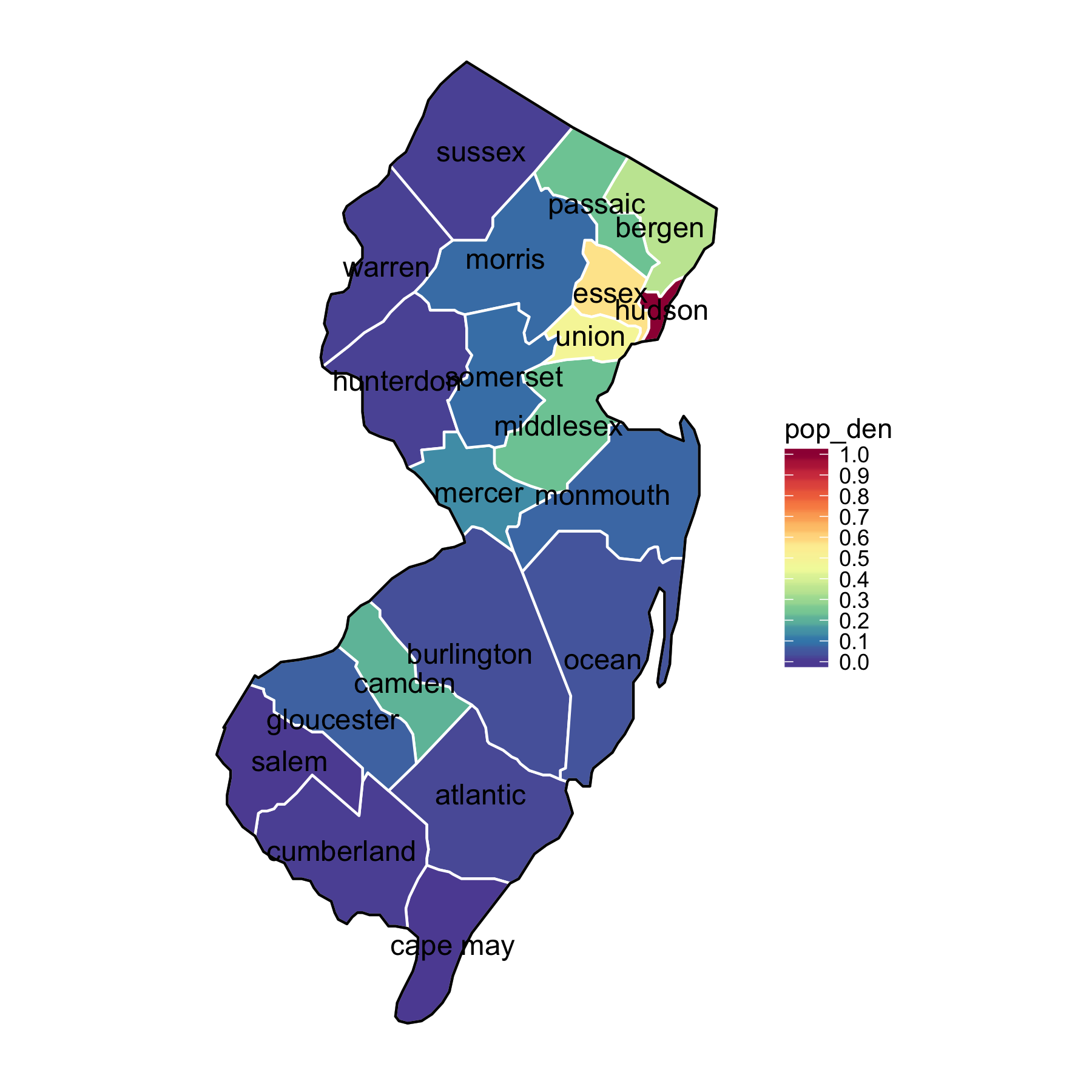}}
\hfil
\caption{Statistics of NJ counties~\cite{population2010}. (a) Population of NJ counties, 2020 (b) Population density of NJ counties, 2010. The population density is the population per square mile. The values shown in the map of both statistics are normalized by $(\frac{x-min}{max-min})$ so that the range is $[0,1]$. The counties located in the eastern NJ have large population size, and the counties around the northeastern area are of especially high population density.}
\label{fig:njstatistics}
\end{figure}

\begin{figure}[t]
\centering
\subfloat[Adjacency matrix of commute flows]{\label{fig:njcommute}\includegraphics[width=2.5in]{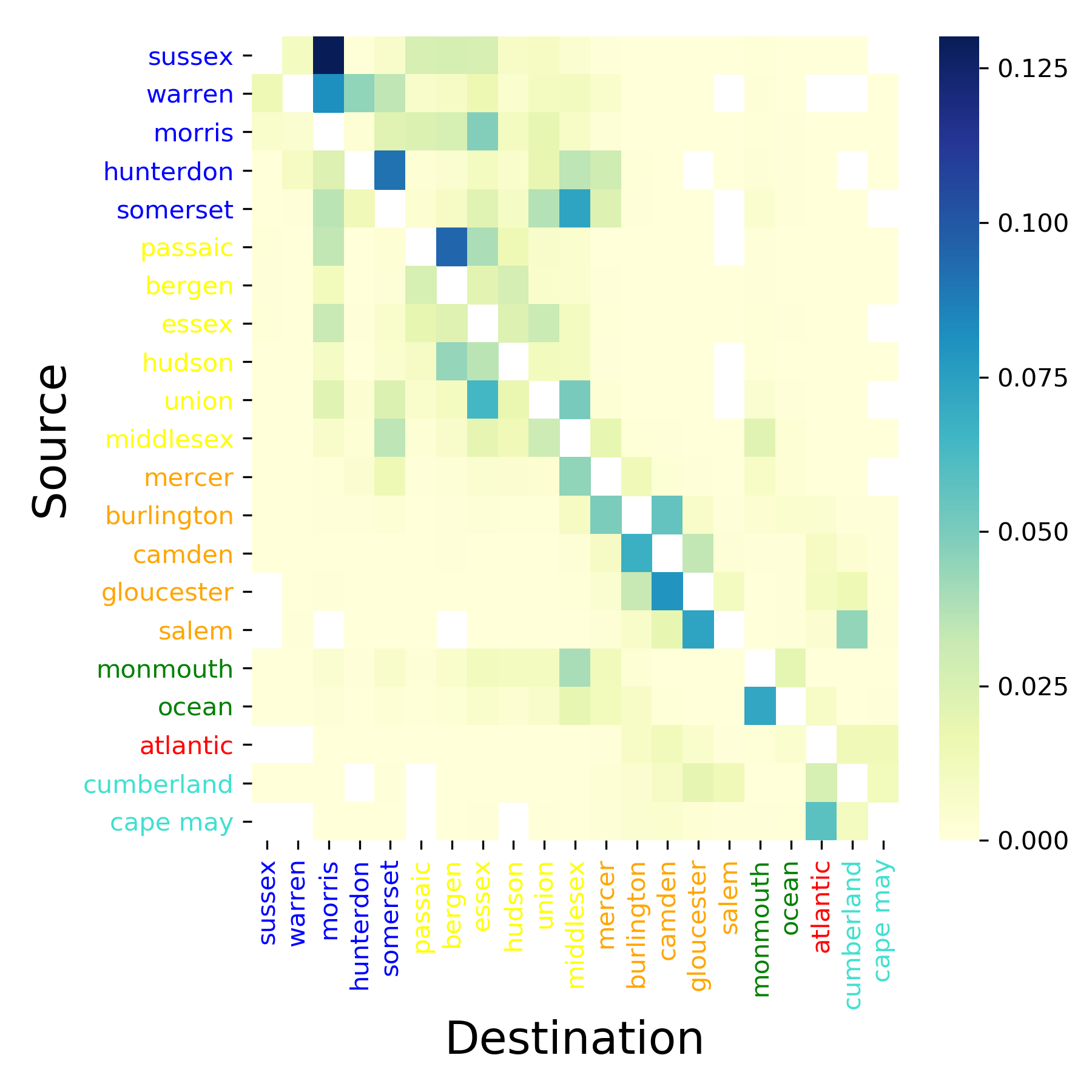}}
\hfil
\subfloat[New Jersey county regions map]{\label{fig:regionmap}\includegraphics[height=2.5in]{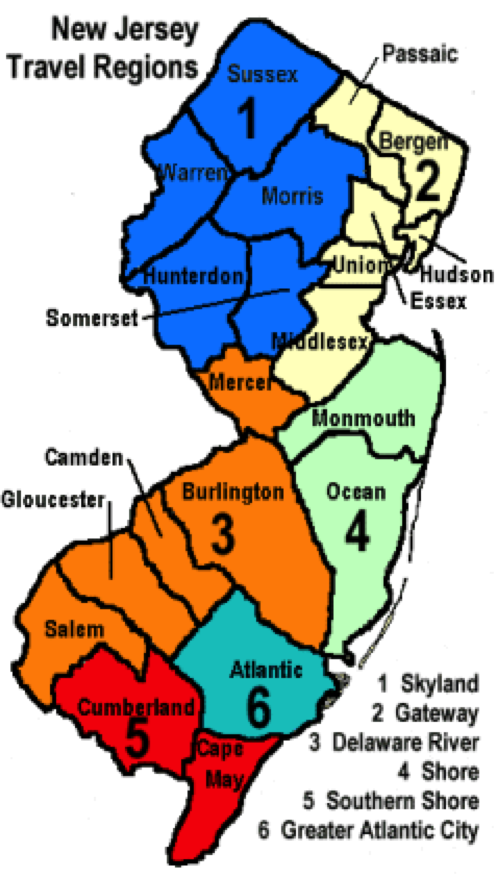}}
\hfil
\caption{(a) Adjacency matrix of commute flows with the counties of NJ arranged according to spatial neighborhood. We use the Regions of New Jersey as defined by the New Jersey State Department of Tourism (color of the labels match the map). Note that the visualization shows the normalized commute counts by county population size. A larger value means a larger commute flow between the two counties. The self loop flows are not shown, since they're an order of magnitude larger than the rest. (b) New Jersey county regions map (downloaded from~\cite{njmap}).}
\label{fig:commute}
\end{figure}

\subsection{Exploratory Analysis of Spatial Dynamics of NJ County Level Dataset}
The spatiotemporal spread of influenza in a state depends on the social demographic attributes (e.g. population density) of the counties as well as the individual behavior and movement between counties. In this subsection we explore the county demographic data and between county commute data using visualization, and discuss their association with the disease spread spatially over time.



In Figure~\ref{fig:njstatistics}, we show the statistics for NJ counties including population and population density (i.e. population per square mile). Values are normalized by using $(\frac{x-min}{max-min})$ so that the range is $[0,1]$. In general the counties located in northwestern NJ and southwestern NJ have small population and population density, while the counties concentrated in northeastern NJ have large ones.

From county-to-county commute counts data from the American Community Survey (ACS) 2009-2013~\cite{commute}, we extract commute counts of which both source and destination are NJ counties.
In Figure~\ref{fig:commute}, we show the adjacency matrix of commute flows with the counties of NJ arranged according to spatial neighborhood. The flow in the figure is the normalized commute counts by the population size of the source county. A larger value means a larger commute flow between the two counties. The figure shows larger commute flows between counties which are physically close to each other. Nevertheless, there is substantial flow between counties that are far away from each other --- this small-world like flow is a hall-mark of human mobility patterns.
During an epidemic, counties with large populations and high connectivity serve as hubs --- these counties often start the epidemic early and also aid the spread to other counties. 

\begin{figure}[t]
\centering
\includegraphics[width=0.8\textwidth]{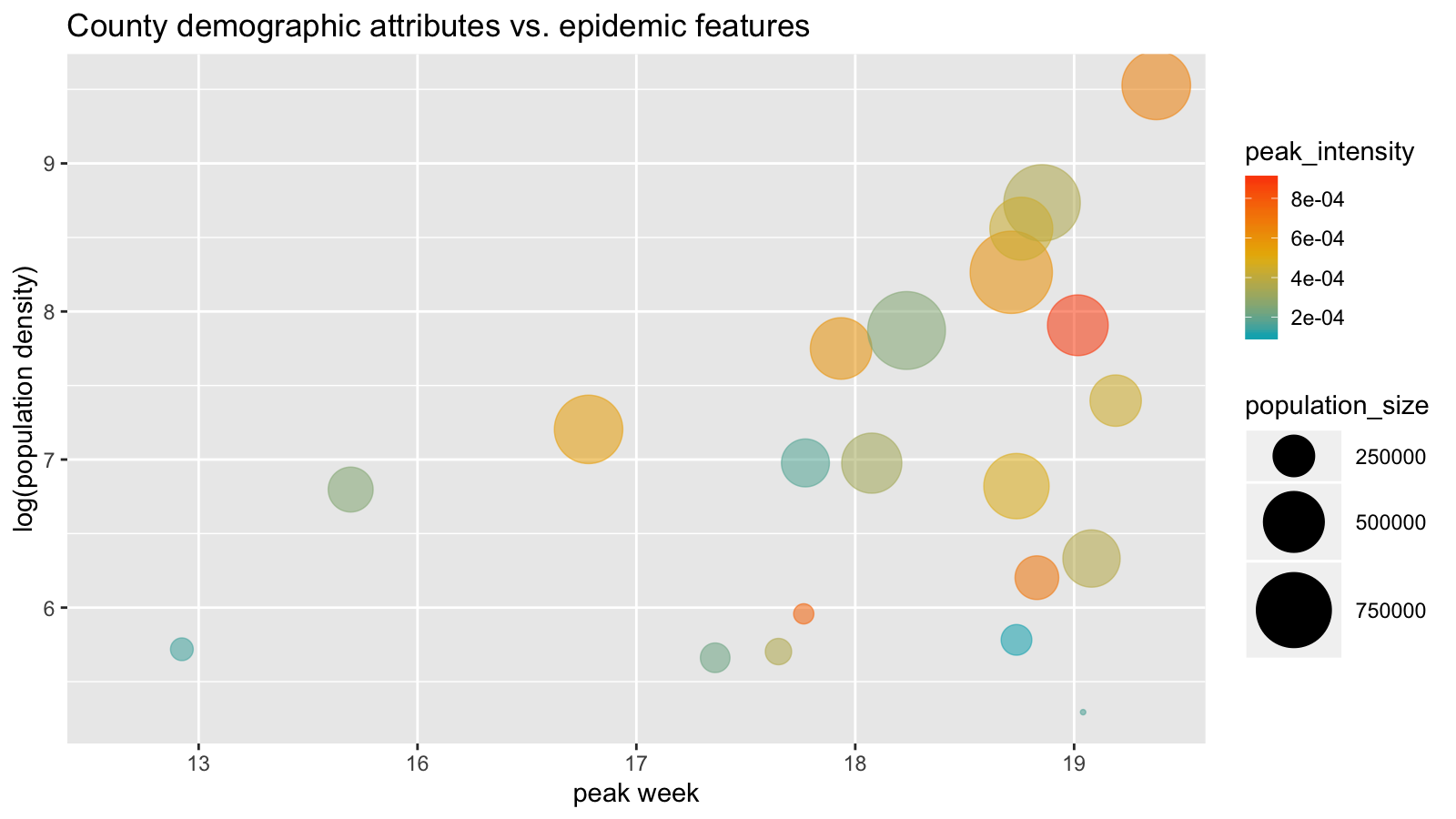}
  \caption{The correlation between the peak (including peak week and peak intensity) and the population density of NJ counties. x-axis denotes peak week ordered by $sw$, y-axis represents log value of the population density. The bubble color and size denote peak intensity and population size. The peak week and peak intensity are only partially correlated with county population size/density.}
  \label{fig:bubble}
\end{figure}

In Figure~\ref{fig:bubble} we visualize the correlation between county demographic attributes (population size and density) and county epidemic features (peak timing and peak intensity) in the ground truth data. While counties with larger populations or higher population density seem to peak later in the season, this is not always true: there are small, low density counties that peak late. But there is no high density county that peaks early. This suggests that the spatial features, e.g. the conventional geographic distance or {\em effective distance} (defined based on the commute flow matrix)~\cite{brockmann2013hidden} to the source county (where the epidemic starts), may play an important role in determining the disease spread trajectory among the counties of the state.

\begin{figure}[t]
\centering
\subfloat[Week 10]{\label{fig:w10}\includegraphics[width=1in]{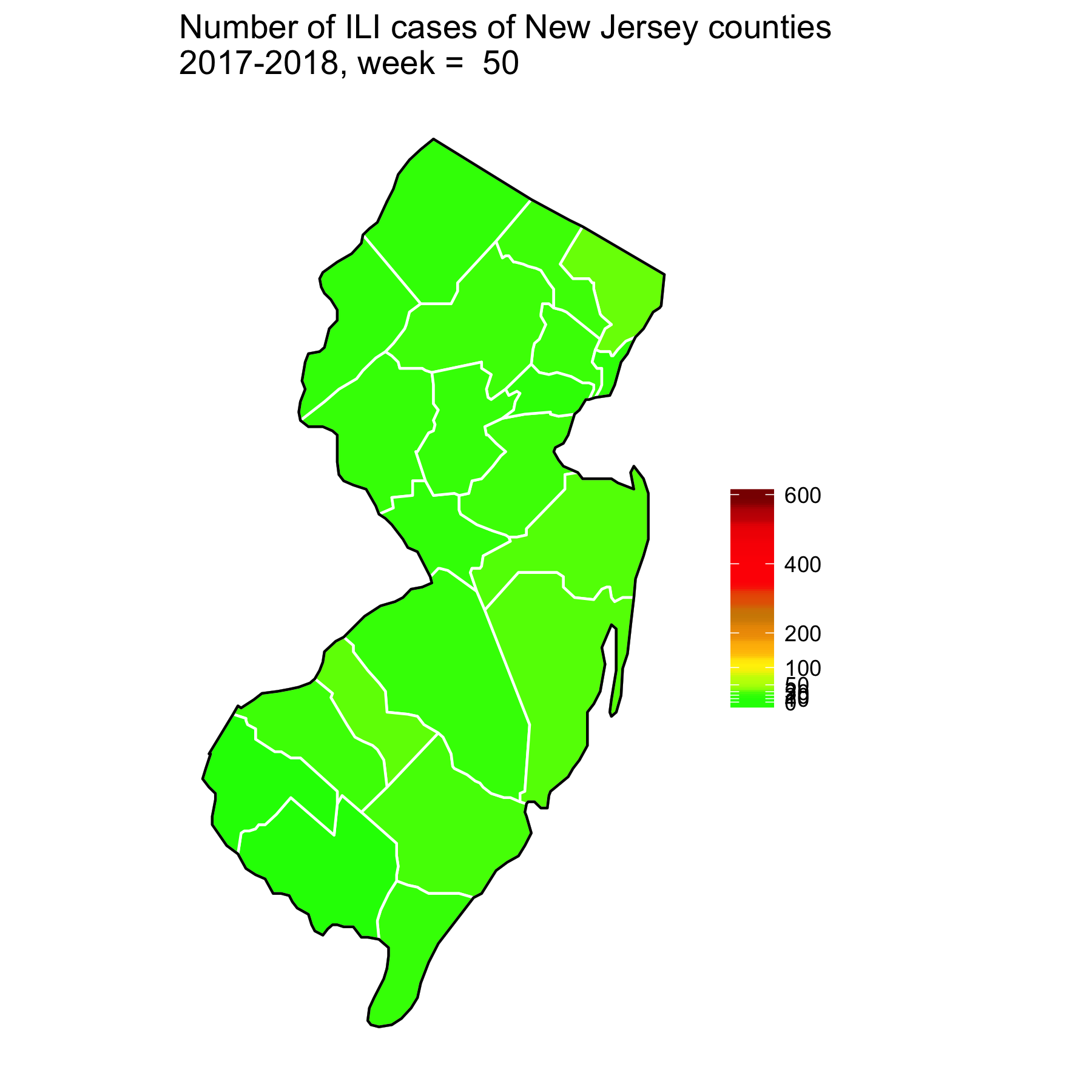}}
\hfil
\subfloat[Week 13]{\label{fig:w13}\includegraphics[width=1in]{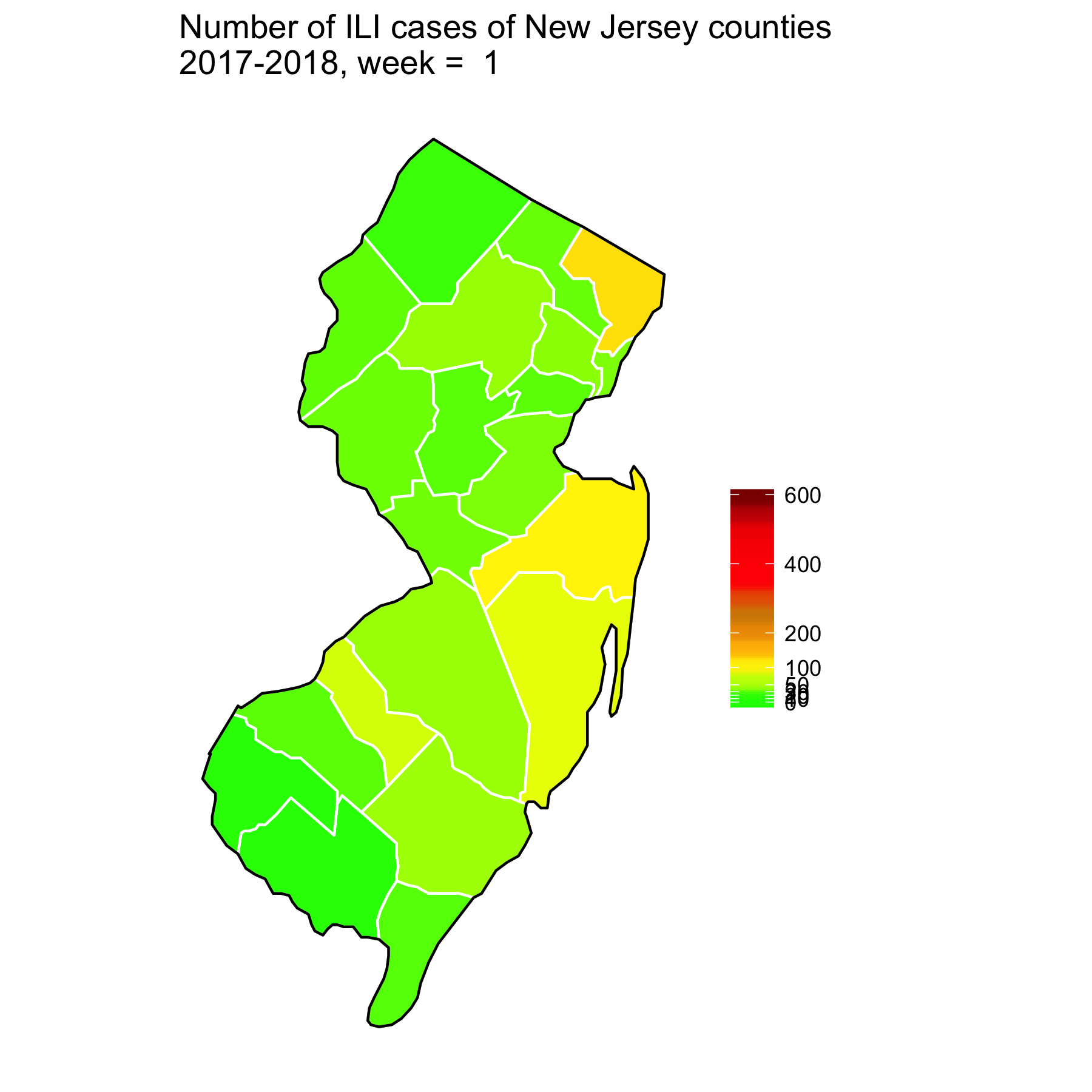}}
\hfil
\subfloat[Week 18]{\label{fig:w18}\includegraphics[width=1in]{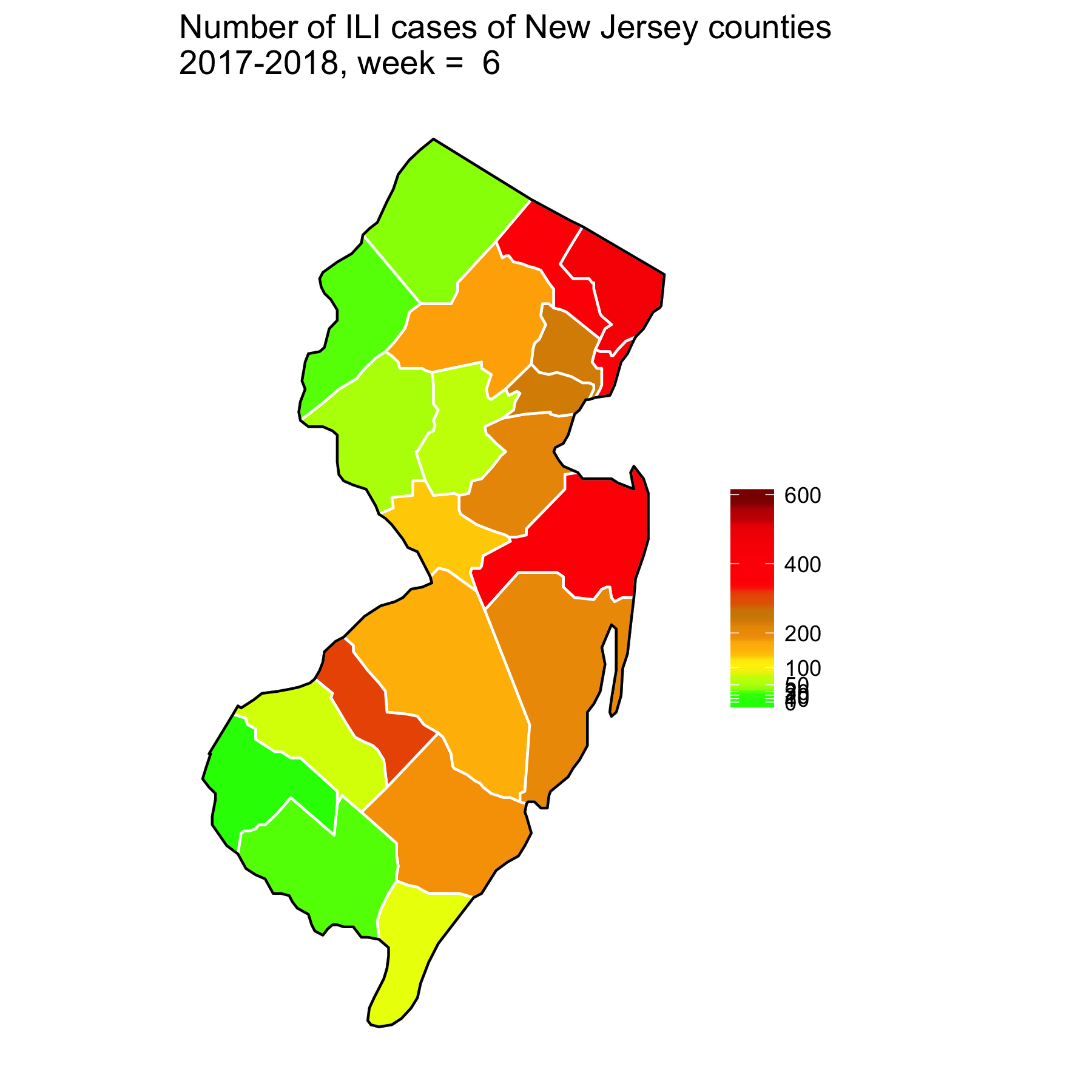}}
\hfil
\subfloat[Week 21]{\label{fig:w21}\includegraphics[width=1in]{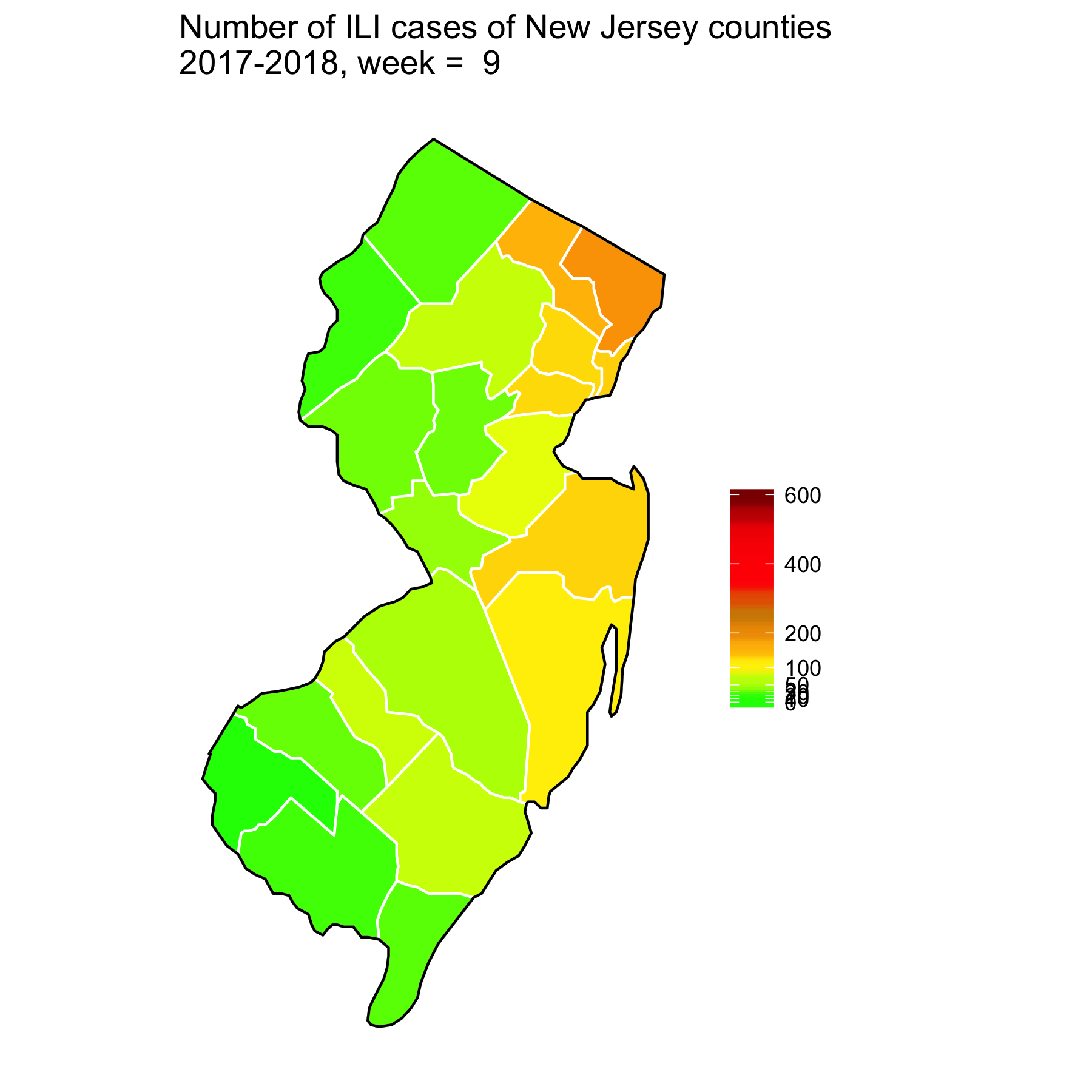}}
\hfil
\subfloat[Week 25]{\label{fig:w25}\includegraphics[width=1in]{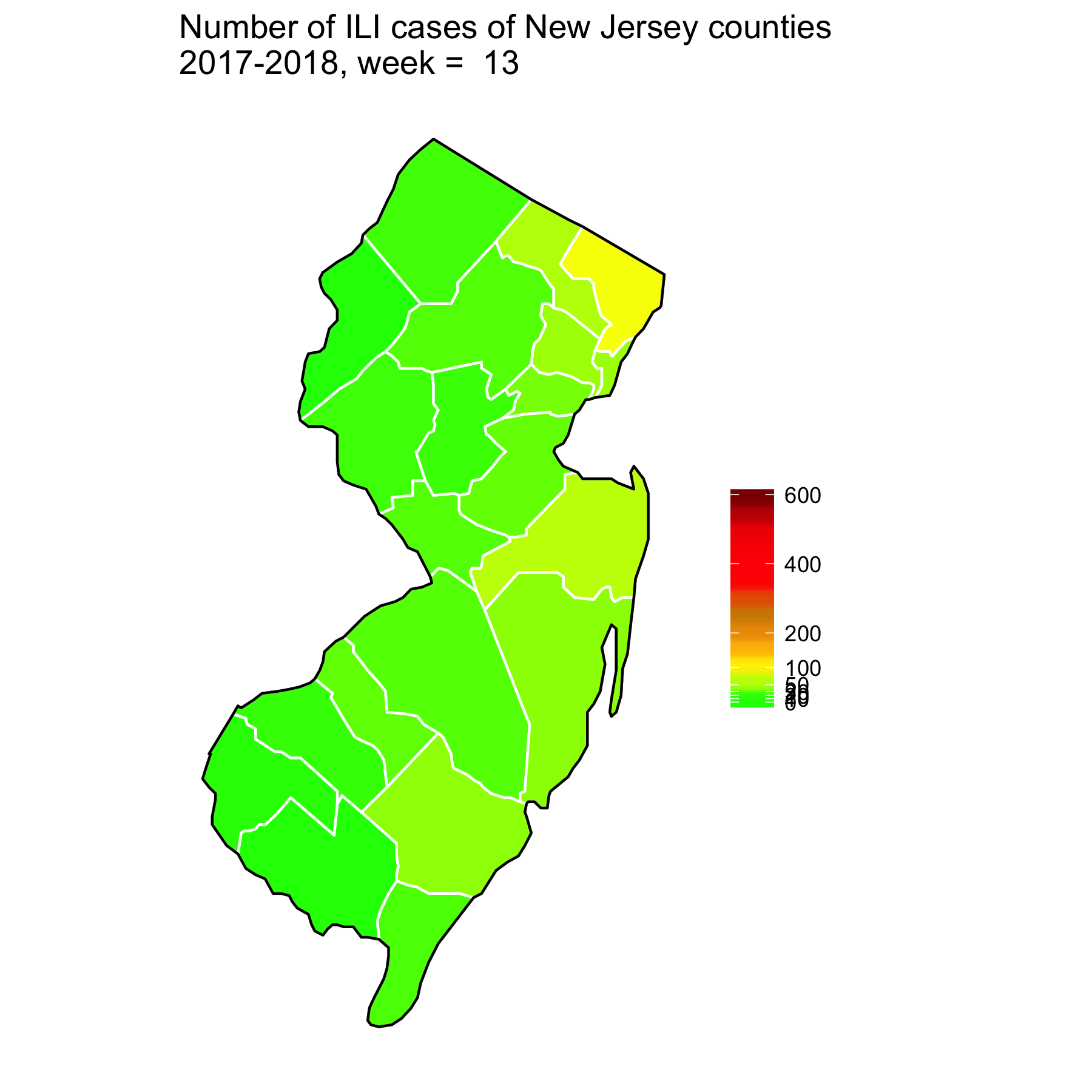}}
\hfil
\caption{The number of ILI cases of New Jersey counties, season 2017-2018. (a) Week 10. (b) Week 13. (c) Week 18. (d) Week 21. (e) Week 25. Note that the flu starts to spread rapidly in region 2, 3, 4, 6 that have counties with large population.}
\label{fig:njdynamic}
\end{figure}

In Figure~\ref{fig:njdynamic} we show the change of the ILI case numbers of New Jersey counties through the weeks {\bf sw} of the 2017-2018 influenza season, at week 10, 13, 18, 21, and 25. For this season, one can note that the flu starts to spread rapidly in the east part of the state where the counties have large populations.  Interested readers can find a week-by-week animation at~\cite{gif2019}. The spreading process shows spatial heterogeneity over the counties and is correlated to the population size and commute flow. 


%% file: sections/subsections/exp-performance-simulated.tex
\subsection{Performance on Simulated Testing Dataset}
\label{subsec:simtest}
\begin{figure}[t]
\centering
\subfloat[horizon 1]{\label{fig:vahorizon1}\includegraphics[width=1.7in]{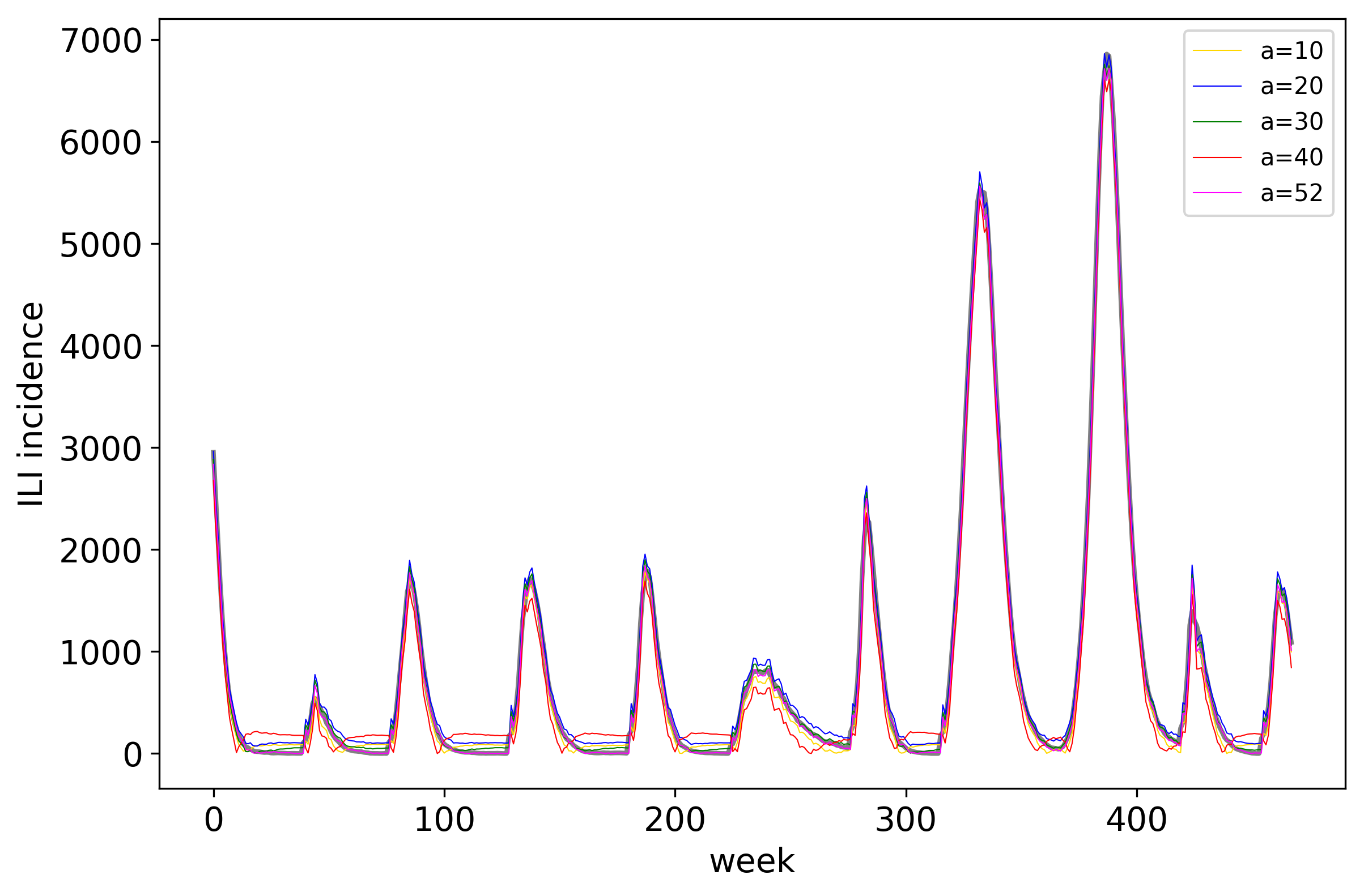}}
\hfil
\subfloat[horizon 5]{\label{fig:vahorizon5}\includegraphics[width=1.7in]{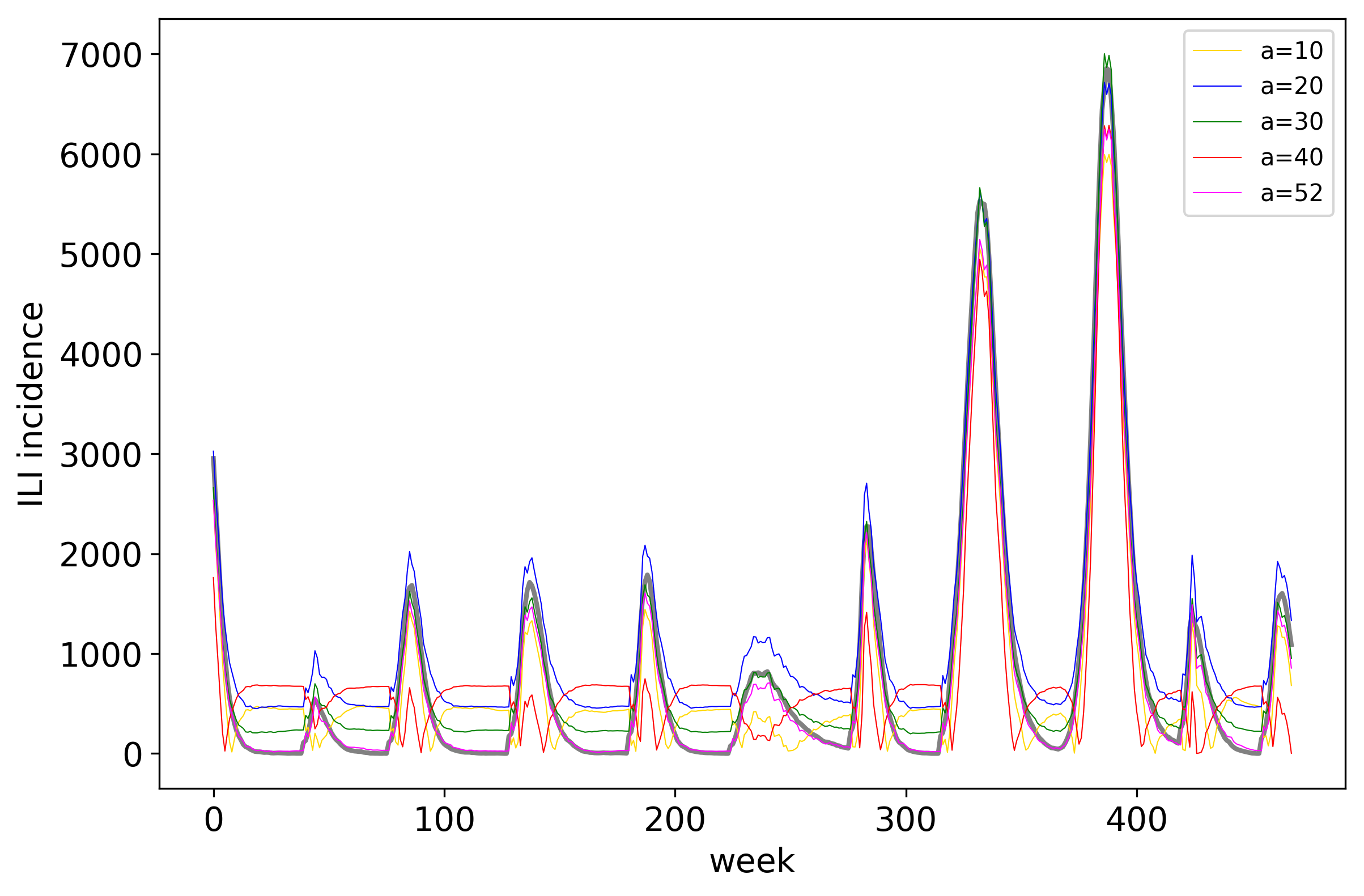}}
\hfil
\subfloat[horizon 10]{\label{fig:vahorizon10}\includegraphics[width=1.7in]{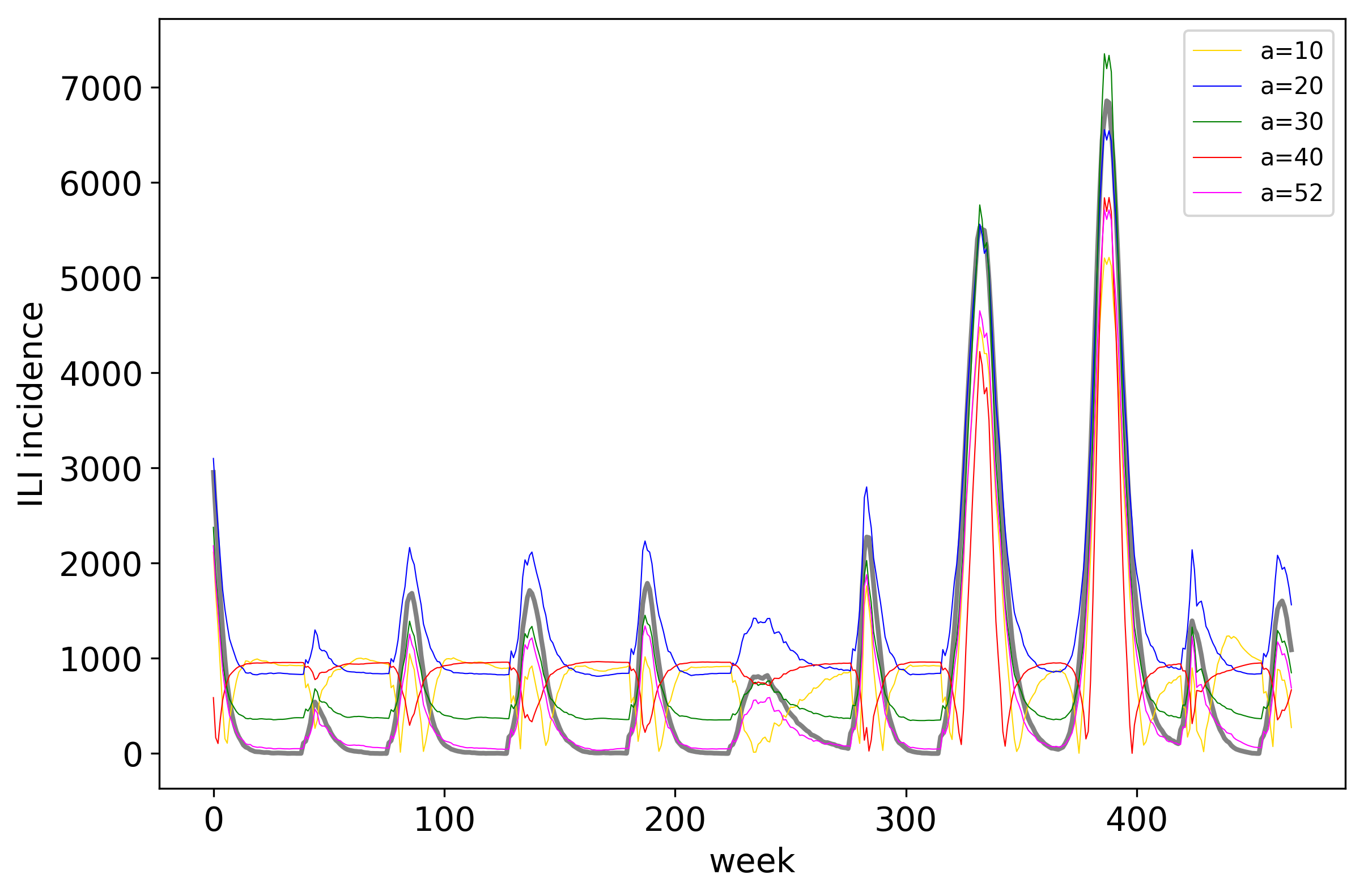}}
\hfil
\caption{State level forecasting curves on sim-testing dataset with (a) horizon 1; (b) horizon 5; (c) horizon 10. The x-axis is the week number of ten simulated curves. Various settings of $a$ are compared. The black curve is the ground truth, while the other colors correspond to models with different values of $a$. It is observable that the predictive power of the model weakens as the horizon increases. The model (magenta curve) with $a=52$ performs the best.
}
\label{fig:vahorizon}
\end{figure}

In this experiment, we tested TDEFSI on sim-testing dataset for VA. We set $\lambda = \mu = 0$, and set $b=5$, then conduct sensitivity analysis on the length of within-season observations, i.e. $a = \{10, 20, 30, 40, 52\}$. 
Figure~\ref{fig:vahorizon} shows the state level forecasting curves (partial curves of sim-testing) in horizon 1, 5, 10 using various $a$. The black curve is the ground truth, while the other colors correspond to different $a$ values. By comparing across (a), (b), (c), we find that the predictive power of the model weakens as the horizon increases. In addition, the model with a seasonal length of $a=52$ performs the best.

\begin{figure}
\centering
\includegraphics[width=5.3in]{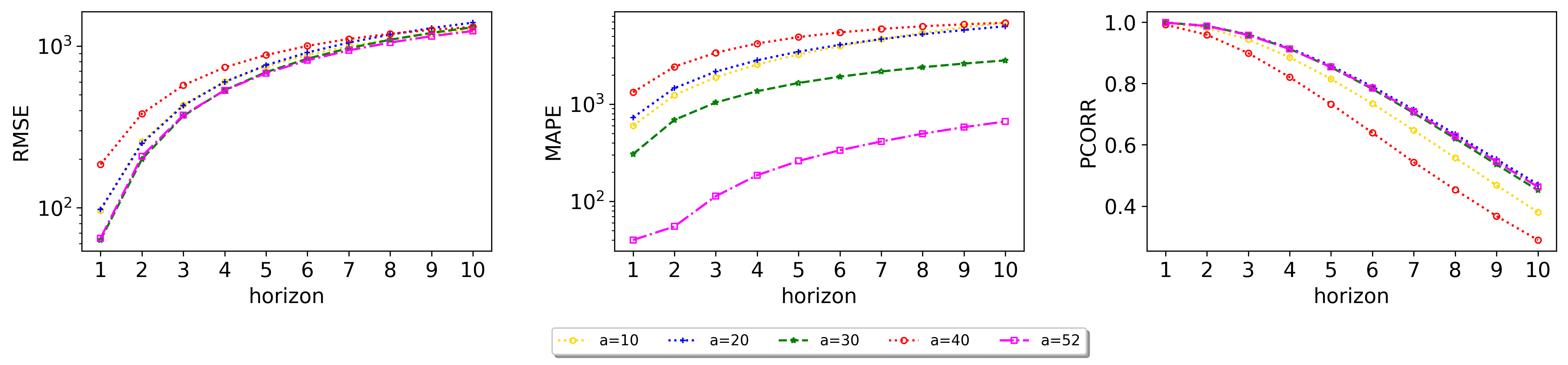}
\caption{State level forecasting performance on sim-testing dataset of VA with various length of within-season observations $a=\{10,20,30,40,52\}$, which is evaluated by RMSE (left), MAPE (middle), PCORR (right). The x-axis represents horizons from 1 to 10. The value is averaged on all weeks of testing curves. A log y-scale is used in RMSE and MAPE. Across different horizons and metrics, the best model is always the model with $a=52$.}
\label{fig:vasimulatedstate}
\end{figure}

\begin{figure}
\centering
\includegraphics[width=5.3in]{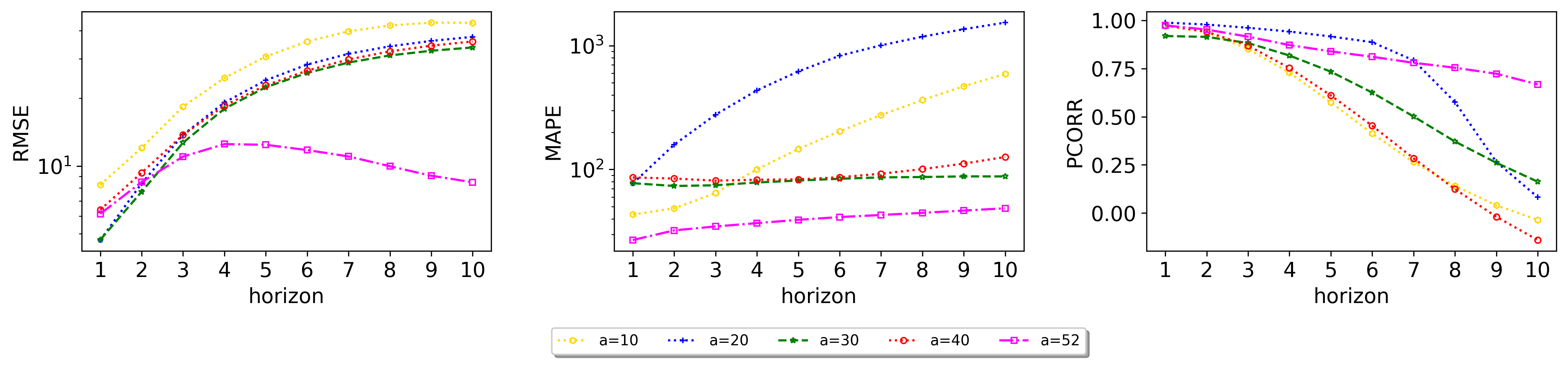}
\caption{County level forecasting performance on sim-testing dataset of VA with various length of within-season observations $a=\{10,20,30,40,52\}$, which is evaluated by RMSE (left), MAPE (middle), PCORR (right). The x-axis represents horizons from 1 to 10. The value is averaged on all weeks of testing curves. A log y-scale is used in RMSE and MAPE. Across different horizons and metrics, the best model is always the model with $a=52$, especially with larger horizons.}
\label{fig:vasimulatedcounty}
\end{figure}

To verify our observations, we evaluate the model performances with metrics RMSE, MAPE, PCORR at both the state level (shown in Figure~\ref{fig:vasimulatedstate}) and county level (shown in Figure~\ref{fig:vasimulatedcounty}).
The best model is always the one with $a=52$. This is not random. It is the manifestation of a flu season normally consisting of 52/53 weeks, i.e. the seasonality of the time-series data.
Thus, we suggest setting $a$ to a multiple of 52 in practice. Unless explicitly stated, we fix $a=52, b=5$ in the rest of our experiments. 

%% file: sections/subsections/exp-performance-real-new.tex
\subsection{Performance on Real Seasonal ILI Testing Dataset}
\label{subsec:realtest}
\subsubsection{Performance of Flat-resolution Forecasting}

\input{tables/exp-state-eval.tex}

\begin{figure}[t]
  \includegraphics[width=0.8\textwidth]{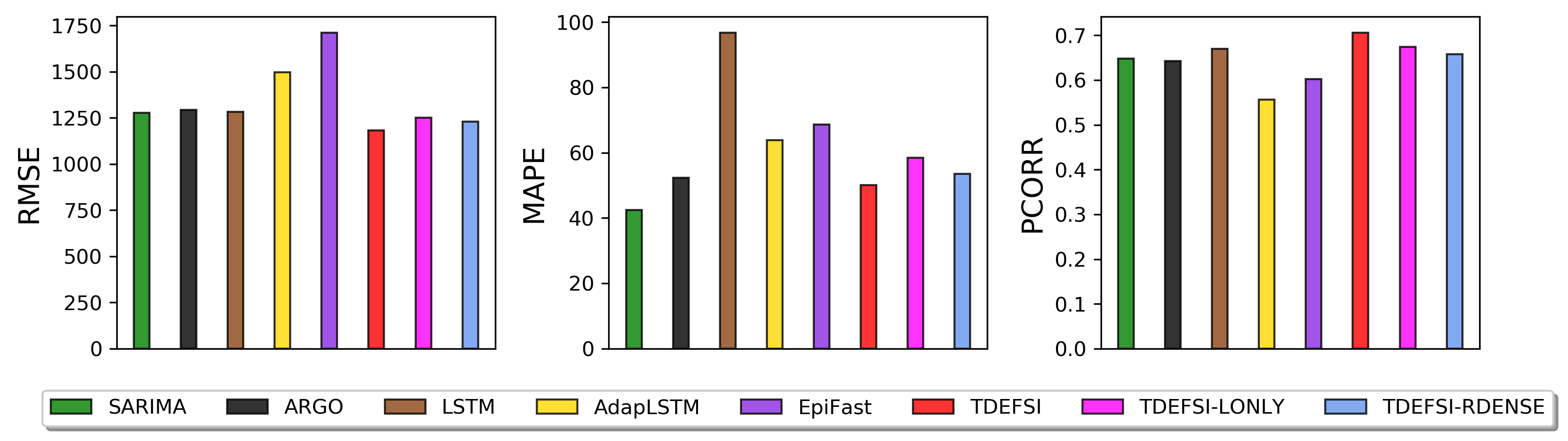}
  \caption{State level performance (RMSE, MAPE, PCORR). The value is averaged across two states, two seasons, and 5 horizons. }
  \label{fig:state-performance}
\end{figure}

We forecast state level ILI incidence on real-testing dataset for VA and NJ. 
Table~\ref{tab:state-performance} shows the performance on RMSE, MAPE, PCORR for (a) VA and (b) NJ with horizon=$\{1,2,3,4,5\}$. Figure~\ref{fig:state-performance} presents the overall performance across all states, weeks, horizons. 
($i$) \textit{Performance on RMSE}:
In VA, TDEFSI, TDEFSI-LONLY, TDEFSI-RDENSE, SARIMA, ARGO, and LSTM achieve similar performance that is better than EpiFast and AdapLSTM. Compared with other methods, AdapLSTM does not perform well with small horizons while EpiFast has poor performance with large horizons. 
In NJ, TDEFSI, TDEFSI-LONLY, and TDEFSI-RDENSE consistently outperform others across the horizon.
Overall, TDEFSI and its variants slightly outperform comparison methods in RMSE.  
($ii$) \textit{Performance on MAPE}:
In VA, SARIMA performs the best overall among all methods.
In NJ, TDEFSI-RDENSE achieves the best performance closely followed by SARIMA. 
Overall, SARIMA outperforms others, and TDEFSI and its variants achieve similar performance with ARGO which are better than LSTM, AdapLSTM, EpiFast.
($iii$) \textit{Performance on PCORR}:
In VA, ARGO performs the best with horizon 1,2,3 and TDEFSI achieves better performance with horizon 4,5.
In NJ, TDEFSI performs the best and TDEFSI-LONLY, TDEFSI-RDENSE achieve similar performance.
Overall, TDEFSI and its variants slightly outperform SARIMA, ARGO, LSTM, while they are much better than AdapLSTM and EpiFast.

\begin{figure}[t]
\centering
\subfloat[VA, 2017-2018]{\label{fig:vastate2017week}\includegraphics[width=2.5in]{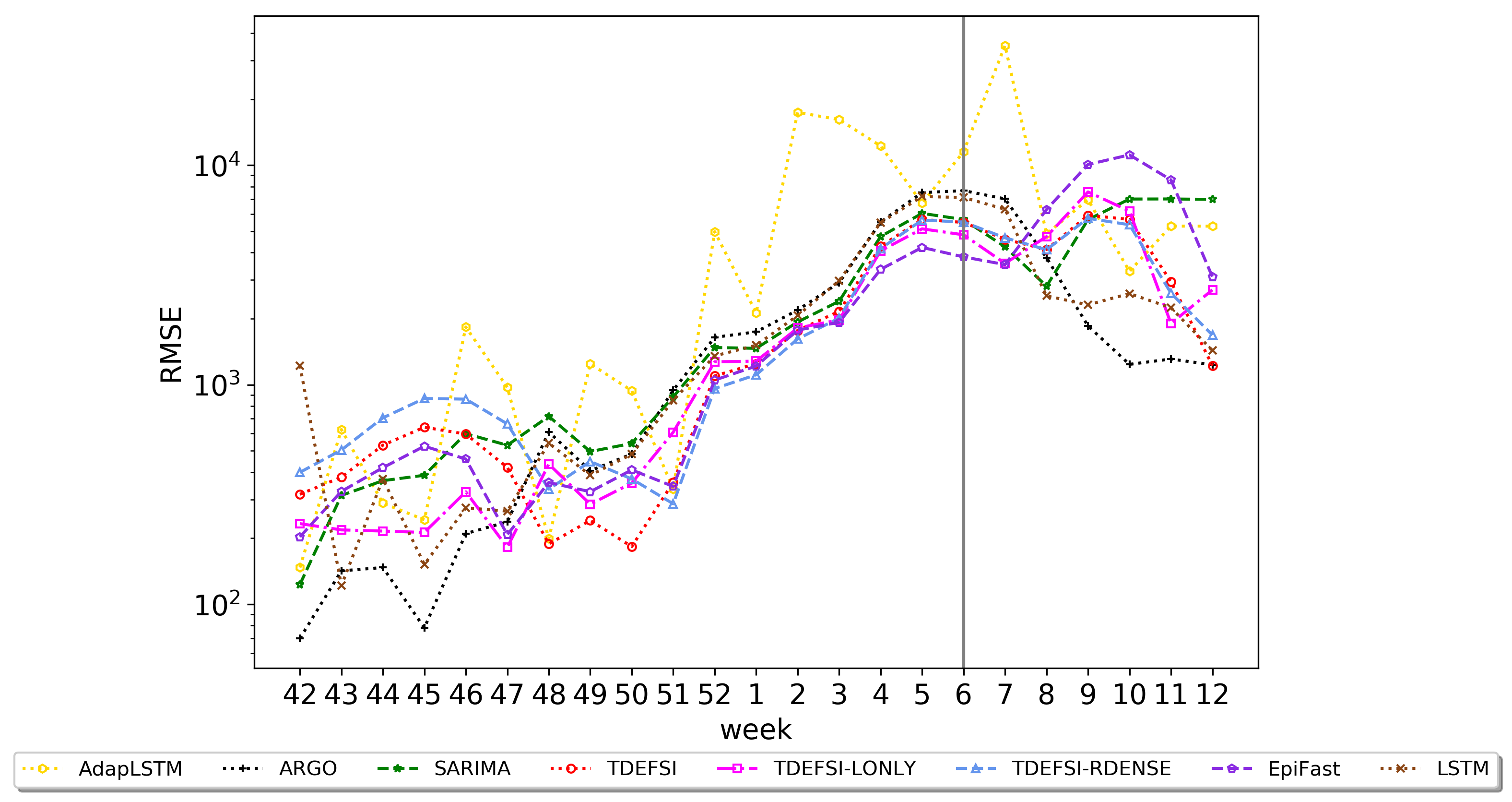}}
\hfil
\subfloat[NJ, 2017-2018]{\label{fig:njstate2017week}\includegraphics[width=2.5in]{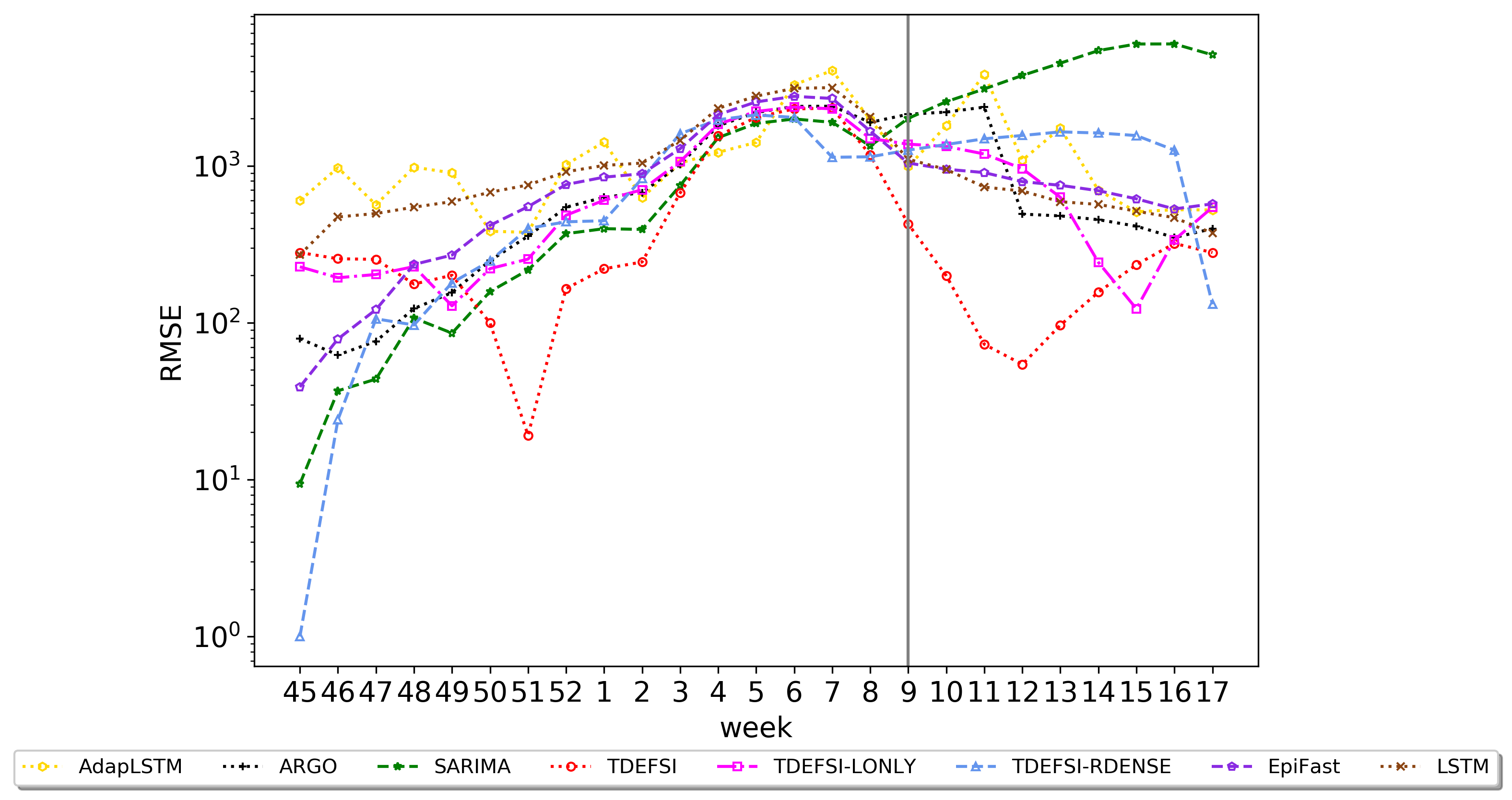}}
\hfil
\caption{State level performance by weeks (RMSE). (a) VA, 2017-2018; (b) NJ, 2017-2018. TDEFSI and its variants, and all comparison methods are evaluated and compared. The x-axis denotes $ew$ number, the value is averaged on 5 horizons. A log y-scale is used. The black vertical line marks the peak week of the season in the state. }
\label{fig:state2017week}
\end{figure}

Figure~\ref{fig:state2017week} shows the \emph{weekly} state level model performance measured on season 2017-2018 using RMSE: 
The x-axis denotes $ew$ number, the value is averaged over 5 horizons. A log y-scale is used. The black vertical line marks the peak week of the season. We observe that these models perform with great variance around the beginning and the end of a season than in weeks near the peak. 

The above discussion can be summarized as follows:
\begin{itemize}
    \item 
Our TDEFSI and its variants achieve comparable/better performance than the other methods on the state level ILI forecasting. 
\item 
EpiFast and AdapLSTM perform relatively worse than other methods in our experiments. 
\end{itemize}

\subsubsection{Performance of High-resolution Forecasting}

\input{tables/exp-county-eval.tex}

\begin{figure}[t]
  \includegraphics[width=0.8\textwidth]{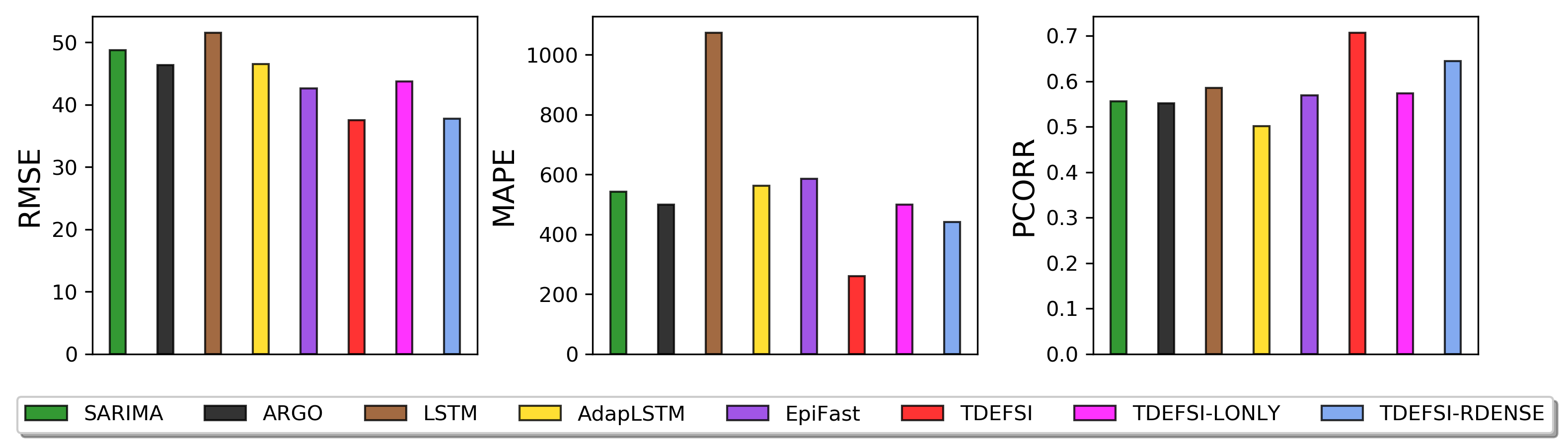}
  \caption{County level performance (RMSE, MAPE, PCORR). The value is averaged on two seasons, 5 horizons and 21 counties of NJ. }
  \label{fig:county-performance}
\end{figure}

The performance of county level forecasts is evaluated on NJ counties. Note that EpiFast, TDEFSI, TDEFSI-LONLY, TDEFSI-RDENSE make county level predictions directly from models, while the other baselines obtain county level predictions by multiplying state level prediction with county population proportions. 
Table~\ref{tab:county-performance} shows the forecasting performance on RMSE, MAPE, PCORR with horizon=$\{1,2,3,4,5\}$. The value is the average across weeks and counties. Figure~\ref{fig:county-performance} presents the overall performance across all counties, weeks, horizons. 
From the table we observe that SARIMA performs well with horizon = 1. TDEFSI consistently outperforms others across horizons, followed by TDEFSI-RDENSE. Among TDEFSI variants, TDEFSI and TDEFSI-RDENSE perform better than TDEFSI-LONLY, which indicates that the between-season observations are helpful for improving forecasting accuracy. 
The figure shows consistent results with the table. 
Overall, our method outperforms the comparison methods on the county level forecasting. 

\begin{figure}[t]
\centering
\subfloat[NJ counties, RMSE-ratio]{\label{fig:ratermse}\includegraphics[width=1.7in]{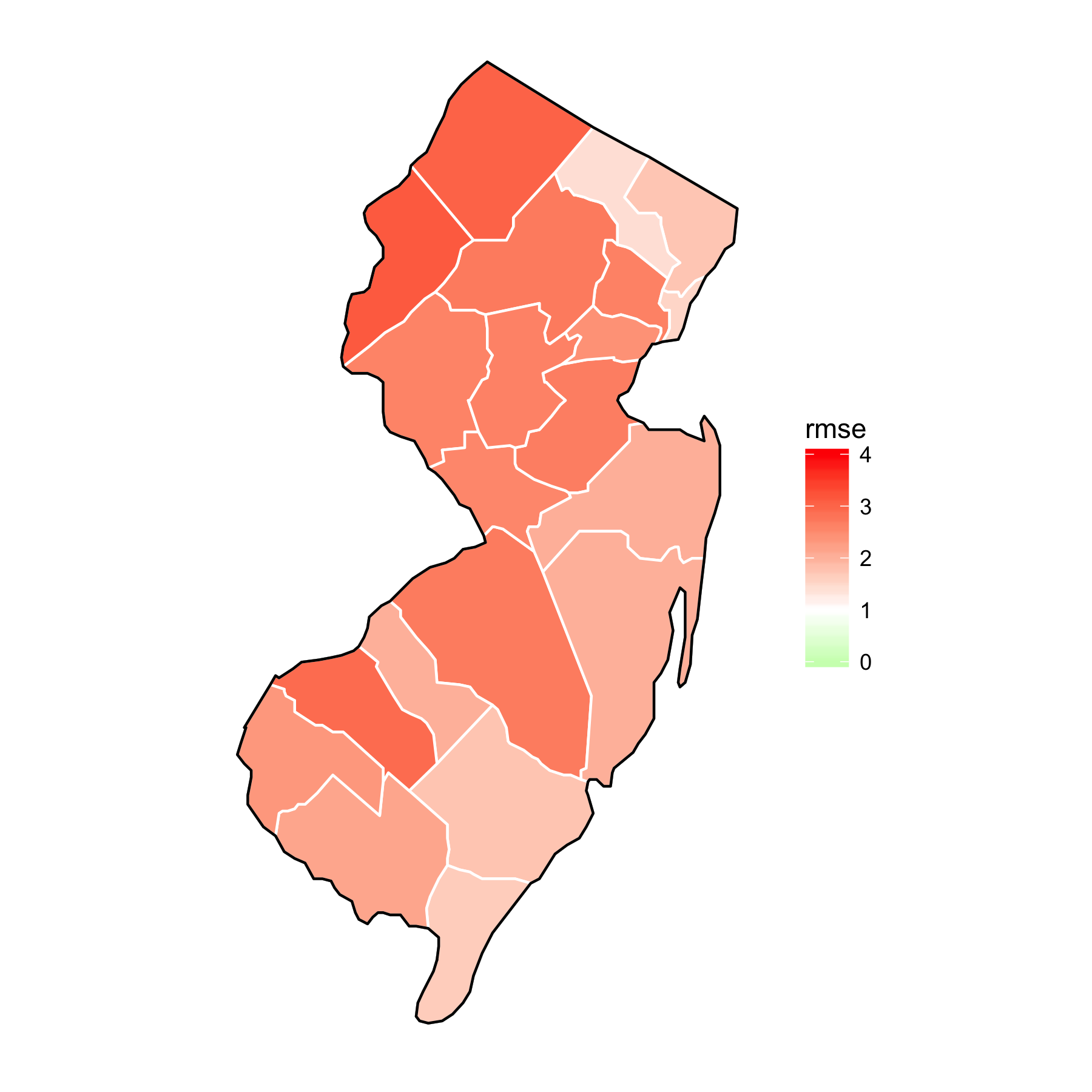}}
\hfil
\subfloat[NJ counties, MAPE-ratio]{\label{fig:ratemape}\includegraphics[width=1.7in]{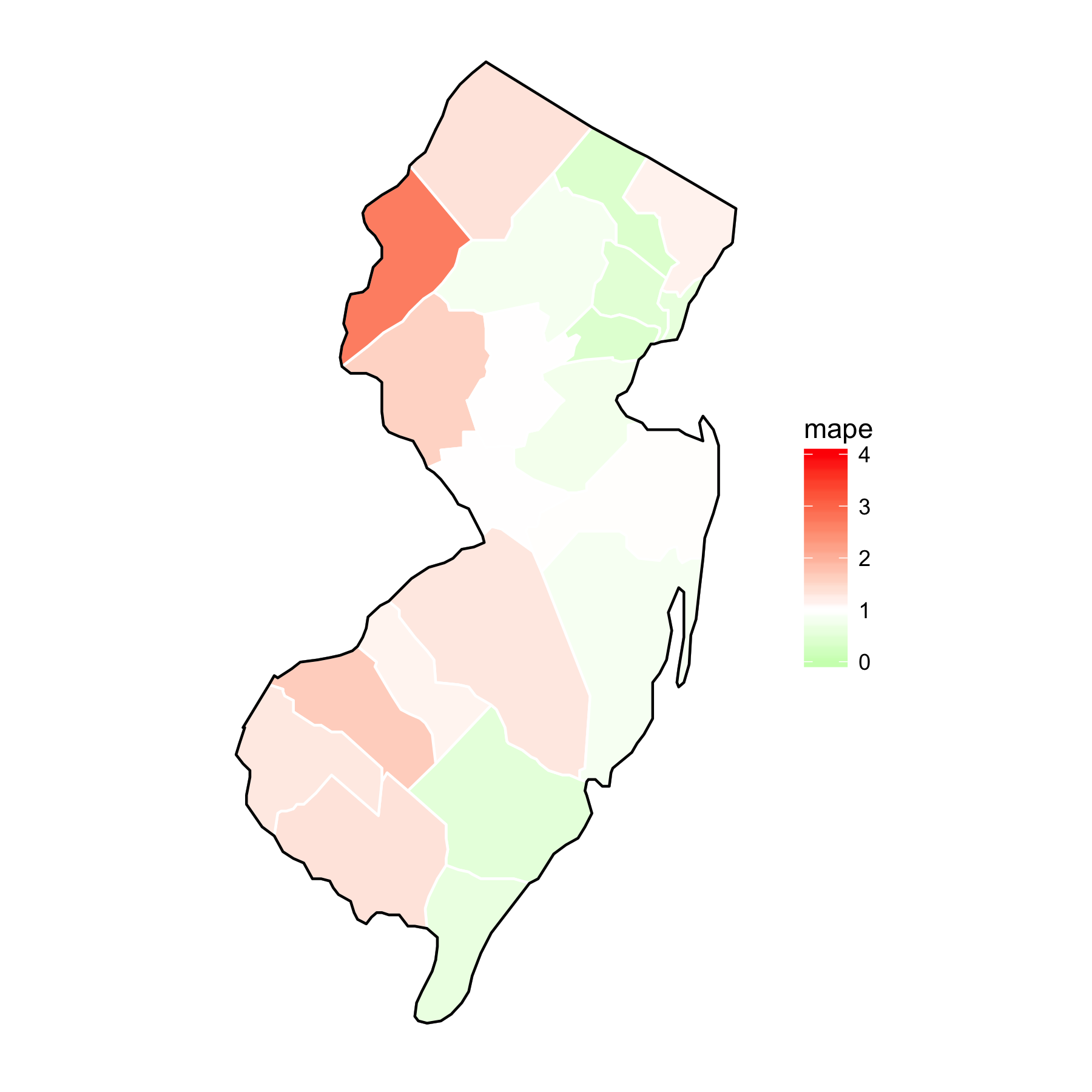}}
\hfil
\subfloat[NJ counties, PCORR-ratio]{\label{fig:ratepcorr}\includegraphics[width=1.7in]{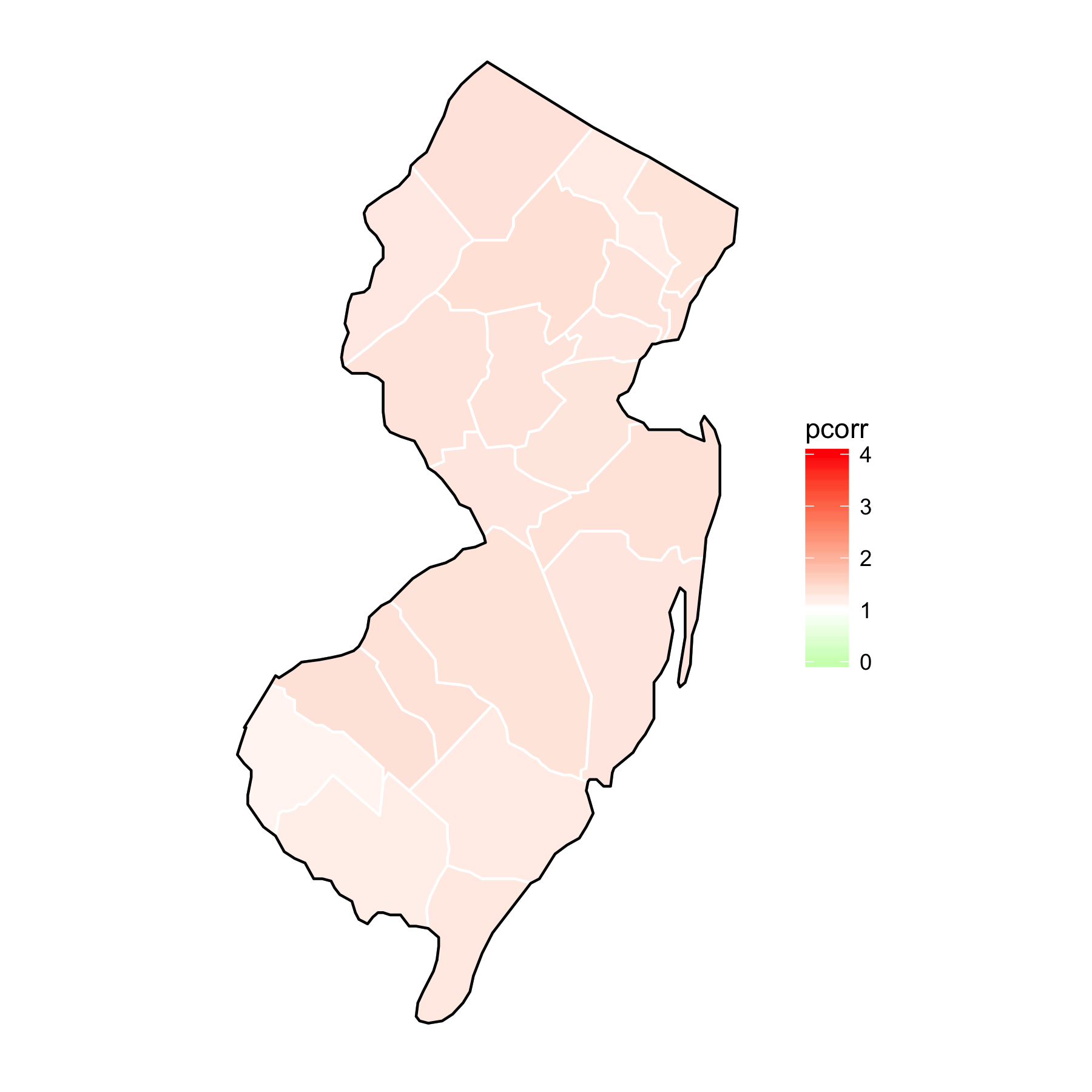}}
\hfil
\caption{Comparison of the county level spatial forecasting performance between TDEFSI and EpiFast for NJ, season 2017-2018. (a) RMSE-ratio; (b) MAPE-ratio; (c) PCORR-ratio. For each county in NJ, the ratio value of the county is computed using equations~\ref{equ:njrate}, which is the average value across horizons. A value larger than 1 (red color) means TDEFSI outperforms EpiFast, a value equal to 1 (white color) means they both perform equally, and a value smaller than 1 (green color) means EpiFast performs better than TDEFSI. The absolute magnitude of the value denotes the significance of the difference of the two models' performance. The comparison results exhibit that TDEFSI performs better than EpiFast in the counties located in western NJ.}
\label{fig:njrate}
\end{figure}

\noindent{\textbf{Heterogeneous high-resolution forecasting.}} 
To better understand the results from a spatial perspective, we compare results between TDEFSI and EpiFast in Figure~\ref{fig:njrate}. The reason we choose to compare these two methods is that they both can make high-resolution predictions directly from the models. For each county in NJ, we compare TDEFSI and EpiFast using a ratio value for each of three metrics defined as:
\begin{equation}
\begin{aligned}
RMSE-ratio &= \frac{\frac{1}{m}\sum_{i=1}^{m}RMSE_i(EpiFast)}{\frac{1}{m}\sum_{i=1}^{m}{RMSE_i(TDEFSI)}}\\
MAPE-ratio &= \frac{\frac{1}{m}\sum_{i=1}^{m}MAPE_i(EpiFast)}{\frac{1}{m}\sum_{i=1}^{m}{MAPE_i(TDEFSI)}}\\
PCORR-ratio &= \frac{\frac{1}{m}\sum_{i=1}^{m}{(PCORR_i(TDEFSI)+1)}}{\frac{1}{m}\sum_{i=1}^{m}{(PCORR_i(EpiFast)+1)}}
\end{aligned}
\label{equ:njrate}
\end{equation}
where $m$ is the number of horizons. The ratio is averaged across all horizons. For any of these ratios, a value larger than 1 means TDEFSI outperforms EpiFast; a value close to 1 means they have similar performance; and a value smaller than 1 means EpiFast performs better than TDEFSI. 

From Figure~\ref{fig:njrate} (a) RMSE-ratio, we observe that TDEFSI significantly outperforms EpiFast in all counties (all counties show red colors) especially in the western counties of NJ. In (b) MAPE-ratio, TDEFSI performs better than EpiFast in eleven out of twenty one counties, most of which are located in the west side of NJ.  And (c) PCORR-ratio shows that TDEFSI constantly outperforms EpiFast in all counties (all in red colors). 
The comparison results exhibit that TDEFSI performs better than EpiFast in the counties located in western NJ. EpiFast tries to find a model that best matches the state level observations, and use it to make predictions. However, the identified model is usually locally optimal due to the limitation of the searching algorithm and the computational efficiency. In our experiments, we run the searching algorithm once and then find a locally optimal model which performs fairly well in eastern NJ counties but not in western NJ counties. If we run the searching algorithm again, we will find another locally optimal model which might perform well in western NJ counties instead. In TDEFSI model, the deep neural network model allows TDEFSI to learn from many models. What is learned is an ensemble of all models. Thus, TDEFSI is more robust than EpiFast in different runs of the flu forecasting experiment. 

\begin{figure}[t]
  \includegraphics[width=0.6\textwidth]{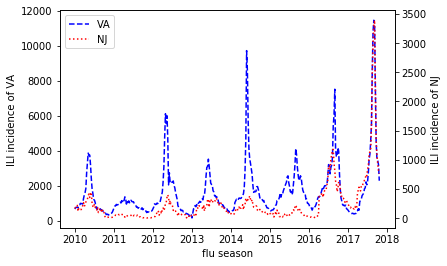}
  \caption{CDC surveillance ILI incidence of VA (blue dash line) and NJ (red dot line). It is observable that, for testing season 2017-2018, a similar epi-curve (i.e. similar curve shape and the peak size) occurs at season 2014-2015 in VA, while no similar seasons could be found in NJ.}
  \label{fig:history}
\end{figure}

\subsubsection{Discussion} \label{subsec:discussion}
In general, for state level, AdapLSTM and EpiFast do not perform very well in our experiments compared with other methods.
For AdapLSTM, weather features are considered for post adjustment of LSTM outputs. 
As stated in~\cite{venna2019novel}, the weather factors are estimated using time delays computed by apriori associations and selected by the largest confidence. However, in our experiment, they all show very low confidences (less than 0.3). This may cause arbitrary adjustment for predictions and consequently poor performance.
For EpiFast, one possible reason is that we did not find a good estimate of the underlying disease model for a specific region and season due to the noisy CDC observations.
If we rank the performance of all methods, ARGO performs slightly better on VA than on NJ. The possible reason is that about $80\%$ of the top 100 Google correlated terms for NJ are irrelevant to flu and most of them have zero frequencies, while the top 100 correlated terms for VA are of good quality. This will give ARGO a better performance on VA than on NJ.
Similarly, LSTM performs relatively better on VA than on NJ. 
One possible reason is that LSTM cannot learn a pattern that has never occurred in the  historical observations. So its performance depends on whether a similar epicurve occurred in previous seasons.
As shown in Figure~\ref{fig:history}, the epicurve of VA 2017-2018 is similar to that of VA 2014-2015, and 2016-2017 is similar to 2012-2013. However, the epicurve of NJ 2017-2018 seems to be much higher than all previous ones, as well as 2016-2017. Actually, this is the limitation of all data driven models. 
On the contrary, TDEFSI models have stable performance on both VA and NJ. They manage to avoid overfitting through training on a large volume of synthetic training data. In addition, the simulated training dataset includes many realistic simulated patterns that are unseen in the real world, thus provides a better generalizability to our models.

As seen through the results, TDEFSI enables high-resolution forecasting that outperforms baselines. Meanwhile, it achieves comparable/better performance than the comparison methods at state level forecasting. And in our framework, the large volume of realistic simulated data allows us to train a more complex DNN model and reduces the risk of overfitting. Our experiments demonstrate that TDEFSI integrates the strengths of ANN methods and causal methods to improve epidemic forecasting.

%% file: tables/exp-state-eval.tex
\begin{table*}[!t]
\footnotesize
\centering
  \caption{State level performance across season 2016-2017 and 2017-2018 for VA and NJ with horizon = 1, 2, 3, 4, 5. The best value is marked in bold, and the second best value is marked with underline.}
  \label{tab:state-performance}
  \begin{tabular}{|r|ccccc|ccccc|}
    \hline
     & \multicolumn{5}{c|}{VA} & \multicolumn{5}{c|}{NJ} \\
     \hline
\textbf{RMSE} & 1 & 2 & 3 & 4 & 5 & 1 & 2 & 3 & 4 & 5 \\
    \hline
SARIMA & \textbf{824} & \underline{1463} & 2059 & 2440 & 2682 & 218 & 464 & 690 & 891 & 1050 \\
ARGO & 1073 & 1592 & 2072 & 2444 & 2580 & 313 & 512 & 717 & 760 & 874 \\
LSTM & 1083 & 1629 & \textbf{2013} & 2273 & \textbf{2438} & 240 & 470 & 699 & 902 & 1070 \\
AdapLSTM & 2012 & 2038 & 2264 & \underline{2382} & \underline{2449} & 586 & 729 & 640 & 871 & 1006 \\
EpiFast & 1300 & 2087 & 2989 & 3674 & 4284 & 238 & 382 & 567 & 725 & 871 \\
TDEFSI & 1000 & \textbf{1447} & \underline{2014} & \textbf{2358} & 2544 & \textbf{174} & \textbf{344} & \underline{511} & \underline{665} & \underline{757} \\
TDEFSI-LONLY & \underline{900} & 1572 & 2119 & 2582 & 2742 & 197 & 373 & 531 & 696 & 801 \\
TDEFSI-RDENSE & 1109 & 1686 & 2136 & 2421 & 2540 & \underline{193} & \underline{358} & \textbf{506} & \textbf{630} & \textbf{711} \\
    \hline
\textbf{MAPE} & 1 & 2 & 3 & 4 & 5 & 1 & 2 & 3 & 4 & 5 \\
    \hline
SARIMA & \textbf{15.96} & \textbf{32.57} & \textbf{50.62} & \textbf{65.60} & 77.94 & \textbf{13.28} & \underline{24.32} & \underline{35.62} & \underline{48.32} & 59.99 \\
ARGO & 31.06 & 54.00 & 73.69 & 78.97 & 77.85 & 24.96 & 33.14 & 44.52 & 50.05 & \underline{54.60} \\
LSTM & 38.40 & 49.29 & 58.80 & 67.98 & \underline{71.00} & 39.44 & 78.53 & 131.19 & 189.79 & 243.40 \\
AdapLSTM & 42.67 & 51.22 & 61.02 & \underline{67.33} & \textbf{70.60} & 64.30 & 64.77 & 65.56 & 74.14 & 76.50 \\
EpiFast & 31.14 & 53.45 & 84.32 & 124.05 & 167.44 & 30.32 & 32.40 & 50.75 & 64.61 & 76.27 \\
TDEFSI & 25.75 & 40.69 & \underline{58.61} & 74.06 & 88.95 & 18.16 & 29.74 & 43.49 & 55.12 & 66.09 \\
TDEFSI-LONLY & \underline{22.40} & \underline{35.18} & 59.27 & 89.95 & 123.70 & 15.56 & 32.21 & 45.74 & 60.46 & 72.13 \\
TDEFSI-RDENSE & 31.89 & 51.69 & 76.94 & 101.38 & 125.23 & \underline{15.17} & \textbf{21.74} & \textbf{29.19} & \textbf{37.95} & \textbf{44.14} \\
    \hline
\textbf{PCORR} & 1 & 2 & 3 & 4 & 5 & 1 & 2 & 3 & 4 & 5 \\
    \hline
SARIMA & \underline{0.9461} & 0.8271 & 0.6468 & 0.4925 & 0.3788 & 0.9541 & 0.8173 & 0.6421 & 0.4611 & 0.3195 \\
ARGO & \textbf{0.9590} & \underline{0.8728} & \textbf{0.7219} & 0.4518 & 0.3218 & 0.9444 & 0.8005 & 0.6043 & 0.4530 & 0.2921 \\
LSTM & 0.9223 & 0.7890 & 0.6350 & 0.5050 & \underline{0.4101} & 0.9603 & 0.8542 & 0.6995 & 0.5340 & 0.3939 \\
AdapLSTM & 0.7048 & 0.6397 & 0.5174 & 0.4307 & 0.3818 & 0.8113 & 0.5912 & \textbf{0.7686} & 0.4477 & 0.2753 \\
EpiFast & 0.8876 & 0.7665 & 0.5616 & 0.3906 & 0.2340 & 0.9573 & 0.8535 & 0.7044 & 0.3835 & 0.2841 \\
TDEFSI & 0.9358 & 0.8487 & 0.6892 & \textbf{0.5555} & \textbf{0.4647} & \textbf{0.9683} & \textbf{0.8773} & \underline{0.7348} & \textbf{0.5639} & \underline{0.4247} \\
TDEFSI-LONLY & 0.9460 & \textbf{0.8776} & \underline{0.7037} & \underline{0.5074} & 0.3266 & \underline{0.9659} & \underline{0.8697} & 0.7288 & 0.4946 & 0.3245 \\
TDEFSI-RDENSE & 0.9043 & 0.7824 & 0.6182 & 0.4409 & 0.2826 & 0.9654 & 0.8692 & 0.7280 & \underline{0.5630} & \textbf{0.4248} \\
    \hline
\end{tabular}
\begin{tablenotes}
  \small
  \item 
\end{tablenotes}
\end{table*}

%% file: tables/exp-county-eval.tex
\begin{table*}[!t]
\footnotesize
\centering
  \caption{County level performance for counties of NJ with horizon = 1, 2, 3, 4, 5. The value is the average of 21 counties of NJ across season 2016-2017 and 2017-2018. The best value is marked in bold, and the second best value is marked with underline.}
  \label{tab:county-performance}
  \begin{tabular}{|r|ccccc|}
    \hline
     & \multicolumn{5}{c|}{NJ-Counties}\\
     \hline
\textbf{RMSE} & 1 & 2 & 3 & 4 & 5  \\
    \hline
SARIMA & \textbf{30.58} & 38.02 & 48.60 & 58.92 & 67.68 \\
ARGO & 33.69 & 39.89 & 49.61 & 51.46 & 57.35 \\
LSTM & 33.80 & 41.95 & 52.25 & 61.56 & 68.30 \\
AdapLSTM & 36.67 & 45.30 & 39.46 & 51.70 & 59.60 \\
EpiFast & 34.34 & 36.74 & 40.51 & 47.40 & 54.09 \\
TDEFSI & 35.17 & \textbf{31.40} & \textbf{34.70} & \textbf{40.44} & \textbf{45.95} \\
TDEFSI-LONLY & \underline{33.13} & 36.45 & 42.41 & 50.63 & 56.22 \\
TDEFSI-RDENSE & 34.79 & \underline{31.59} & \underline{35.22} & \underline{40.98} & \underline{46.35} \\
    \hline
\textbf{MAPE} & 1 & 2 & 3 & 4 & 5  \\
    \hline
SARIMA & \underline{575.19} & 550.74 & 540.04 & 525.20 & 525.57 \\
ARGO & 649.32 & 552.18 & 498.42 & 430.74 & 366.89 \\
LSTM & 745.52 & 876.56 & 1066.80 & 1264.64 & 1417.91 \\
AdapLSTM & 584.18 & \underline{489.51} & 417.72 & 599.53 & 717.61 \\
EpiFast & 712.97 & 632.96 & 577.74 & 519.37 & 487.54 \\
TDEFSI & \textbf{260.95} & \textbf{247.70} & \textbf{209.69} & \textbf{270.58} & \textbf{308.95} \\
TDEFSI-LONLY & 603.33 & 528.62 & 478.08 & 454.52 & 435.50 \\
TDEFSI-RDENSE & 614.95 & 499.13 & \underline{412.68} & \underline{360.99} & \underline{315.78} \\
    \hline
\textbf{PCORR} & 1 & 2 & 3 & 4 & 5  \\
    \hline
SARIMA & \textbf{0.8645} & 0.7474 & 0.5678 & 0.3806 & 0.2211 \\
ARGO & 0.8606 & 0.7388 & 0.5455 & 0.3922 & 0.2211 \\
LSTM & \underline{0.8611} & 0.7699 & 0.6132 & 0.4234 & 0.2597 \\
AdapLSTM & 0.7260 & 0.5150 & 0.6717 & 0.3710 & 0.2205 \\
EpiFast & 0.8555 & 0.7762 & 0.6450 & 0.3530 & 0.2133 \\
TDEFSI & 0.7877 & \textbf{0.8500} & \textbf{0.7835} & \textbf{0.6425} & \textbf{0.4710} \\
TDEFSI-LONLY & 0.8499 & 0.7669 & 0.6184 & 0.4146 & 0.2176 \\
TDEFSI-RDENSE & 0.7860 & \underline{0.8063} & \underline{0.7056} & \underline{0.5467} & \underline{0.3774} \\
    \hline
\end{tabular}
\begin{tablenotes}
  \small
  \item 
\end{tablenotes}
\end{table*}

%% file: sections/subsections/exp-consistency.tex
\subsection{Physical Consistency Constraints}
\label{subsec:consistency}
In this section, we conduct sensitivity analysis on two regularizer coefficients $\mu$ and $\lambda$ in equation \eqref{equ:defsi}, which control the weights of the spatial constraint $\phi$ and non-negative constraint $\delta$ in the loss function. $\mu = 0$ means no spatial constraint and $\lambda = 0$ means no non-negative constraint.
We train TDEFSI by setting $a = 52, b = 5$ with various $\mu, \lambda$ values shown in Table~\ref{tab:hyper}. 
We then use the trained models to make predictions for Season 2017-2018 of VA and NJ. 
The performance is evaluated using RMSE.

\begin{figure}[t]
\centering
\subfloat[VA, 2017-2018]{\includegraphics[width=2.5in]{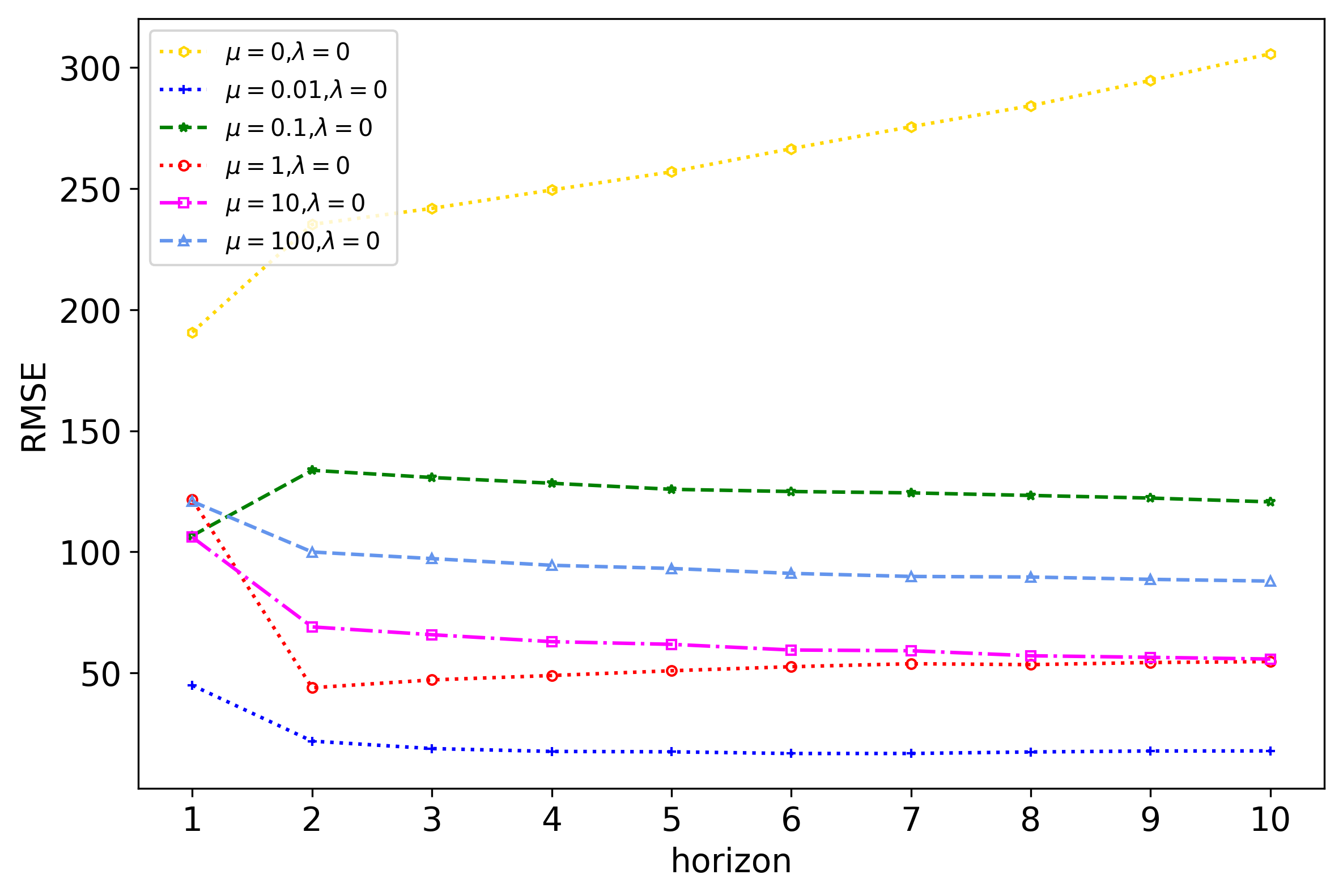}
\label{fig:vaspatialconsistency}
}
\hfil
\subfloat[NJ, 2017-2018]{\includegraphics[width=2.5in]{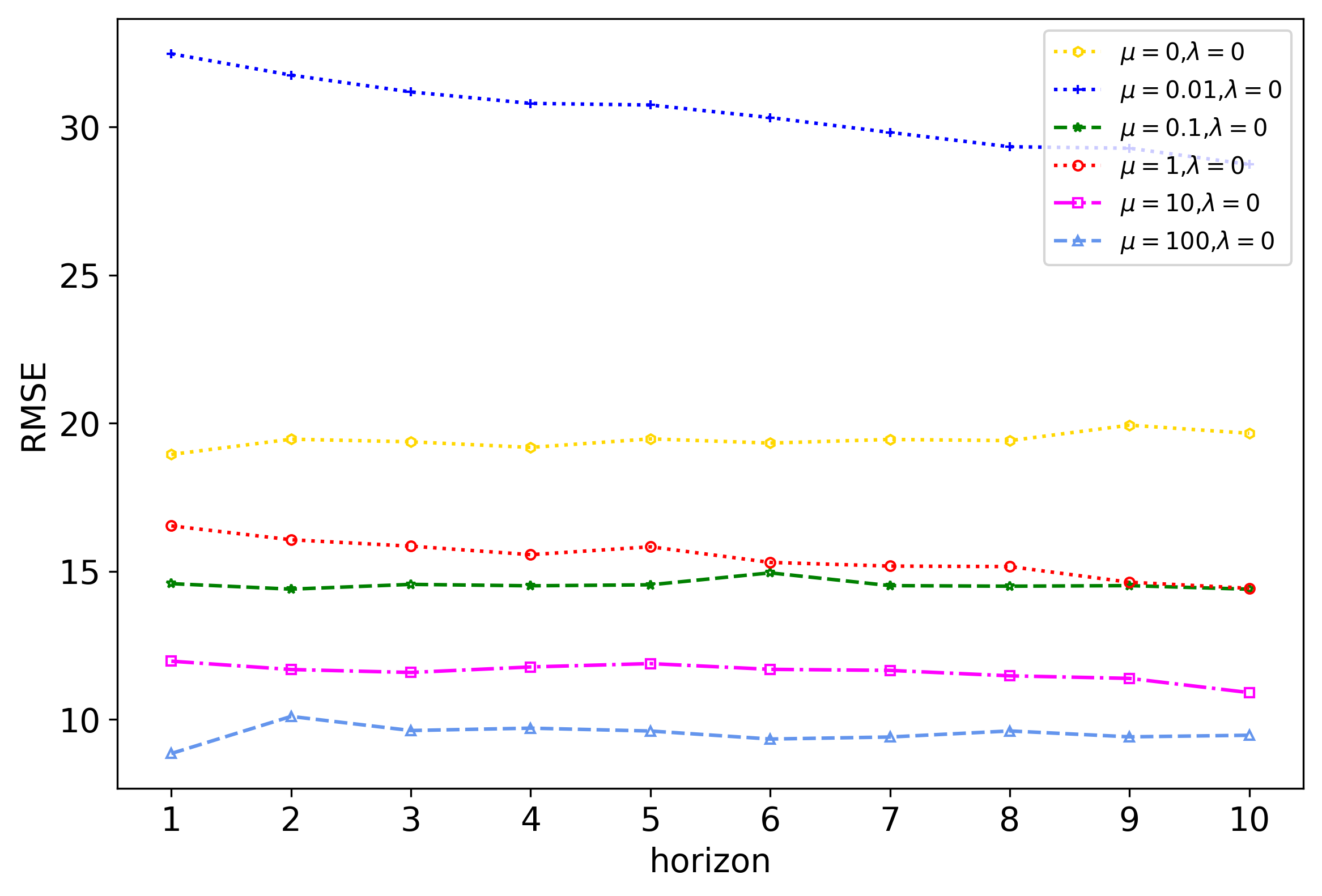}
\label{fig:njspatialconsistency}
}
\caption{Spatial consistency error (computed as $\sqrt{\frac{1}{n}\sum_{i=1}^{n}(\hat{y}_i - \sum \limits_{C \in \mathcal{D}} \hat{y}^C_i)^2}$) on (a) VA, 2017-2018; (b) NJ, 2017-2018. The coefficient of the spatial consistency regularizer is set to $\mu = \{0, 0.001, 0.01, 0.1, 1, 10, 100\}$. The results show that the spatial consistency error does not vary much with horizon, but significantly depends on $\mu$. The optimal $\mu$ differs between states.}
\label{fig:spatialconsistency}
\end{figure}

\textbf{Spatial consistency}  
The experiments are conducted using $\lambda = 0$ and $\mu = $ $\{0,$ $0.001,$ $0.01,$ $0.1,$ $1,$ $10,$ $100\}$.
We evaluate the spatial consistency by computing RMSE of the predicted state level ILI incidence and the summation of the predicted county level ILI incidence, i.e. $\sqrt{\frac{1}{n}\sum_{i=1}^{n}(\hat{y}_i - \sum \limits_{C \in \mathcal{D}} \hat{y}^C_i)^2}$.
Figure~\ref{fig:spatialconsistency} shows the spatial consistency error measured by RMSE on (a) VA, 2017-2018 and (b) NJ, 2017-2018. The results show that the spatial consistency error does not vary much with horizon, but significantly depends on $\mu$. 
The possible reason is that ,in TDEFSI model, the input is only state level data, so the LSTM layers learn the temporal pattern on state level time sequence which closely relates to model performance with horizons. However, spatial information is not propagated along the cells during training, but only compounds in the last step of outputs, thus is not impacted by horizons.
The optimal $\mu$ differs between states. The results indicate that TDEFSI enables the spatial consistency with a proper $\mu$ value. However, a better spatial consistency does not mean a better model forecasting performance. In practice, we need to keep balance between keeping good spatial consistency and maintaining good model performance. 

\begin{figure}[t]
\centering
\subfloat[VA, 2017-2018]{\includegraphics[width=2.5in]{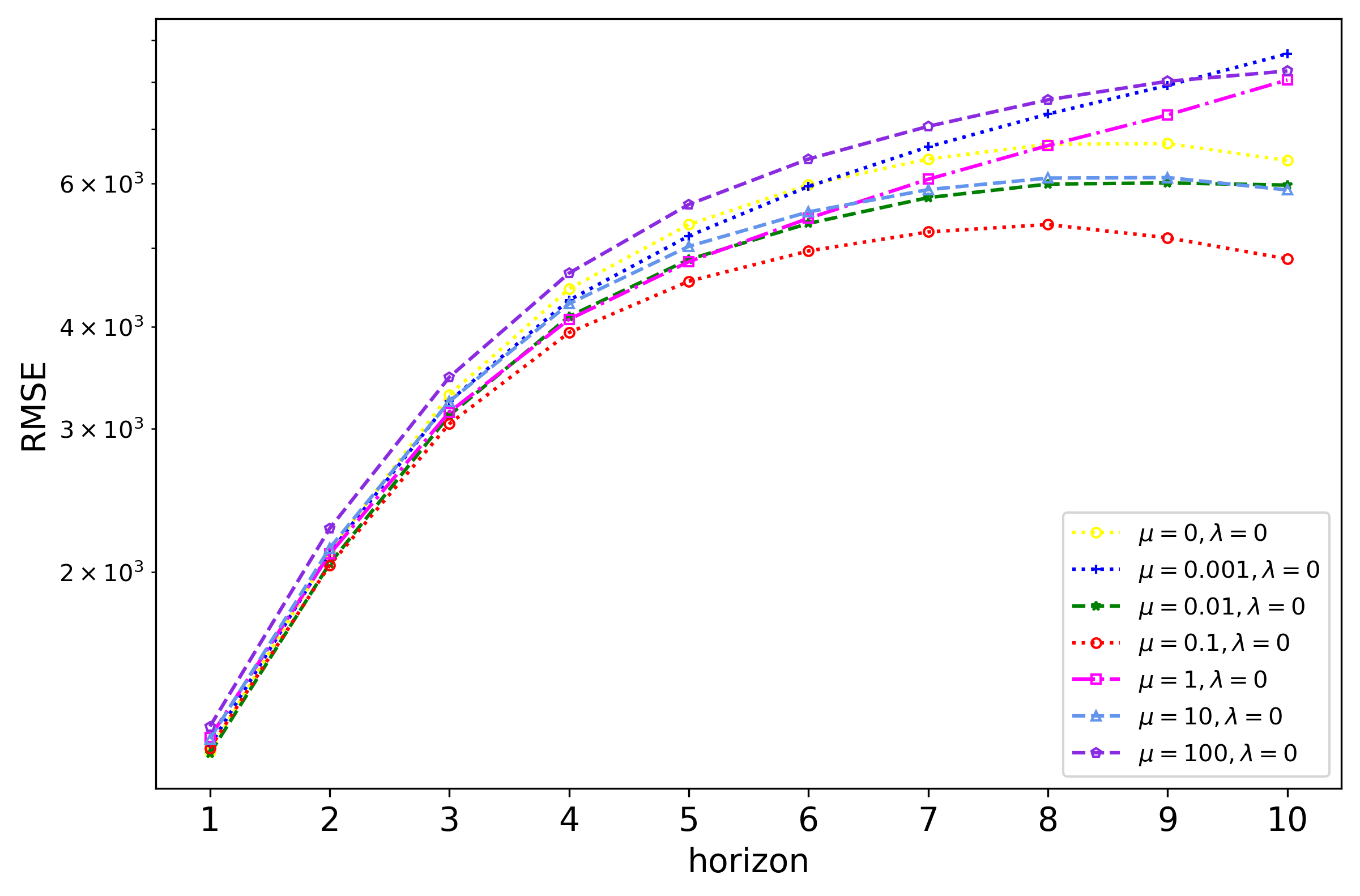}
\label{fig:vaspatial}
}
\hfil
\subfloat[NJ, 2017-2018]{\includegraphics[width=2.5in]{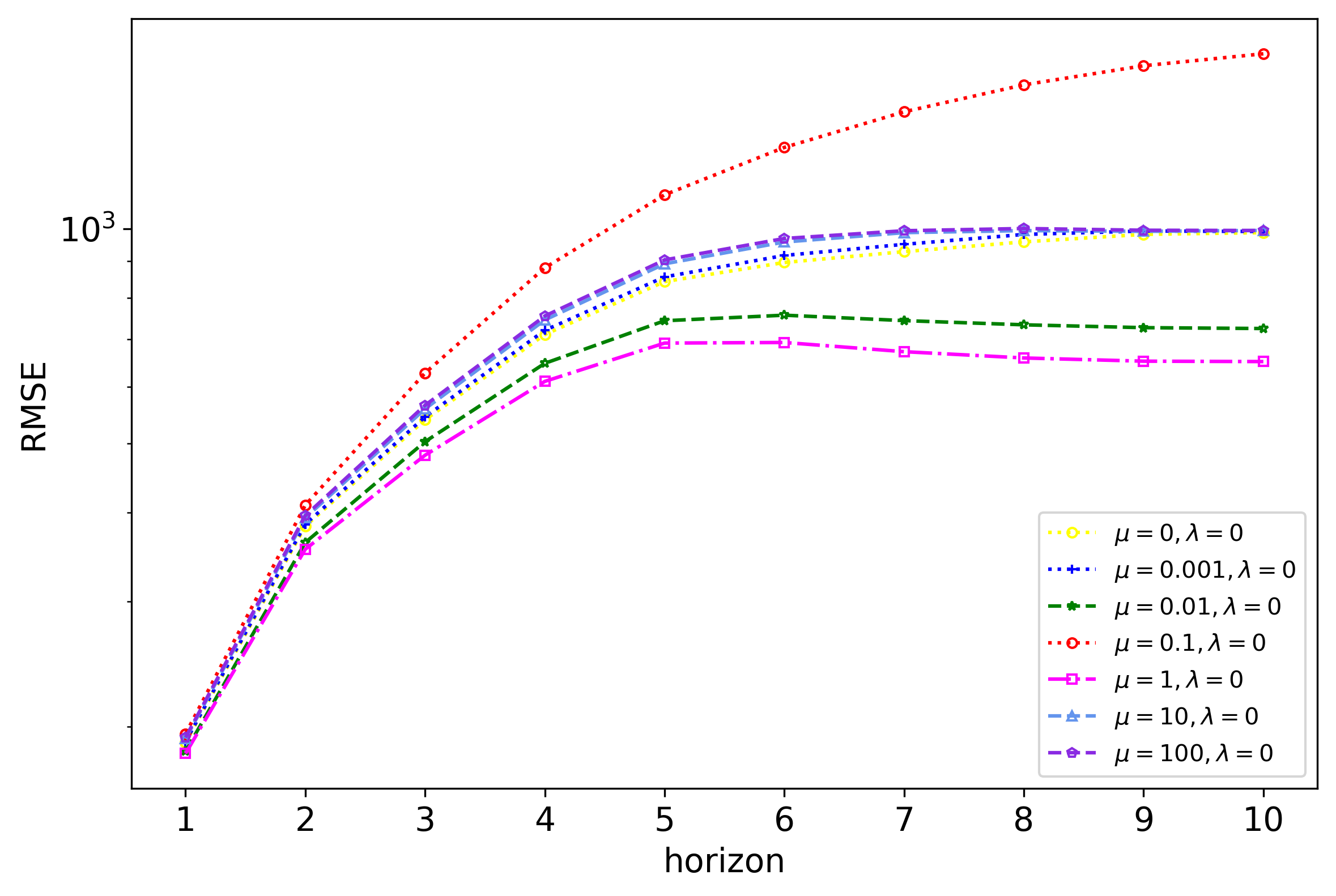}
\label{fig:njspatial}
}
\caption{TDEFSI performance with spatial consistency constraints of different coefficients $\mu =$ $\{0,$ $0.001,$ $0.01,$ $0.1,$ $1,$ $10,$ $100\}$. The performance is evaluated on (a) VA 2017-2018 season and (b) NJ 2017-2018 season. The results show that the coefficient $\mu$ has significant influence on the model forecasting performance especially with large horizons. The optimal value of $\mu$ should be chosen independently in different regions. A log y-scale is used in RMSE and MAPE.}
\label{fig:spatial}
\end{figure}

To evaluate the significance of the spatial consistency constraint for model forecasting power, we compare the forecasting performance of models on real seasonal data with various $\mu$ using RMSE (shown in Figure~\ref{fig:spatial}). 
For VA, the best performance is the model with $\mu=0.1$.
For NJ, the best performance is the model with $\mu=1$.
Overall, the spatial consistency constraint with a proper coefficient, which may vary between different regions, helps improve the forecasting performance.

\begin{figure}[t]
\centering
\subfloat[VA, 2017-2018]{\includegraphics[width=2.5in]{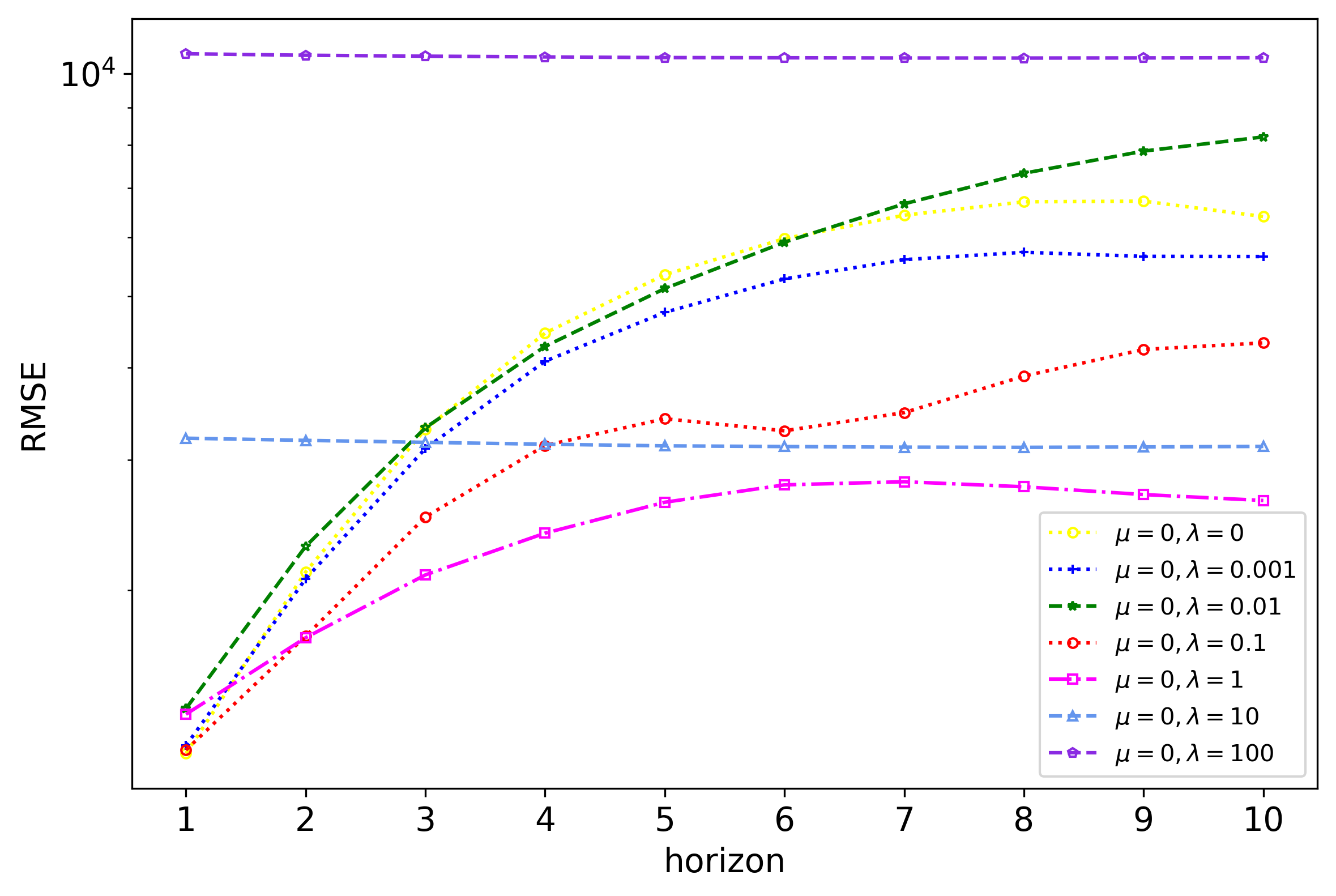}
\label{fig:vanonneg}
}
\hfil
\subfloat[NJ, 2017-2018]{\includegraphics[width=2.5in]{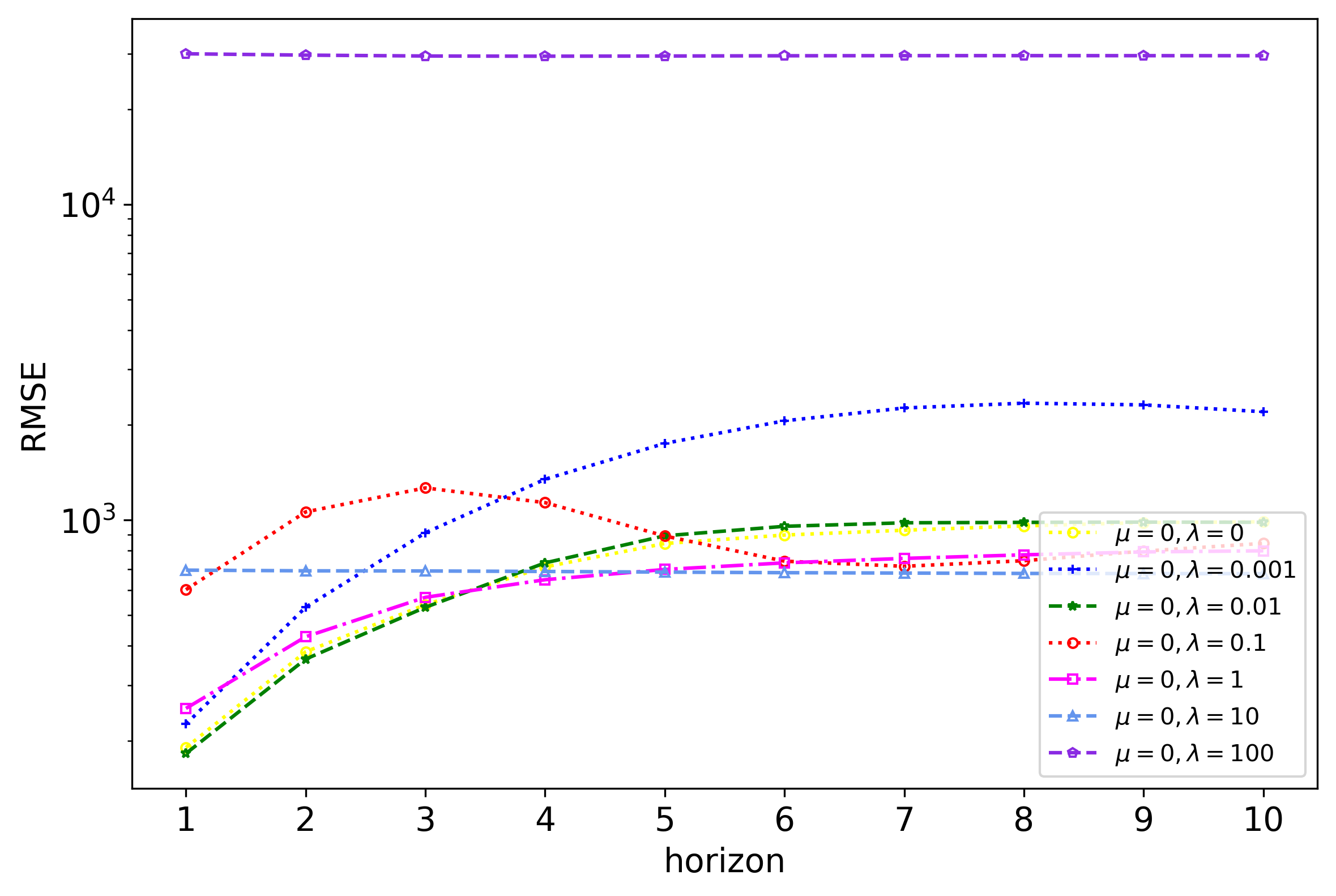}
\label{fig:njnonneg}
}
\caption{TDEFSI performance with non-negative consistency constraints of different coefficients $\lambda = \{0, 0.001, 0.01, 0.1, 1, 10, 100\}$. The performance is evaluated on (a) VA 2017-2018 season; (b) NJ 2017-2018 season. The results show that the coefficient $\lambda$ has significant influence on the model forecasting performance. The optimal value of $\lambda$ should be chosen independently in different regions. A log y-scale is used in RMSE.}
\label{fig:nonneg}
\end{figure}

\textbf{Non-negative consistency}  
The experiments are conducted using $\mu = 0$ and $\lambda =$ $\{0,$ $0.001,$ $0.01,$ $0.1,$ $1,$ $10,$ $100\}$.
Similar to the spatial consistency evaluation, we compare the performance of models with various $\lambda$ using RMSE (shown in Figure~\ref{fig:nonneg}). 
For VA, the best performance is the model with $\lambda=1$, and the models with the non-negative consistency constraint ($\lambda \leq 1$) outperform the model without the constraint. 
For NJ, the best performance is the model with $\lambda=1$. 
For both VA and NJ, from the figures we observe that the models with $\lambda$ equal or larger than 10 will have no predicting power (i.e. they are almost horizontal lines with high RMSE).
The possible reason is that a strong penalty (large $\lambda$) may cause the weights of the hidden units to shrink towards zero. When $\mathbf{W,U}$ in Equation~\ref{equ:cell} become zero the LSTM layer gives a constant output. This will make the network stop learning and output constant predictions.
Overall, the non-negative consistency constraint with a proper coefficient, which may vary between different regions, helps improve the forecasting performance.

\textbf{Implications} 
Three types of physical consistency were incorporated in our TDEFSI models. 
Computational experiments show that these constraints can lead to a better domain consistency as well as improve the forecasting performance.
By incorporating physical consistency, TDEFSI enables theory guided deep learning for epidemic forecasting. Spatial and non-negative consistency constraints also positively influence the overall performance. However we note that no single parameter setting works across all scenarios thus context specific tuning is needed.

%% file: sections/subsections/exp-vaccine-analysis.tex
\subsection{Vaccination-based Interventions} 
\label{subsec:interv}

When TDEFSI framework uses an agent-based SEIR model to generate a simulated training dataset, it is straightforward to implement various interventions in the simulations. E.g., in our parameter space $\mathcal{P}(p_E,p_I,\tau,N_I,I_V)$, $I_V$ represents the vaccination-based intervention. 
We investigate how $I_V$ affects the performance of TDEFSI by generating two synthetic training datasets: ($i$) \textbf{\textit{vaccine-case}}: generated by simulations with $I_V$ (TDEFSI and its variants in previous experiments of Section~\ref{sec:experiment} are trained on vaccine-case simulated training dataset); and ($ii$) \textbf{\textit{base-case}}: generated by simulations that share the common settings of $p_E,p_I,\tau,N_I$ with vaccine-case except $I_V = \emptyset$. 
We train TDEFSI on the vaccine-case and base-case with the same settings described in Section~\ref{subsec:setup}, and denote the trained models as \textbf{\textit{TDEFSI-vac}} and \textbf{\textit{TDEFSI-base}}, respectively. 
Note that here TDEFSI-vac is the same as TDEFSI in the previous experiments. 

\begin{figure}[t]
\centering
\subfloat[VA, 2017-2018]{\label{fig:base-vac-va}\includegraphics[width=5.3in]{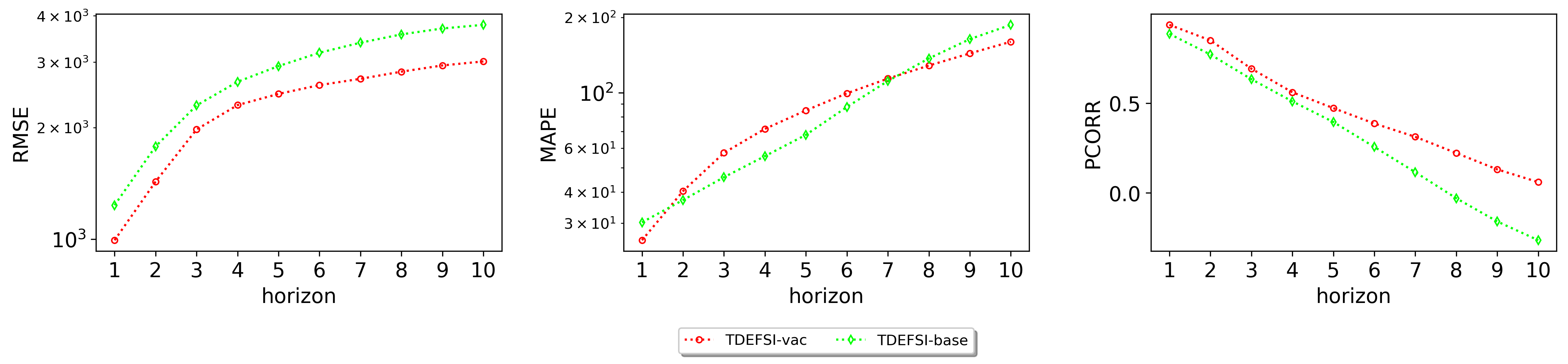}}
\hfil
\subfloat[NJ, 2017-2018]{\label{fig:base-vac-nj}\includegraphics[width=5.3in]{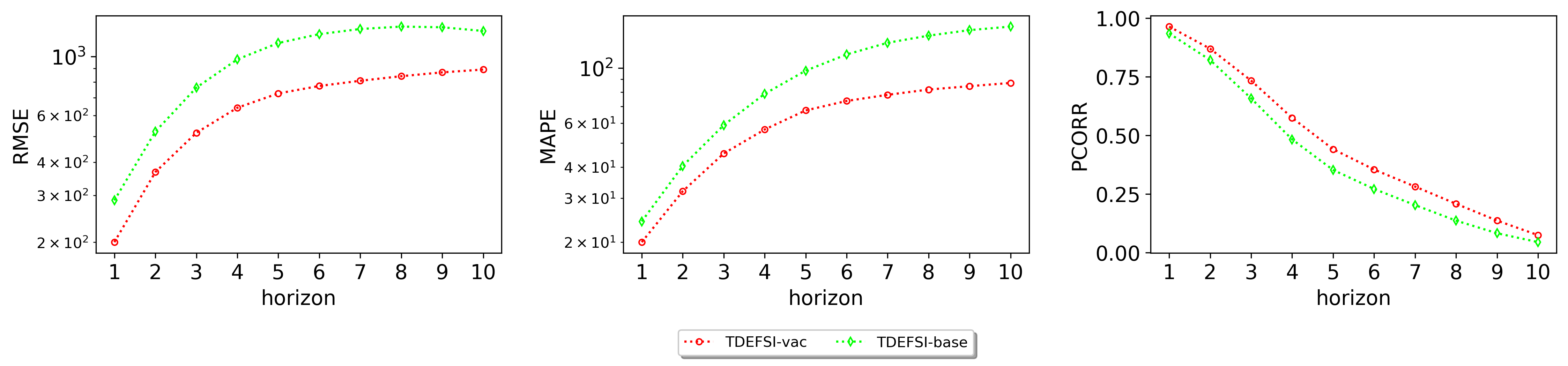}}
\hfil
\caption{State level forecasting performance comparison between TDEFSI models trained on the base-case simulated training dataset (TDEFSI-base) and the vaccine-case simulated training dataset (TDEFSI-vac). They test on VA, 2017-2018 with a horizon up to ten weeks ahead. TDEFSI-vac outperforms TDEFSI-base across three metrics. A log y-scale is used in RMSE and MAPE.}
\label{fig:njstate-vacVSbase}
\end{figure}

\begin{figure}[t]
\centering
\subfloat[NJ counties, RMSE-ratio]{\label{fig:ratermse-v}\includegraphics[width=1.7in]{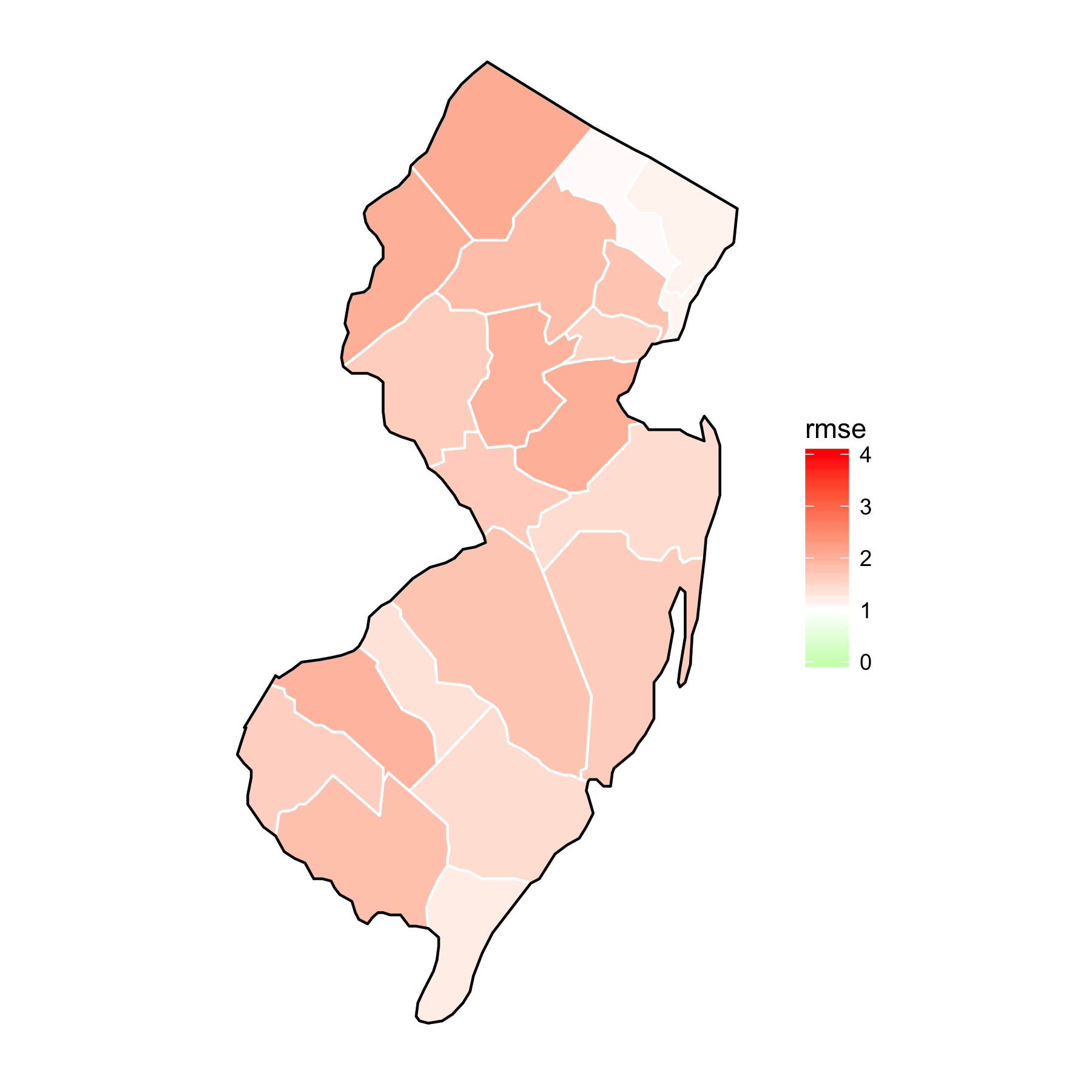}}
\hfil
\subfloat[NJ counties, MAPE-ratio]{\label{fig:ratemape-v}\includegraphics[width=1.7in]{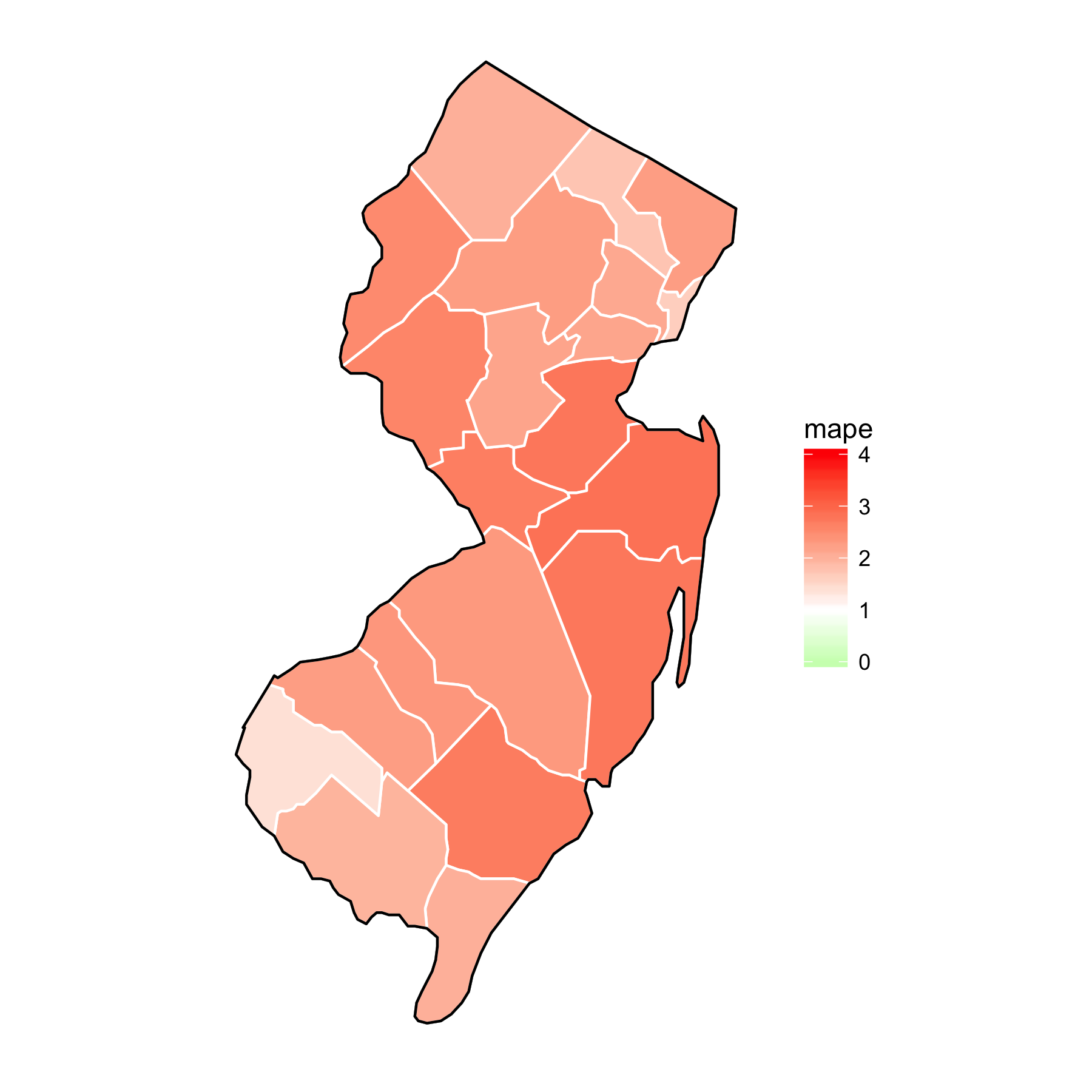}}
\hfil
\subfloat[NJ counties, PCORR-ratio]{\label{fig:ratepcorr-v}\includegraphics[width=1.7in]{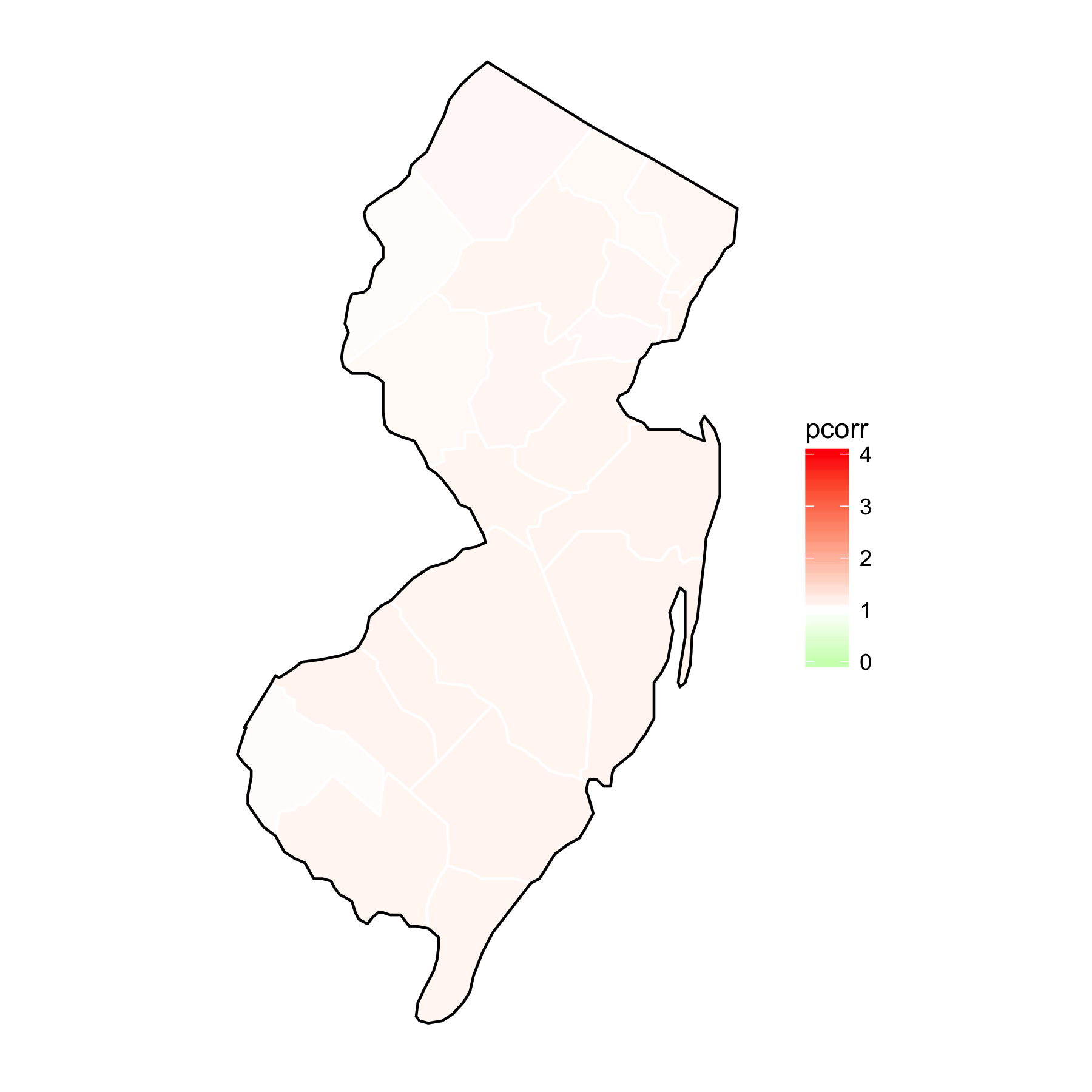}}
\hfil
\caption{NJ, 2017-2018 county level spatial forecasting performance comparison between TDEFSI-vac and TDEFSI-base for NJ, season 2017-2018. (a) RMSE-ratio; (b) MAPE-ratio; (c) PCORR-ratio. For each county in NJ, the ratio value of the county is computed using equations~\ref{equ:njrate}, which is the average value across horizons. A value larger than 1 (red color) means TDEFSI-vac outperforms TDEFSI-base, a value equal to 1 (white color) means they both perform equally, and a value smaller than 1 (green color) means TDEFSI-base performs better than TDEFSI-vac. The absolute magnitude of the value denotes the significance of the difference of the two models' performance. It is observable that TDEFSI-vac performs better than TDEFSI-base in all counties of NJ.}
\label{fig:njrate-vacVSbase}
\end{figure}

Figure~\ref{fig:base-vac-nj} and Figure~\ref{fig:base-vac-va} show the state level forecasting performance of NJ and VA on RMSE, MAPE, and PCORR using real-testing dataset. We observe that TDEFSI-vac significantly outperforms TDEFSI-base for all metrics on both states except that for the MAPE result of VA, TDEFSI-vac is compatible with TDEFSI-base.
In Figure~\ref{fig:njrate-vacVSbase}, we present the comparison ratio between two models from the spatial dimension of NJ counties. It is observable that TDEFSI-vac performs better than TDEFSI-base in all counties of NJ. 
The results indicate that vaccination-based interventions applied in the simulations to generate training datasets can significantly improve the forecasting performance. 

The models learned from the vaccine-case datasets are more generalizable to unseen surveillance data.
Our experiments show the significance of vaccination-based interventions applied in the simulations on the forecasting performance.
The proposed framework is extensible for other realistic interventions, such as school closure or antivirals, to further improve the forecasting performance.

%% file: sections/subsections/exp-uncertainty.tex
\subsection{Prediction Uncertainty Estimation}
\label{subsec:uncertainty}
\begin{figure}[t]
  \includegraphics[width=0.6\textwidth]{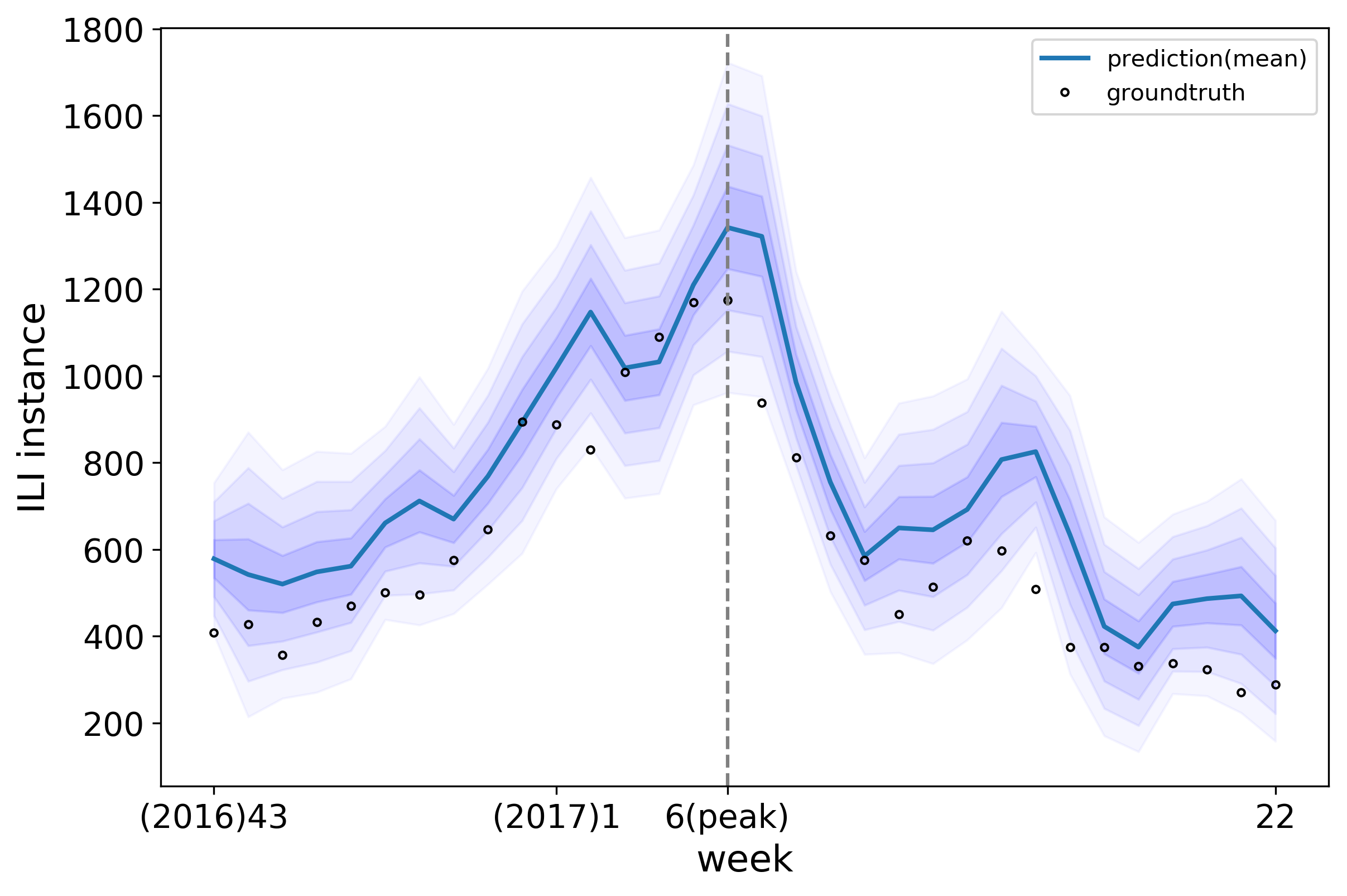}
  \caption{NJ state level mean predicted curve with predictive intervals of $(mean \pm k*std)$ where $k=\{0.5, 1, 1.5, 2\}$. The black circles are ground truths. We can observe that all ground truths are within 2 standard deviations. }
  \label{fig:uncertainty}
\end{figure}

In the epidemic forecasting domain, probabilistic forecasting is important for capturing the uncertainty of the disease dynamics and to better support public health decision making. Probabilistic forecasting with deep learning models is challenging due to the lack of interpretability of such models. Most works on this are based on Bayesian Neural Networks. Gal et al.~\cite{gal2016dropout} in 2016 proved that using dropout technique is equivalent to Bayesian NN's and proposed Monte Carlo Dropout (MC Dropout) to estimate uncertainty in deep learning. The proposed method is computationally efficient. We implement MC Dropout in TDEFSI and demonstrate estimation of prediction uncertainty with a case study of state level forecasting for NJ season 2016-2017. The model setting is the same as that described in~\ref{subsec:setup}, and the MC number is 20. Figure~\ref{fig:uncertainty} shows the curve of mean predictions with predictive intervals of $(mean \pm k*std)$ where $k=\{0.5, 1, 1.5, 2\}$. We can observe that all ground truths are within 2 standard deviations. 

%% file: sections/conclusion.tex
\section{Concluding remarks and directions for future work}\label{sec:conclusion}
We described TDEFSI -- a novel epidemic forecasting framework that combines deep learning methods with high performance computing oriented simulations of epidemic processes over realistic social contact networks.
TDEFSI and its variants use a two-branch LSTM based neural network model and are designed to combine within-season and between-season observations. 
TDEFSI incorporates domain knowledge into deep neural network models by considering temporal, spatial, and non-negative consistency constraints as well as natural constraints imposed by the use of epidemic simulations. 

The models are trained on a region-specific simulated dataset constructed at multiple spatially fine-grained scales. 
The trained models can provide high-resolution forecasts using flat-resolution surveillance data.
We carried out extensive computational experiments on NJ and VA, using synthetic as well as state level real surveillance data. 
The results show that TDEFSI combined with epidemic simulations achieve comparable/better performance than the state-of-the-art methods for ILI forecasting at the state level. For high-resolution forecasting at the county level, TDEFSI significantly outperforms the comparison methods.
Through sensitivity analysis on spatial and non-negative consistency constraints, we discuss the influence of these constraints on model performance.
A case study of probabilistic forecasting on NJ state shows the model's ability to provide prediction uncertainty using MC Dropout technique. 
Experiments involving more states and more seasons are desirable to show that the performance comparison of TDEFSI against other methods is robust, but due to the limitation on the availability of high resolution data and historical data of flu seasons we only tested the framework on two states and two seasons. In future work, we plan to look for more datasets so that the robustness of our observations can be tested. 

\noindent\textbf{Future work.} 
A direction for future work is to investigate the use of synthetic data generated by social, epidemiological, and behavioral models in conjunction with observed data to improve epidemic forecasts.
($i$) In this work, we try to reduce the gap between simulated and real world data distributions by simulating with parameter settings learned from observations so that the generated epi-curves are realistic. In future work, we plan to further reduce the gap by using synthetic data based on real-time observations to train the neural networks. 
($ii$) We also plan to explore the capability of TDEFSI on {\em what-if} forecasts. What-if forecasts capture various what-if scenarios due to expected or unexpected public health interventions or individual level behavioral reactions as the epidemic evolves. They provide insights on possible trajectories of the ongoing epidemic under different assumptions. They can help public health decision making with risk/benefit predictions. The data-driven methods can only provide passive forecasts, while what-if forecasts are natural in TDEFSI thanks to the causal model behind it. A possible way to make what-if forecasts with TDEFSI works as follows:
based on the current status of the epidemic, make a few assumptions about what may happen in the future that will change the epidemic dynamics; implement each assumption as a set of interventions (e.g. school closure from $ew(51)$ to $ew(52)$) in the simulations and generate synthetic epi-curves; re-train the deep neural network with the updated synthetic curves; and make predictions which describe future dynamics with this particular assumption.
Note that one what-if scenario can be associated with multiple interventions.

%% file: sections/appendix.tex
\section{Appendix}
\label{sec:appendix}
\subsection{Synthetic social contact network}
A synthetic population and the corresponding social contact network are used to simulate the spread of the disease. In our work, we use the synthetic social contact network of Virginia and New Jersey. Below we briefly describe the methodology used for constructing the synthetic population and the social network{\footnote{The description is similar as the one we described in our previous work~\cite{wang2018framework} since they use the same synthetic dataset.}}. Interested readers can find more details about this methodology in~\cite{barrett2009generation,beckman1996,bisset2009, eubank2004,ndssl:opendata2014}. 

To construct the social network, first a statistical representation of each individual in the population is built using US Census data. This synthetic population is statistically equivalent to the real population when aggregated to the census block group level.  Individuals in the synthetic population are assigned a complete range of demographic attributes as available in the Census~\cite{beckman1996,bisset2009cyber}, including age, gender, household location, and household income.

Next, a set of activity templates are extracted from American time-use surveys~\cite{ATUSsurvey} and the National Household Travel Survey. Each of these activity templates provides a daily sequence of activities for individuals and the time of day they are performed. Each synthetic household is matched with one of the survey households, using a decision tree based on demographics
such as the size of the household, number of workers in the household, number of children, etc. The synthetic household members are then assigned the activity templates of the matching survey household members, giving each synthetic individual a daily sequence of activities.
For each activity of each individual, a geographic location is identified based on land-use patterns, transportation network, and data from commercially available databases such as Dun and BradStreet.

A social network is constructed by connecting individuals simultaneously present at the same location. The co-location based social network is dynamic and changes as people visit different locations and come in contact with individuals at these locations.

\subsection{Surveillance ratio}\label{subsec:hosp}
In our experiments, we scale the ILINet case count to the population case count using a surveillance ratio. We assume that the ratio between ILI cases captured by CDC ILINet (denoted ILITOTAL) and ILI cases in the population (ILIPOP) is the same as that between patients of all diseases captured by CDC ILINet (TOTALPATIENT) and patients of all diseases in the population (PATIENTPOP). We approximate PATIENTPOP with all doctor visit data from AHRQ~\cite{doctorratio2017}. The doctor visit data provides county level counts for total hospital visits in a year which is aggregated to state level counts later. Note that it is an underestimate. From surveillance ratio = $\frac{\mbox{\footnotesize ILITOTAL}}{\mbox{\footnotesize ILIPOP}}=\frac{\mbox{\footnotesize TOTALPATIENT}}{\mbox{\footnotesize PATIENTPOP}}$, we can derive the only unknown ILIPOP. 
Table~\ref{tab:hosp} presents the surveillance ratios for all the states.  

\input{tables/apdx-hosp-ratio.tex}

\subsection{Disease parameter space} 
\label{subsec:fit}
Among $\mathcal{P}$, $p_I, p_E$ are from literature, $I_V$ is derived from historical data and we assume $I_V$ follows a discrete uniform distribution. The distributions of $\tau$ and $N_I$ are fitted distributions using KS-test on collected samples. The samples used to fit a distribution are collected from historical training seasons. For example, given a state New Jersey, the training data includes 6 seasons from 2010-2011 to 2015-2016, its neighbors are Delaware, New York, and Pennsylvania. Then we can collect $6 * 4 = 24$ samples of $ar$ or $N_I$ for NJ. 
We calibrate $\tau$ using Nelder-Mead~\cite{nelder1965} algorithm based on each collected pair of $(ar, N_I)$. For each of $ar, N_I, \tau$, we obtained 36 data points for VA and 24 for NJ.
At the fitting step, normal and uniform distributions are included. We run KS-test (the null hypothesis being that the sample is drawn from the reference distribution) to choose a distribution with the highest significance (p-value). 
The learned parameter space is shown in Table~\ref{tab:distr}. Note that each parameter in $\mathcal{P}$ follows a marginal distribution. 

\input{tables/apdx-param-space.tex}

\subsection{Baseline Model Settings}
\label{subsec:baselinesettings}
In this section, we elaborate the details of model setting of the baselines. Note that, in the experiments, we choose the final model with the best validation accuracy by grid searching. Unless explicitly noted, the hyperparameters are set with default values from python libraries. 
\begin{itemize}
\item \textit{Single layer LSTM model (LSTM)}: It consists of one LSTM layer and one dense layer. The input is the sequence of state level ILI incidence and the output is the state level prediction of the current week. By grid searching, we set the look back window size to 52 and LSTM hidden units to 128. The Adam optimizer is used.

\item \textit{AdapLSTM}~\cite{venna2019novel}: This method makes predictions using a simple LSTM model, then adjusts the predictions by applying impacts of weather factors and spatiotemporal factors. The LSTM model has the same setting with single layer LSTM model described above. In~\cite{venna2019novel}, the weather features include maximum temperature, minimum temperature, humidity, and precipitation. However, humidity is not used in our experiments since it is not publicly available in the collected weather dataset. The confidences of symbol pairs (the climatic variable time series and the flu count time series) in our experiment are less than 0.3, which will lead to arbitrary adjustment for predictions. The neighbors of each state used for spatiotemporal adjustment factor are geographical adjacent states that are the same with those used in constructing disease parameter space. For more details please refer to the original paper~\cite{venna2019novel}.

\item \textit{Simple SARIMA model (SARIMA)}: We use the Seasonal ARIMA model, denoted as $SARIMA$ $(p,d,q)$$\times$$(P,D,Q)_m$, where $p$ is the order (number of time lags) of the autoregressive model, $d$ is the degree of differencing (the number of times the data have had past values subtracted), $q$ is the order of the moving-average model, $m$ refers to the number of periods in each season, and the uppercase $P,D,Q$ refer to the autoregressive, differencing, and moving average terms for the seasonal part of the SARIMA model. By grid searching, the selected model is $SARIMA(8, 1, 0)\times(5, 0, 0)_{52}$. No exogenous variables are used in this model. 

\item \textit{AutoRegression with Google search data (ARGO)}~\cite{yang2015accurate}: The method uses an autoregression model utilizing Google search data. We use the publicly available tool from~\cite{yang2015accurate}. In our experiment, we set the look back window size to 52 and the training window to 104. In the Google data we collected, all of the top 100 Google correlate terms of VA are flu related, while only one out of the top 100 Google correlated terms of NJ are flu related. This may cause ARGO to perform better on VA than on NJ as discussed in Section~\ref{subsec:discussion}.

\item \textit{EpiFast}~\cite{beckman2014}:  This method takes the same setting of $p_E$ and $p_I$ as shown in Table\ref{tab:distr}, and searches for $N_I, \tau$ by minimizing the dissimilarity between the predicted and the actual ILI incidence using the Nelder-Mead algorithm~\cite{nelder1965}. 
\end{itemize}


%% file: tables/apdx-hosp-ratio.tex
\begin{table}
    \begin{center}
    \caption{Surveillance ratios for each state in the US}
    \label{tab:hosp}
    \begin{tabular}{l|l|l}
    \toprule
Alabama: 0.0759 & Kansas: 0.1093 & New York: 0.1204 \\
Alaska: 0.1143 & Kentucky: 0.1114 & North Carolina: 0.0875 \\
Arizona: 0.0723 & Louisiana: 0.0931 & North Dakota: 0.1960 \\
Arkansas: 0.0894 & Maine: 0.1931 & Ohio: 0.1339 \\
California: 0.0628 & Maryland: 0.0755 & Oklahoma: 0.1039 \\
Colorado: 0.0764 & Massachusetts: 0.1380 & Oregon: 0.1050 \\
Connecticut: 0.1047 & Michigan: 0.1356 & Pennsylvania: 0.1299 \\
Delaware: 0.1030 & Minnesota: 0.0898 & Rhode Island: 0.0932 \\
District of Columbia: 0.1852 & Mississippi: 0.0874 & South Carolina: 0.0663 \\
Florida: 0.0582 & Missouri: 0.1492 & South Dakota: 0.1882 \\
Georgia: 0.0701 & Montana: 0.1739 & Tennessee: 0.0811 \\
Hawaii: 0.0705 & Nebraska: 0.1329 & Texas: 0.0738 \\
Idaho: 0.1190 & Nevada: 0.0643 & Utah: 0.0913 \\
Illinois: 0.1066 & New Hampshire: 0.1566 & Vermont: 0.2111 \\
Indiana: 0.1215 & New Jersey: 0.0692 & Virginia: 0.0914 \\
Iowa: 0.1420 & New Mexico: 0.1258 & Washington: 0.0885 \\
Kansas: 0.1093 & New York: 0.1204 & West Virginia: 0.1684 \\
    \bottomrule
    \end{tabular}
    \begin{tablenotes}
    \item  
    \end{tablenotes}
    \end{center}
\end{table}

%% file: tables/apdx-param-space.tex
  \begin{table}
    \begin{center}
    \caption{Marginal Distributions of the Parameter Spaces for VA and NJ. $\mathcal{N}$ denotes normal distribution, $\mathcal{U}$ denotes uniform distribution.}
    \label{tab:distr}
    \begin{tabular}{c|c|r|r|r}
\hline
       \textbf{Parameter} & \textbf{State} & \textbf{Name} & \textbf{Distribution} & \textbf{P-value} \\
\hline 
\multirow{2}{*}{$p_E$} 
&VA &Discrete distribution&(1:0.3, 2:0.5, 3:0.2)~\cite{achla2011b,wang2018framework}&- \\
&NJ &Discrete distribution&(1:0.3, 2:0.5, 3:0.2)~\cite{achla2011b,wang2018framework}&- \\
\hline

\multirow{2}{*}{$p_I$} 
&VA &Discrete distribution&(3:0.3, 4:0.4, 5:0.2, 6:0.1)~\cite{achla2011b,wang2018framework}&- \\
&NJ &Discrete distribution&(3:0.3, 4:0.4, 5:0.2, 6:0.1)~\cite{achla2011b,wang2018framework}&- \\
\hline

\multirow{2}{*}{$\tau$} 
&VA &Normal&$\mathcal{N}(\mu=4.88\mathrm{e}{-5},\delta=9.33\mathrm{e}{-7})$&0.74 \\
&NJ &Normal&$\mathcal{N}(\mu=4.63\mathrm{e}{-5},\delta=1.05\mathrm{e}{-6})$&0.85 \\
\hline

\multirow{2}{*}{$N_I$} 
&VA &Uniform&$\mathcal{U}(7355,16278)$&0.85 \\
&NJ &Uniform&$\mathcal{U}(567,7647)$&0.40 \\
\hline

\multirow{2}{*}{$I_V$} 
&VA &Discrete uniform& 6 vaccination schedules~\cite{vacschedule}&- \\
&NJ &Discrete uniform& 6 vaccination schedules~\cite{vacschedule}&- \\
\hline
    \end{tabular}
    \begin{tablenotes}
      \item The null hypothesis for the two-sample KS test is that both groups were sampled from populations with identical distributions. If the p-value returned by the KS test is less than a significance level, we reject the null hypothesis. In our experiments, we do not specify a significance level but instead choose the distribution with the largest p-value among multiple assumed distributions.
    \end{tablenotes}
    \end{center}
  \end{table}

%% file: main.bbl

\begin{thebibliography}{103}


\ifx \showCODEN    \undefined \def \showCODEN     #1{\unskip}     \fi
\ifx \showDOI      \undefined \def \showDOI       #1{#1}\fi
\ifx \showISBNx    \undefined \def \showISBNx     #1{\unskip}     \fi
\ifx \showISBNxiii \undefined \def \showISBNxiii  #1{\unskip}     \fi
\ifx \showISSN     \undefined \def \showISSN      #1{\unskip}     \fi
\ifx \showLCCN     \undefined \def \showLCCN      #1{\unskip}     \fi
\ifx \shownote     \undefined \def \shownote      #1{#1}          \fi
\ifx \showarticletitle \undefined \def \showarticletitle #1{#1}   \fi
\ifx \showURL      \undefined \def \showURL       {\relax}        \fi
\providecommand\bibfield[2]{#2}
\providecommand\bibinfo[2]{#2}
\providecommand\natexlab[1]{#1}
\providecommand\showeprint[2][]{arXiv:#2}

\bibitem[\protect\citeauthoryear{ACS}{ACS}{2013}]%
        {commute}
\bibfield{author}{\bibinfo{person}{ACS}.} \bibinfo{year}{2009-2013}\natexlab{}.
\newblock \bibinfo{title}{{2009-2013 5-Year American Community Survey Commuting
  Flows}}.
\newblock
  \bibinfo{howpublished}{\url{https://www.census.gov/data/tables/time-series/demo/commuting/commuting-flows.html}}.
\newblock


\bibitem[\protect\citeauthoryear{AHRQ}{AHRQ}{2017}]%
        {doctorratio2017}
\bibfield{author}{\bibinfo{person}{AHRQ}.} \bibinfo{year}{2017}\natexlab{}.
\newblock \bibinfo{title}{Hospital visits for a population}.
\newblock
  \bibinfo{howpublished}{\url{https://www.ahrq.gov/data/resources/index.html}}.
\newblock
\newblock
\shownote{Accessed June 01, 2017.}


\bibitem[\protect\citeauthoryear{Alessa and Faezipour}{Alessa and
  Faezipour}{2018}]%
        {alessa2018}
\bibfield{author}{\bibinfo{person}{Ali Alessa} {and} \bibinfo{person}{Miad
  Faezipour}.} \bibinfo{year}{2018}\natexlab{}.
\newblock \showarticletitle{A review of influenza detection and prediction
  through social networking sites}.
\newblock \bibinfo{journal}{\emph{Theoretical Biology \& Medical Modelling}}
  \bibinfo{volume}{15} (\bibinfo{year}{2018}), \bibinfo{pages}{2}.
\newblock


\bibitem[\protect\citeauthoryear{Bailey et~al\mbox{.}}{Bailey
  et~al\mbox{.}}{1975}]%
        {bailey1975mathematical}
\bibfield{author}{\bibinfo{person}{Norman~TJ Bailey} {et~al\mbox{.}}}
  \bibinfo{year}{1975}\natexlab{}.
\newblock \bibinfo{booktitle}{\emph{The mathematical theory of infectious
  diseases and its applications}}.
\newblock Number 2nd ediition. \bibinfo{publisher}{Charles Griffin \& Company
  Ltd 5a Crendon Street, High Wycombe, Bucks HP13 6LE.}
\newblock


\bibitem[\protect\citeauthoryear{Bardak and Tan}{Bardak and Tan}{2015}]%
        {bardak2015prediction}
\bibfield{author}{\bibinfo{person}{Batuhan Bardak} {and}
  \bibinfo{person}{Mehmet Tan}.} \bibinfo{year}{2015}\natexlab{}.
\newblock \showarticletitle{Prediction of influenza outbreaks by integrating
  Wikipedia article access logs and Google flu trend data}. In
  \bibinfo{booktitle}{\emph{2015 IEEE 15th International Conference on
  Bioinformatics and Bioengineering (BIBE)}}. IEEE, \bibinfo{pages}{1--6}.
\newblock


\bibitem[\protect\citeauthoryear{Barrett, Beckman, Khan, Anil~Kumar, Marathe,
  Stretz, Dutta, and Lewis}{Barrett et~al\mbox{.}}{2009}]%
        {barrett2009generation}
\bibfield{author}{\bibinfo{person}{Christopher~L Barrett},
  \bibinfo{person}{Richard~J Beckman}, \bibinfo{person}{Maleq Khan},
  \bibinfo{person}{VS Anil~Kumar}, \bibinfo{person}{Madhav~V Marathe},
  \bibinfo{person}{Paula~E Stretz}, \bibinfo{person}{Tridib Dutta}, {and}
  \bibinfo{person}{Bryan Lewis}.} \bibinfo{year}{2009}\natexlab{}.
\newblock \showarticletitle{Generation and analysis of large synthetic social
  contact networks}. In \bibinfo{booktitle}{\emph{Winter Simulation
  Conference}}. Winter Simulation Conference, \bibinfo{pages}{1003--1014}.
\newblock


\bibitem[\protect\citeauthoryear{Beckman, Bisset, Chen, Lewis, Marathe, and
  Stretz}{Beckman et~al\mbox{.}}{2014}]%
        {beckman2014}
\bibfield{author}{\bibinfo{person}{Richard Beckman}, \bibinfo{person}{Keith~R
  Bisset}, \bibinfo{person}{Jiangzhuo Chen}, \bibinfo{person}{Bryan Lewis},
  \bibinfo{person}{Madhav Marathe}, {and} \bibinfo{person}{Paula Stretz}.}
  \bibinfo{year}{2014}\natexlab{}.
\newblock \showarticletitle{Isis: A networked-epidemiology based pervasive web
  app for infectious disease pandemic planning and response}. In
  \bibinfo{booktitle}{\emph{Proceedings of the 20th ACM SIGKDD international
  conference on Knowledge discovery and data mining}}. ACM,
  \bibinfo{pages}{1847--1856}.
\newblock


\bibitem[\protect\citeauthoryear{Beckman, Baggerly, and McKay}{Beckman
  et~al\mbox{.}}{1996}]%
        {beckman1996}
\bibfield{author}{\bibinfo{person}{Richard~J Beckman}, \bibinfo{person}{Keith~A
  Baggerly}, {and} \bibinfo{person}{Michael~D McKay}.}
  \bibinfo{year}{1996}\natexlab{}.
\newblock \showarticletitle{Creating synthetic baseline populations}.
\newblock \bibinfo{journal}{\emph{Transportation Research Part A: Policy and
  Practice}} \bibinfo{volume}{30}, \bibinfo{number}{6} (\bibinfo{year}{1996}),
  \bibinfo{pages}{415--429}.
\newblock


\bibitem[\protect\citeauthoryear{Benjamin, Rigby, and Stasinopoulos}{Benjamin
  et~al\mbox{.}}{2003}]%
        {benjamin2003generalized}
\bibfield{author}{\bibinfo{person}{Michael~A Benjamin},
  \bibinfo{person}{Robert~A Rigby}, {and} \bibinfo{person}{D~Mikis
  Stasinopoulos}.} \bibinfo{year}{2003}\natexlab{}.
\newblock \showarticletitle{Generalized autoregressive moving average models}.
\newblock \bibinfo{journal}{\emph{Journal of the American Statistical
  association}} \bibinfo{volume}{98}, \bibinfo{number}{461}
  (\bibinfo{year}{2003}), \bibinfo{pages}{214--223}.
\newblock


\bibitem[\protect\citeauthoryear{Bergmeir, Hyndman, and Ben{\'\i}tez}{Bergmeir
  et~al\mbox{.}}{2016}]%
        {bergmeir2016bagging}
\bibfield{author}{\bibinfo{person}{Christoph Bergmeir}, \bibinfo{person}{Rob~J
  Hyndman}, {and} \bibinfo{person}{Jos{\'e}~M Ben{\'\i}tez}.}
  \bibinfo{year}{2016}\natexlab{}.
\newblock \showarticletitle{Bagging exponential smoothing methods using STL
  decomposition and Box-Cox transformation}.
\newblock \bibinfo{journal}{\emph{International journal of forecasting}}
  \bibinfo{volume}{32}, \bibinfo{number}{2} (\bibinfo{year}{2016}),
  \bibinfo{pages}{303--312}.
\newblock


\bibitem[\protect\citeauthoryear{Biggerstaff, Alper, Dredze, Fox, Fung,
  Hickmann, Lewis, Rosenfeld, Shaman, Tsou, et~al\mbox{.}}{Biggerstaff
  et~al\mbox{.}}{2016}]%
        {biggerstaff2016results}
\bibfield{author}{\bibinfo{person}{Matthew Biggerstaff}, \bibinfo{person}{David
  Alper}, \bibinfo{person}{Mark Dredze}, \bibinfo{person}{Spencer Fox},
  \bibinfo{person}{Isaac Chun-Hai Fung}, \bibinfo{person}{Kyle~S Hickmann},
  \bibinfo{person}{Bryan Lewis}, \bibinfo{person}{Roni Rosenfeld},
  \bibinfo{person}{Jeffrey Shaman}, \bibinfo{person}{Ming-Hsiang Tsou},
  {et~al\mbox{.}}} \bibinfo{year}{2016}\natexlab{}.
\newblock \showarticletitle{Results from the centers for disease control and
  prevention's predict the 2013-2014 Influenza Season Challenge}.
\newblock \bibinfo{journal}{\emph{BMC infectious diseases}}
  \bibinfo{volume}{16}, \bibinfo{number}{1} (\bibinfo{year}{2016}),
  \bibinfo{pages}{357}.
\newblock


\bibitem[\protect\citeauthoryear{Biggerstaff, Johansson, Alper, Brooks,
  Chakraborty, Farrow, Hyun, Kandula, McGowan, Ramakrishnan,
  et~al\mbox{.}}{Biggerstaff et~al\mbox{.}}{2018}]%
        {biggerstaff2018results}
\bibfield{author}{\bibinfo{person}{Matthew Biggerstaff},
  \bibinfo{person}{Michael Johansson}, \bibinfo{person}{David Alper},
  \bibinfo{person}{Logan~C Brooks}, \bibinfo{person}{Prithwish Chakraborty},
  \bibinfo{person}{David~C Farrow}, \bibinfo{person}{Sangwon Hyun},
  \bibinfo{person}{Sasikiran Kandula}, \bibinfo{person}{Craig McGowan},
  \bibinfo{person}{Naren Ramakrishnan}, {et~al\mbox{.}}}
  \bibinfo{year}{2018}\natexlab{}.
\newblock \showarticletitle{Results from the second year of a collaborative
  effort to forecast influenza seasons in the United States}.
\newblock \bibinfo{journal}{\emph{Epidemics}}  \bibinfo{volume}{24}
  (\bibinfo{year}{2018}), \bibinfo{pages}{26--33}.
\newblock


\bibitem[\protect\citeauthoryear{Bisset and Marathe}{Bisset and
  Marathe}{2009}]%
        {bisset2009cyber}
\bibfield{author}{\bibinfo{person}{Keith Bisset} {and} \bibinfo{person}{Madhav
  Marathe}.} \bibinfo{year}{2009}\natexlab{}.
\newblock \showarticletitle{A cyber-environment to support pandemic planning
  and response}.
\newblock \bibinfo{journal}{\emph{DOE SciDAC Magazine}}  \bibinfo{volume}{13}
  (\bibinfo{year}{2009}), \bibinfo{pages}{36--47}.
\newblock


\bibitem[\protect\citeauthoryear{Bisset, Chen, Feng, Kumar, and Marathe}{Bisset
  et~al\mbox{.}}{2009}]%
        {bisset2009}
\bibfield{author}{\bibinfo{person}{Keith~R. Bisset}, \bibinfo{person}{Jiangzhuo
  Chen}, \bibinfo{person}{Xizhou Feng}, \bibinfo{person}{V.S.~Anil Kumar},
  {and} \bibinfo{person}{Madhav~V. Marathe}.} \bibinfo{year}{2009}\natexlab{}.
\newblock \showarticletitle{{EpiFast}: A Fast Algorithm for Large Scale
  Realistic Epidemic Simulations on Distributed Memory Systems}. In
  \bibinfo{booktitle}{\emph{Proceedings of the 23rd international conference on
  Supercomputing}}. \bibinfo{publisher}{ACM}, \bibinfo{pages}{430--439}.
\newblock


\bibitem[\protect\citeauthoryear{Brockmann and Helbing}{Brockmann and
  Helbing}{2013}]%
        {brockmann2013hidden}
\bibfield{author}{\bibinfo{person}{Dirk Brockmann} {and} \bibinfo{person}{Dirk
  Helbing}.} \bibinfo{year}{2013}\natexlab{}.
\newblock \showarticletitle{The hidden geometry of complex, network-driven
  contagion phenomena}.
\newblock \bibinfo{journal}{\emph{Science}} \bibinfo{volume}{342},
  \bibinfo{number}{6164} (\bibinfo{year}{2013}), \bibinfo{pages}{1337--1342}.
\newblock


\bibitem[\protect\citeauthoryear{Brooks, Farrow, Hyun, Tibshirani, and
  Rosenfeld}{Brooks et~al\mbox{.}}{2018}]%
        {brooks2018nonmechanistic}
\bibfield{author}{\bibinfo{person}{Logan~C Brooks}, \bibinfo{person}{David~C
  Farrow}, \bibinfo{person}{Sangwon Hyun}, \bibinfo{person}{Ryan~J Tibshirani},
  {and} \bibinfo{person}{Roni Rosenfeld}.} \bibinfo{year}{2018}\natexlab{}.
\newblock \showarticletitle{Nonmechanistic forecasts of seasonal influenza with
  iterative one-week-ahead distributions}.
\newblock \bibinfo{journal}{\emph{PLoS computational biology}}
  \bibinfo{volume}{14}, \bibinfo{number}{6} (\bibinfo{year}{2018}),
  \bibinfo{pages}{e1006134}.
\newblock


\bibitem[\protect\citeauthoryear{{Bureau of Labor Statistics}}{{Bureau of Labor
  Statistics}}{2017}]%
        {ATUSsurvey}
\bibfield{author}{\bibinfo{person}{{Bureau of Labor Statistics}}.}
  \bibinfo{year}{2017}\natexlab{}.
\newblock \bibinfo{title}{American Time Use Survey}.
\newblock \bibinfo{howpublished}{\url{https://www.bls.gov/tus/}}.
\newblock


\bibitem[\protect\citeauthoryear{CDC}{CDC}{2018}]%
        {vacschedule}
\bibfield{author}{\bibinfo{person}{CDC}.} \bibinfo{year}{2018}\natexlab{}.
\newblock \bibinfo{title}{Historical Seasonal Influenza Vaccine Schedule}.
\newblock
  \bibinfo{howpublished}{\url{https://www.cdc.gov/flu/professionals/vaccination/vaccinesupply.htm}}.
\newblock
\newblock
\shownote{Accessed June 01, 2018.}


\bibitem[\protect\citeauthoryear{CDC}{CDC}{2019a}]%
        {cdcburden}
\bibfield{author}{\bibinfo{person}{CDC}.} \bibinfo{year}{2019}\natexlab{a}.
\newblock \bibinfo{title}{Disease Burden of Influenza.}
\newblock
  \bibinfo{howpublished}{\url{https://www.cdc.gov/flu/about/disease/burden.htm}}.
\newblock
\newblock
\shownote{Accessed April 01, 2019.}


\bibitem[\protect\citeauthoryear{CDC}{CDC}{2019b}]%
        {cdcfluview}
\bibfield{author}{\bibinfo{person}{CDC}.} \bibinfo{year}{2019}\natexlab{b}.
\newblock \bibinfo{title}{Fluview Interactive.}
\newblock
  \bibinfo{howpublished}{\url{https://www.cdc.gov/flu/weekly/fluviewinteractive.htm}}.
\newblock
\newblock
\shownote{Accessed April 20, 2019.}


\bibitem[\protect\citeauthoryear{CDO}{CDO}{2018}]%
        {cdo}
\bibfield{author}{\bibinfo{person}{CDO}.} \bibinfo{year}{2018}\natexlab{}.
\newblock \bibinfo{title}{Climate Data Online.}
\newblock
  \bibinfo{howpublished}{\url{https://www.ncdc.noaa.gov/cdo-web/datasets}}.
\newblock
\newblock
\shownote{Accessed August 28, 2018.}


\bibitem[\protect\citeauthoryear{Chao, Halloran, Obenchain, and
  Longini~Jr}{Chao et~al\mbox{.}}{2010}]%
        {chao2010}
\bibfield{author}{\bibinfo{person}{Dennis~L Chao}, \bibinfo{person}{M~Elizabeth
  Halloran}, \bibinfo{person}{Valerie~J Obenchain}, {and}
  \bibinfo{person}{Ira~M Longini~Jr}.} \bibinfo{year}{2010}\natexlab{}.
\newblock \showarticletitle{{FluTE}, a publicly available stochastic influenza
  epidemic simulation model}.
\newblock \bibinfo{journal}{\emph{PLoS computational biology}}
  \bibinfo{volume}{6}, \bibinfo{number}{1} (\bibinfo{year}{2010}),
  \bibinfo{pages}{e1000656}.
\newblock


\bibitem[\protect\citeauthoryear{Chretien, George, Shaman, Chitale, and
  McKenzie}{Chretien et~al\mbox{.}}{2014}]%
        {chretien2014influenza}
\bibfield{author}{\bibinfo{person}{Jean-Paul Chretien}, \bibinfo{person}{Dylan
  George}, \bibinfo{person}{Jeffrey Shaman}, \bibinfo{person}{Rohit~A Chitale},
  {and} \bibinfo{person}{F~Ellis McKenzie}.} \bibinfo{year}{2014}\natexlab{}.
\newblock \showarticletitle{Influenza forecasting in human populations: a
  scoping review}.
\newblock \bibinfo{journal}{\emph{PloS one}} \bibinfo{volume}{9},
  \bibinfo{number}{4} (\bibinfo{year}{2014}), \bibinfo{pages}{e94130}.
\newblock


\bibitem[\protect\citeauthoryear{Cui, Chen, and Chen}{Cui
  et~al\mbox{.}}{2016}]%
        {cui2016multi}
\bibfield{author}{\bibinfo{person}{Zhicheng Cui}, \bibinfo{person}{Wenlin
  Chen}, {and} \bibinfo{person}{Yixin Chen}.} \bibinfo{year}{2016}\natexlab{}.
\newblock \showarticletitle{Multi-scale convolutional neural networks for time
  series classification}.
\newblock \bibinfo{journal}{\emph{arXiv preprint arXiv:1603.06995}}
  (\bibinfo{year}{2016}).
\newblock


\bibitem[\protect\citeauthoryear{Deng, Wang, Rangwala, Wang, and Ning}{Deng
  et~al\mbox{.}}{2019}]%
        {deng2019graph}
\bibfield{author}{\bibinfo{person}{Songgaojun Deng}, \bibinfo{person}{Shusen
  Wang}, \bibinfo{person}{Huzefa Rangwala}, \bibinfo{person}{Lijing Wang},
  {and} \bibinfo{person}{Yue Ning}.} \bibinfo{year}{2019}\natexlab{}.
\newblock \showarticletitle{Graph Message Passing with Cross-location
  Attentions for Long-term ILI Prediction}.
\newblock \bibinfo{journal}{\emph{arXiv preprint arXiv:1912.10202}}
  (\bibinfo{year}{2019}).
\newblock


\bibitem[\protect\citeauthoryear{DOH}{DOH}{2019}]%
        {nj2018}
\bibfield{author}{\bibinfo{person}{DOH}.} \bibinfo{year}{2019}\natexlab{}.
\newblock \bibinfo{title}{{ILI Weekly Reports}}.
\newblock
  \bibinfo{howpublished}{\url{http://www.nj.gov/health/cd/statistics/flu-stats/}}.
\newblock
\newblock
\shownote{Accessed April 20, 2019.}


\bibitem[\protect\citeauthoryear{Doms, Kramer, and Shaman}{Doms
  et~al\mbox{.}}{2018}]%
        {doms2018assessing}
\bibfield{author}{\bibinfo{person}{Colin Doms}, \bibinfo{person}{Sarah~C
  Kramer}, {and} \bibinfo{person}{Jeffrey Shaman}.}
  \bibinfo{year}{2018}\natexlab{}.
\newblock \showarticletitle{Assessing the Use of Influenza Forecasts and
  Epidemiological Modeling in Public Health Decision Making in the United
  States}.
\newblock \bibinfo{journal}{\emph{Scientific reports}} \bibinfo{volume}{8},
  \bibinfo{number}{1} (\bibinfo{year}{2018}), \bibinfo{pages}{12406}.
\newblock


\bibitem[\protect\citeauthoryear{Dugas, Jalalpour, Gel, Levin, Torcaso, Igusa,
  and Rothman}{Dugas et~al\mbox{.}}{2013}]%
        {dugas2013}
\bibfield{author}{\bibinfo{person}{Andrea~Freyer Dugas}, \bibinfo{person}{Mehdi
  Jalalpour}, \bibinfo{person}{Yulia Gel}, \bibinfo{person}{Scott Levin},
  \bibinfo{person}{Fred Torcaso}, \bibinfo{person}{Takeru Igusa}, {and}
  \bibinfo{person}{Richard~E Rothman}.} \bibinfo{year}{2013}\natexlab{}.
\newblock \showarticletitle{Influenza forecasting with Google flu trends}.
\newblock \bibinfo{journal}{\emph{PloS one}} \bibinfo{volume}{8},
  \bibinfo{number}{2} (\bibinfo{year}{2013}), \bibinfo{pages}{e56176}.
\newblock


\bibitem[\protect\citeauthoryear{Eksin, Paarporn, and Weitz}{Eksin
  et~al\mbox{.}}{2019}]%
        {eksin2019systematic}
\bibfield{author}{\bibinfo{person}{Ceyhun Eksin}, \bibinfo{person}{Keith
  Paarporn}, {and} \bibinfo{person}{Joshua~S Weitz}.}
  \bibinfo{year}{2019}\natexlab{}.
\newblock \showarticletitle{Systematic biases in disease forecasting-the role
  of behavior change}.
\newblock \bibinfo{journal}{\emph{Epidemics}} (\bibinfo{year}{2019}).
\newblock


\bibitem[\protect\citeauthoryear{Eubank, Guclu, Kumar, Marathe, Srinivasan,
  Toroczkai, and Wang}{Eubank et~al\mbox{.}}{2004}]%
        {eubank2004}
\bibfield{author}{\bibinfo{person}{Stephen Eubank}, \bibinfo{person}{Hasan
  Guclu}, \bibinfo{person}{VS~Anil Kumar}, \bibinfo{person}{Madhav~V Marathe},
  \bibinfo{person}{Aravind Srinivasan}, \bibinfo{person}{Zoltan Toroczkai},
  {and} \bibinfo{person}{Nan Wang}.} \bibinfo{year}{2004}\natexlab{}.
\newblock \showarticletitle{Modelling disease outbreaks in realistic urban
  social networks}.
\newblock \bibinfo{journal}{\emph{Nature}} \bibinfo{volume}{429},
  \bibinfo{number}{6988} (\bibinfo{year}{2004}), \bibinfo{pages}{180--184}.
\newblock


\bibitem[\protect\citeauthoryear{Faghmous, Nguyen, Le, and Kumar}{Faghmous
  et~al\mbox{.}}{2014}]%
        {faghmous2014}
\bibfield{author}{\bibinfo{person}{James Faghmous}, \bibinfo{person}{Hung
  Nguyen}, \bibinfo{person}{Matthew Le}, {and} \bibinfo{person}{Vipin Kumar}.}
  \bibinfo{year}{2014}\natexlab{}.
\newblock \showarticletitle{Spatio-Temporal Consistency as a Means to Identify
  Unlabeled Objects in a Continuous Data Field}. In
  \bibinfo{booktitle}{\emph{AAAI Conference on Artificial Intelligence}}.
\newblock


\bibitem[\protect\citeauthoryear{Fischer, Tibbetts, Morgan, and Ceder}{Fischer
  et~al\mbox{.}}{2006}]%
        {fischer2006}
\bibfield{author}{\bibinfo{person}{Christopher~C Fischer},
  \bibinfo{person}{Kevin~J Tibbetts}, \bibinfo{person}{Dane Morgan}, {and}
  \bibinfo{person}{Gerbrand Ceder}.} \bibinfo{year}{2006}\natexlab{}.
\newblock \showarticletitle{Predicting crystal structure by merging data mining
  with quantum mechanics}.
\newblock \bibinfo{journal}{\emph{Nature Materials}}  \bibinfo{volume}{5}
  (\bibinfo{date}{07} \bibinfo{year}{2006}), \bibinfo{pages}{641}.
\newblock


\bibitem[\protect\citeauthoryear{Flahault, Vergu, Coudeville, and
  Grais}{Flahault et~al\mbox{.}}{2006}]%
        {flahault2006}
\bibfield{author}{\bibinfo{person}{Antoine Flahault},
  \bibinfo{person}{Elisabeta Vergu}, \bibinfo{person}{Laurent Coudeville},
  {and} \bibinfo{person}{Rebecca~F Grais}.} \bibinfo{year}{2006}\natexlab{}.
\newblock \showarticletitle{{Strategies for containing a global influenza
  pandemic}}.
\newblock \bibinfo{journal}{\emph{Vaccine}} \bibinfo{volume}{24},
  \bibinfo{number}{44} (\bibinfo{year}{2006}), \bibinfo{pages}{6751--6755}.
\newblock


\bibitem[\protect\citeauthoryear{Forestier, Petitjean, Dau, Webb, and
  Keogh}{Forestier et~al\mbox{.}}{2017}]%
        {forestier2017generating}
\bibfield{author}{\bibinfo{person}{Germain Forestier},
  \bibinfo{person}{Fran{\c{c}}ois Petitjean}, \bibinfo{person}{Hoang~Anh Dau},
  \bibinfo{person}{Geoffrey~I Webb}, {and} \bibinfo{person}{Eamonn Keogh}.}
  \bibinfo{year}{2017}\natexlab{}.
\newblock \showarticletitle{Generating synthetic time series to augment sparse
  datasets}. In \bibinfo{booktitle}{\emph{2017 IEEE international conference on
  data mining (ICDM)}}. IEEE, \bibinfo{pages}{865--870}.
\newblock


\bibitem[\protect\citeauthoryear{Gal and Ghahramani}{Gal and
  Ghahramani}{2016}]%
        {gal2016dropout}
\bibfield{author}{\bibinfo{person}{Yarin Gal} {and} \bibinfo{person}{Zoubin
  Ghahramani}.} \bibinfo{year}{2016}\natexlab{}.
\newblock \showarticletitle{Dropout as a bayesian approximation: Representing
  model uncertainty in deep learning}. In
  \bibinfo{booktitle}{\emph{international conference on machine learning}}.
  \bibinfo{pages}{1050--1059}.
\newblock


\bibitem[\protect\citeauthoryear{GHT}{GHT}{2018}]%
        {ght}
\bibfield{author}{\bibinfo{person}{GHT}.} \bibinfo{year}{2018}\natexlab{}.
\newblock \bibinfo{title}{Google Health Trends.}
\newblock \bibinfo{howpublished}{\url{https://trends.google.com/trends}}.
\newblock
\newblock
\shownote{Accessed August 28, 2018.}


\bibitem[\protect\citeauthoryear{Goldstein, Cobey, Takahashi, Miller, and
  Lipsitch}{Goldstein et~al\mbox{.}}{2011}]%
        {goldstein2011predicting}
\bibfield{author}{\bibinfo{person}{Edward Goldstein}, \bibinfo{person}{Sarah
  Cobey}, \bibinfo{person}{Saki Takahashi}, \bibinfo{person}{Joel~C Miller},
  {and} \bibinfo{person}{Marc Lipsitch}.} \bibinfo{year}{2011}\natexlab{}.
\newblock \showarticletitle{Predicting the epidemic sizes of influenza
  {A/H1N1}, {A/H3N2}, and {B}: a statistical method}.
\newblock \bibinfo{journal}{\emph{PLOS Medicine}} \bibinfo{volume}{8},
  \bibinfo{number}{7} (\bibinfo{year}{2011}), \bibinfo{pages}{e1001051}.
\newblock


\bibitem[\protect\citeauthoryear{Google}{Google}{2018}]%
        {gcorr}
\bibfield{author}{\bibinfo{person}{Google}.} \bibinfo{year}{2018}\natexlab{}.
\newblock \bibinfo{title}{Google Correlate Data.}
\newblock
  \bibinfo{howpublished}{\url{https://www.google.com/trends/correlate}}.
\newblock
\newblock
\shownote{Accessed August 28, 2018.}


\bibitem[\protect\citeauthoryear{Gurumurthy, Kiran~Sarvadevabhatla, and
  Venkatesh~Babu}{Gurumurthy et~al\mbox{.}}{2017}]%
        {gurumurthy2017deligan}
\bibfield{author}{\bibinfo{person}{Swaminathan Gurumurthy},
  \bibinfo{person}{Ravi Kiran~Sarvadevabhatla}, {and} \bibinfo{person}{R
  Venkatesh~Babu}.} \bibinfo{year}{2017}\natexlab{}.
\newblock \showarticletitle{Deligan: Generative adversarial networks for
  diverse and limited data}. In \bibinfo{booktitle}{\emph{Proceedings of the
  IEEE Conference on Computer Vision and Pattern Recognition}}.
  \bibinfo{pages}{166--174}.
\newblock


\bibitem[\protect\citeauthoryear{Hautier, Fischer, Jain, Mueller, and
  Ceder}{Hautier et~al\mbox{.}}{2010}]%
        {hautier2010}
\bibfield{author}{\bibinfo{person}{Geoffroy Hautier},
  \bibinfo{person}{Christopher~C Fischer}, \bibinfo{person}{Anubhav Jain},
  \bibinfo{person}{Tim Mueller}, {and} \bibinfo{person}{Gerbrand Ceder}.}
  \bibinfo{year}{2010}\natexlab{}.
\newblock \showarticletitle{Finding Nature's Missing Ternary Oxide Compounds
  Using Machine Learning and Density Functional Theory}.
\newblock \bibinfo{journal}{\emph{Chemistry of Materials}}
  \bibinfo{volume}{22}, \bibinfo{number}{12} (\bibinfo{year}{2010}),
  \bibinfo{pages}{3762--3767}.
\newblock


\bibitem[\protect\citeauthoryear{Hochreiter and Schmidhuber}{Hochreiter and
  Schmidhuber}{1997}]%
        {hochreiter1997}
\bibfield{author}{\bibinfo{person}{Sepp Hochreiter} {and}
  \bibinfo{person}{J{\"u}rgen Schmidhuber}.} \bibinfo{year}{1997}\natexlab{}.
\newblock \showarticletitle{Long Short-Term Memory}.
\newblock \bibinfo{journal}{\emph{Neural Computation}} \bibinfo{volume}{9},
  \bibinfo{number}{8} (\bibinfo{year}{1997}), \bibinfo{pages}{1735--1780}.
\newblock


\bibitem[\protect\citeauthoryear{Hua, Reddy, Zhang, Wang, Zhao, Lu, and
  Ramakrishnan}{Hua et~al\mbox{.}}{2018}]%
        {hua2018}
\bibfield{author}{\bibinfo{person}{Ting Hua}, \bibinfo{person}{Chandan~K
  Reddy}, \bibinfo{person}{Lei Zhang}, \bibinfo{person}{Lijing Wang},
  \bibinfo{person}{Liang Zhao}, \bibinfo{person}{Chang-Tien Lu}, {and}
  \bibinfo{person}{Naren Ramakrishnan}.} \bibinfo{year}{2018}\natexlab{}.
\newblock \showarticletitle{Social Media based Simulation Models for
  Understanding Disease Dynamics}. In \bibinfo{booktitle}{\emph{Proceedings of
  the Twenty-Seventh International Joint Conference on Artificial Intelligence,
  {IJCAI-18}}}. \bibinfo{publisher}{International Joint Conferences on
  Artificial Intelligence Organization}, \bibinfo{pages}{3797--3804}.
\newblock


\bibitem[\protect\citeauthoryear{IndexMundi}{IndexMundi}{2010}]%
        {population2010}
\bibfield{author}{\bibinfo{person}{IndexMundi}.}
  \bibinfo{year}{2010}\natexlab{}.
\newblock \bibinfo{title}{New Jersey Facts}.
\newblock
  \bibinfo{howpublished}{\url{https://www.indexmundi.com/facts/united-states/quick-facts/new-jersey/}}.
\newblock
\newblock
\shownote{Accessed March 01, 2019.}


\bibitem[\protect\citeauthoryear{Kandula and Shaman}{Kandula and
  Shaman}{2019}]%
        {kandula2019near}
\bibfield{author}{\bibinfo{person}{Sasikiran Kandula} {and}
  \bibinfo{person}{Jeffrey Shaman}.} \bibinfo{year}{2019}\natexlab{}.
\newblock \showarticletitle{Near-term forecasts of influenza-like illness: An
  evaluation of autoregressive time series approaches}.
\newblock \bibinfo{journal}{\emph{Epidemics}} (\bibinfo{year}{2019}).
\newblock


\bibitem[\protect\citeauthoryear{Kandula, Yamana, Pei, Yang, Morita, and
  Shaman}{Kandula et~al\mbox{.}}{2018}]%
        {kandula2018evaluation}
\bibfield{author}{\bibinfo{person}{Sasikiran Kandula}, \bibinfo{person}{Teresa
  Yamana}, \bibinfo{person}{Sen Pei}, \bibinfo{person}{Wan Yang},
  \bibinfo{person}{Haruka Morita}, {and} \bibinfo{person}{Jeffrey Shaman}.}
  \bibinfo{year}{2018}\natexlab{}.
\newblock \showarticletitle{Evaluation of mechanistic and statistical methods
  in forecasting influenza-like illness}.
\newblock \bibinfo{journal}{\emph{Journal of The Royal Society Interface}}
  \bibinfo{volume}{15}, \bibinfo{number}{144} (\bibinfo{year}{2018}),
  \bibinfo{pages}{20180174}.
\newblock


\bibitem[\protect\citeauthoryear{Karpatne, Atluri, Faghmous, Steinbach,
  Banerjee, Ganguly, Shekhar, Samatova, and Kumar}{Karpatne
  et~al\mbox{.}}{2017}]%
        {karpatne2017}
\bibfield{author}{\bibinfo{person}{Anuj Karpatne}, \bibinfo{person}{Gowtham
  Atluri}, \bibinfo{person}{James~H Faghmous}, \bibinfo{person}{Michael
  Steinbach}, \bibinfo{person}{Arindam Banerjee}, \bibinfo{person}{Auroop
  Ganguly}, \bibinfo{person}{Shashi Shekhar}, \bibinfo{person}{Nagiza
  Samatova}, {and} \bibinfo{person}{Vipin Kumar}.}
  \bibinfo{year}{2017}\natexlab{}.
\newblock \showarticletitle{Theory-guided data science: A new paradigm for
  scientific discovery from data}.
\newblock \bibinfo{journal}{\emph{IEEE Transactions on Knowledge and Data
  Engineering}} \bibinfo{volume}{29}, \bibinfo{number}{10}
  (\bibinfo{year}{2017}), \bibinfo{pages}{2318--2331}.
\newblock


\bibitem[\protect\citeauthoryear{Kawale, Liess, Kumar, Steinbach, Snyder,
  Kumar, Ganguly, Samatova, and Semazzi}{Kawale et~al\mbox{.}}{2013}]%
        {kawale2013}
\bibfield{author}{\bibinfo{person}{Jaya Kawale}, \bibinfo{person}{Stefan
  Liess}, \bibinfo{person}{Arjun Kumar}, \bibinfo{person}{Michael Steinbach},
  \bibinfo{person}{Peter Snyder}, \bibinfo{person}{Vipin Kumar},
  \bibinfo{person}{Auroop~R Ganguly}, \bibinfo{person}{Nagiza~F Samatova},
  {and} \bibinfo{person}{Fredrick Semazzi}.} \bibinfo{year}{2013}\natexlab{}.
\newblock \showarticletitle{A graph-based approach to find teleconnections in
  climate data}.
\newblock \bibinfo{journal}{\emph{Statistical Analysis and Data Mining: The ASA
  Data Science Journal}} \bibinfo{volume}{6}, \bibinfo{number}{3}
  (\bibinfo{year}{2013}), \bibinfo{pages}{158--179}.
\newblock


\bibitem[\protect\citeauthoryear{Khandelwal, Karpatne, Marlier, Kim,
  Lettenmaier, and Kumar}{Khandelwal et~al\mbox{.}}{2017}]%
        {khandelwal2017}
\bibfield{author}{\bibinfo{person}{Ankush Khandelwal}, \bibinfo{person}{Anuj
  Karpatne}, \bibinfo{person}{Miriam~E Marlier}, \bibinfo{person}{Jongyoun
  Kim}, \bibinfo{person}{Dennis~P Lettenmaier}, {and} \bibinfo{person}{Vipin
  Kumar}.} \bibinfo{year}{2017}\natexlab{}.
\newblock \showarticletitle{An approach for global monitoring of surface water
  extent variations in reservoirs using MODIS data}.
\newblock \bibinfo{journal}{\emph{Remote sensing of Environment}}
  \bibinfo{volume}{202} (\bibinfo{year}{2017}), \bibinfo{pages}{113--128}.
\newblock


\bibitem[\protect\citeauthoryear{Khandelwal, Mithal, and Kumar}{Khandelwal
  et~al\mbox{.}}{2015}]%
        {khandelwal2015}
\bibfield{author}{\bibinfo{person}{Ankush Khandelwal}, \bibinfo{person}{Varun
  Mithal}, {and} \bibinfo{person}{Vipin Kumar}.}
  \bibinfo{year}{2015}\natexlab{}.
\newblock \showarticletitle{Post classification label refinement using implicit
  ordering constraint among data instances}. In \bibinfo{booktitle}{\emph{2015
  IEEE International Conference on Data Mining}}. IEEE,
  \bibinfo{pages}{799--804}.
\newblock


\bibitem[\protect\citeauthoryear{Kingma and Ba}{Kingma and Ba}{2014}]%
        {kingma2014adam}
\bibfield{author}{\bibinfo{person}{Diederik~P Kingma} {and}
  \bibinfo{person}{Jimmy Ba}.} \bibinfo{year}{2014}\natexlab{}.
\newblock \showarticletitle{Adam: A method for stochastic optimization}.
\newblock \bibinfo{journal}{\emph{arXiv preprint arXiv:1412.6980}}
  (\bibinfo{year}{2014}).
\newblock


\bibitem[\protect\citeauthoryear{Krell, Seeland, and Kim}{Krell
  et~al\mbox{.}}{2018}]%
        {krell2018data}
\bibfield{author}{\bibinfo{person}{Mario~Michael Krell}, \bibinfo{person}{Anett
  Seeland}, {and} \bibinfo{person}{Su~Kyoung Kim}.}
  \bibinfo{year}{2018}\natexlab{}.
\newblock \showarticletitle{Data Augmentation for Brain-Computer Interfaces:
  Analysis on Event-Related Potentials Data}.
\newblock \bibinfo{journal}{\emph{arXiv preprint arXiv:1801.02730}}
  (\bibinfo{year}{2018}).
\newblock


\bibitem[\protect\citeauthoryear{Kuznetsov and Piccardi}{Kuznetsov and
  Piccardi}{1994}]%
        {kuznetsov1994bifurcation}
\bibfield{author}{\bibinfo{person}{Yu~A Kuznetsov} {and} \bibinfo{person}{Carlo
  Piccardi}.} \bibinfo{year}{1994}\natexlab{}.
\newblock \showarticletitle{Bifurcation analysis of periodic SEIR and SIR
  epidemic models}.
\newblock \bibinfo{journal}{\emph{Journal of mathematical biology}}
  \bibinfo{volume}{32}, \bibinfo{number}{2} (\bibinfo{year}{1994}),
  \bibinfo{pages}{109--121}.
\newblock


\bibitem[\protect\citeauthoryear{Kvamme, Sellereite, Aas, and Sjursen}{Kvamme
  et~al\mbox{.}}{2018}]%
        {kvamme2018predicting}
\bibfield{author}{\bibinfo{person}{H{\aa}vard Kvamme}, \bibinfo{person}{Nikolai
  Sellereite}, \bibinfo{person}{Kjersti Aas}, {and} \bibinfo{person}{Steffen
  Sjursen}.} \bibinfo{year}{2018}\natexlab{}.
\newblock \showarticletitle{Predicting mortgage default using convolutional
  neural networks}.
\newblock \bibinfo{journal}{\emph{Expert Systems with Applications}}
  \bibinfo{volume}{102} (\bibinfo{year}{2018}), \bibinfo{pages}{207--217}.
\newblock


\bibitem[\protect\citeauthoryear{Le~Guennec, Malinowski, and
  Tavenard}{Le~Guennec et~al\mbox{.}}{2016}]%
        {le2016data}
\bibfield{author}{\bibinfo{person}{Arthur Le~Guennec}, \bibinfo{person}{Simon
  Malinowski}, {and} \bibinfo{person}{Romain Tavenard}.}
  \bibinfo{year}{2016}\natexlab{}.
\newblock \showarticletitle{Data augmentation for time series classification
  using convolutional neural networks}.
\newblock


\bibitem[\protect\citeauthoryear{Lee, Choi, Cho, and Kim}{Lee
  et~al\mbox{.}}{2012}]%
        {lee2012}
\bibfield{author}{\bibinfo{person}{Jung~Min Lee}, \bibinfo{person}{Donghoon
  Choi}, \bibinfo{person}{Giphil Cho}, {and} \bibinfo{person}{Yongkuk Kim}.}
  \bibinfo{year}{2012}\natexlab{}.
\newblock \showarticletitle{The effect of public health interventions on the
  spread of influenza among cities}.
\newblock \bibinfo{journal}{\emph{Journal of theoretical biology}}
  \bibinfo{volume}{293} (\bibinfo{year}{2012}), \bibinfo{pages}{131--142}.
\newblock


\bibitem[\protect\citeauthoryear{Longini~Jr, Fine, and Thacker}{Longini~Jr
  et~al\mbox{.}}{1986}]%
        {longini1986predicting}
\bibfield{author}{\bibinfo{person}{Ira~M Longini~Jr}, \bibinfo{person}{Paul~EM
  Fine}, {and} \bibinfo{person}{Stephen~B Thacker}.}
  \bibinfo{year}{1986}\natexlab{}.
\newblock \showarticletitle{Predicting the global spread of new infectious
  agents}.
\newblock \bibinfo{journal}{\emph{American journal of epidemiology}}
  \bibinfo{volume}{123}, \bibinfo{number}{3} (\bibinfo{year}{1986}),
  \bibinfo{pages}{383--391}.
\newblock


\bibitem[\protect\citeauthoryear{L{\"o}yt{\"o}nen and Arbona}{L{\"o}yt{\"o}nen
  and Arbona}{1996}]%
        {loytonen1996forecasting}
\bibfield{author}{\bibinfo{person}{Markku L{\"o}yt{\"o}nen} {and}
  \bibinfo{person}{Sonia~I Arbona}.} \bibinfo{year}{1996}\natexlab{}.
\newblock \showarticletitle{Forecasting the AIDS epidemic in Puerto Rico}.
\newblock \bibinfo{journal}{\emph{Social Science \& Medicine}}
  \bibinfo{volume}{42}, \bibinfo{number}{7} (\bibinfo{year}{1996}),
  \bibinfo{pages}{997--1010}.
\newblock


\bibitem[\protect\citeauthoryear{Lunelli, Pugliese, and Rizzo}{Lunelli
  et~al\mbox{.}}{2009}]%
        {lunelli2009}
\bibfield{author}{\bibinfo{person}{Antonella Lunelli}, \bibinfo{person}{Andrea
  Pugliese}, {and} \bibinfo{person}{Caterina Rizzo}.}
  \bibinfo{year}{2009}\natexlab{}.
\newblock \showarticletitle{Epidemic patch models applied to pandemic
  influenza: Contact matrix, stochasticity, robustness of predictions}.
\newblock \bibinfo{journal}{\emph{Mathematical Biosciences}}
  \bibinfo{volume}{220}, \bibinfo{number}{1} (\bibinfo{year}{2009}),
  \bibinfo{pages}{24--33}.
\newblock


\bibitem[\protect\citeauthoryear{Marathe, Lewis, Chen, and Eubank}{Marathe
  et~al\mbox{.}}{2011}]%
        {achla2011b}
\bibfield{author}{\bibinfo{person}{Achla Marathe}, \bibinfo{person}{Bryan
  Lewis}, \bibinfo{person}{Jiangzhuo Chen}, {and} \bibinfo{person}{Stephen
  Eubank}.} \bibinfo{year}{2011}\natexlab{}.
\newblock \showarticletitle{Sensitivity of Household Transmission to Household
  Contact Structure and Size}.
\newblock \bibinfo{journal}{\emph{PLoS ONE}}  \bibinfo{volume}{6}
  (\bibinfo{date}{08} \bibinfo{year}{2011}).
\newblock


\bibitem[\protect\citeauthoryear{Marchesi}{Marchesi}{2017}]%
        {marchesi2017megapixel}
\bibfield{author}{\bibinfo{person}{Marco Marchesi}.}
  \bibinfo{year}{2017}\natexlab{}.
\newblock \showarticletitle{Megapixel size image creation using generative
  adversarial networks}.
\newblock \bibinfo{journal}{\emph{arXiv preprint arXiv:1706.00082}}
  (\bibinfo{year}{2017}).
\newblock


\bibitem[\protect\citeauthoryear{McGowan, Biggerstaff, Johansson, Apfeldorf,
  Ben-Nun, Brooks, Convertino, Erraguntla, Farrow, Freeze,
  et~al\mbox{.}}{McGowan et~al\mbox{.}}{2019}]%
        {mcgowan2019collaborative}
\bibfield{author}{\bibinfo{person}{Craig~J McGowan}, \bibinfo{person}{Matthew
  Biggerstaff}, \bibinfo{person}{Michael Johansson}, \bibinfo{person}{Karyn~M
  Apfeldorf}, \bibinfo{person}{Michal Ben-Nun}, \bibinfo{person}{Logan Brooks},
  \bibinfo{person}{Matteo Convertino}, \bibinfo{person}{Madhav Erraguntla},
  \bibinfo{person}{David~C Farrow}, \bibinfo{person}{John Freeze},
  {et~al\mbox{.}}} \bibinfo{year}{2019}\natexlab{}.
\newblock \showarticletitle{Collaborative efforts to forecast seasonal
  influenza in the United States, 2015--2016}.
\newblock \bibinfo{journal}{\emph{Scientific reports}} \bibinfo{volume}{9},
  \bibinfo{number}{1} (\bibinfo{year}{2019}), \bibinfo{pages}{683}.
\newblock


\bibitem[\protect\citeauthoryear{Molinari, Ortega-Sanchez, Messonnier,
  Thompson, Wortley, Weintraub, and Bridges}{Molinari et~al\mbox{.}}{2007}]%
        {molinari2007}
\bibfield{author}{\bibinfo{person}{Noelle-Angelique~M Molinari},
  \bibinfo{person}{Ismael~R Ortega-Sanchez}, \bibinfo{person}{Mark~L
  Messonnier}, \bibinfo{person}{William~W Thompson}, \bibinfo{person}{Pascale~M
  Wortley}, \bibinfo{person}{Eric Weintraub}, {and} \bibinfo{person}{Carolyn~B
  Bridges}.} \bibinfo{year}{2007}\natexlab{}.
\newblock \showarticletitle{The annual impact of seasonal influenza in the US:
  measuring disease burden and costs}.
\newblock \bibinfo{journal}{\emph{Vaccine}} \bibinfo{volume}{25},
  \bibinfo{number}{27} (\bibinfo{year}{2007}), \bibinfo{pages}{5086--5096}.
\newblock


\bibitem[\protect\citeauthoryear{Morita, Kramer, Heaney, Gil, and
  Shaman}{Morita et~al\mbox{.}}{2018}]%
        {morita2018influenza}
\bibfield{author}{\bibinfo{person}{Haruka Morita}, \bibinfo{person}{Sarah
  Kramer}, \bibinfo{person}{Alexandra Heaney}, \bibinfo{person}{Harold Gil},
  {and} \bibinfo{person}{Jeffrey Shaman}.} \bibinfo{year}{2018}\natexlab{}.
\newblock \showarticletitle{Influenza forecast optimization when using
  different surveillance data types and geographic scale}.
\newblock \bibinfo{journal}{\emph{Influenza and other respiratory viruses}}
  \bibinfo{volume}{12}, \bibinfo{number}{6} (\bibinfo{year}{2018}),
  \bibinfo{pages}{755--764}.
\newblock


\bibitem[\protect\citeauthoryear{NDSSL}{NDSSL}{2014}]%
        {ndssl:opendata2014}
\bibfield{author}{\bibinfo{person}{NDSSL}.} \bibinfo{year}{2014}\natexlab{}.
\newblock \bibinfo{title}{Synthetic Data of {Montgomery} county, {Virginia}}.
\newblock
  \bibinfo{howpublished}{\url{http://ndssl.vbi.vt.edu/synthetic-data/}}.
\newblock


\bibitem[\protect\citeauthoryear{Nelder and Mead}{Nelder and Mead}{1965}]%
        {nelder1965}
\bibfield{author}{\bibinfo{person}{John~A Nelder} {and} \bibinfo{person}{Roger
  Mead}.} \bibinfo{year}{1965}\natexlab{}.
\newblock \showarticletitle{A simplex method for function minimization}.
\newblock \bibinfo{journal}{\emph{The computer journal}} \bibinfo{volume}{7},
  \bibinfo{number}{4} (\bibinfo{year}{1965}), \bibinfo{pages}{308--313}.
\newblock


\bibitem[\protect\citeauthoryear{Nsoesie, Mararthe, and Brownstein}{Nsoesie
  et~al\mbox{.}}{2013b}]%
        {nsoesie2013forecasting}
\bibfield{author}{\bibinfo{person}{Elaine Nsoesie}, \bibinfo{person}{Madhav
  Mararthe}, {and} \bibinfo{person}{John Brownstein}.}
  \bibinfo{year}{2013}\natexlab{b}.
\newblock \showarticletitle{Forecasting peaks of seasonal influenza epidemics}.
\newblock \bibinfo{journal}{\emph{PLoS currents}}  \bibinfo{volume}{5}
  (\bibinfo{year}{2013}).
\newblock


\bibitem[\protect\citeauthoryear{Nsoesie, Beckman, Shashaani, Nagaraj, and
  Marathe}{Nsoesie et~al\mbox{.}}{2013a}]%
        {nsoesie2013simulation}
\bibfield{author}{\bibinfo{person}{Elaine~O Nsoesie},
  \bibinfo{person}{Richard~J Beckman}, \bibinfo{person}{Sara Shashaani},
  \bibinfo{person}{Kalyani~S Nagaraj}, {and} \bibinfo{person}{Madhav~V
  Marathe}.} \bibinfo{year}{2013}\natexlab{a}.
\newblock \showarticletitle{A simulation optimization approach to epidemic
  forecasting}.
\newblock \bibinfo{journal}{\emph{PloS one}} \bibinfo{volume}{8},
  \bibinfo{number}{6} (\bibinfo{year}{2013}), \bibinfo{pages}{e67164}.
\newblock


\bibitem[\protect\citeauthoryear{Nsoesie, Brownstein, Ramakrishnan, and
  Marathe}{Nsoesie et~al\mbox{.}}{2014}]%
        {nsoesie2014systematic}
\bibfield{author}{\bibinfo{person}{Elaine~O Nsoesie}, \bibinfo{person}{John~S
  Brownstein}, \bibinfo{person}{Naren Ramakrishnan}, {and}
  \bibinfo{person}{Madhav~V Marathe}.} \bibinfo{year}{2014}\natexlab{}.
\newblock \showarticletitle{A systematic review of studies on forecasting the
  dynamics of influenza outbreaks}.
\newblock \bibinfo{journal}{\emph{Influenza and other respiratory viruses}}
  \bibinfo{volume}{8}, \bibinfo{number}{3} (\bibinfo{year}{2014}),
  \bibinfo{pages}{309--316}.
\newblock


\bibitem[\protect\citeauthoryear{Osthus, Gattiker, Priedhorsky, Del~Valle,
  et~al\mbox{.}}{Osthus et~al\mbox{.}}{2019}]%
        {osthus2019dynamic}
\bibfield{author}{\bibinfo{person}{Dave Osthus}, \bibinfo{person}{James
  Gattiker}, \bibinfo{person}{Reid Priedhorsky}, \bibinfo{person}{Sara~Y
  Del~Valle}, {et~al\mbox{.}}} \bibinfo{year}{2019}\natexlab{}.
\newblock \showarticletitle{Dynamic Bayesian Influenza Forecasting in the
  United States with Hierarchical Discrepancy (with Discussion)}.
\newblock \bibinfo{journal}{\emph{Bayesian Analysis}} \bibinfo{volume}{14},
  \bibinfo{number}{1} (\bibinfo{year}{2019}), \bibinfo{pages}{261--312}.
\newblock


\bibitem[\protect\citeauthoryear{Parker and Epstein}{Parker and
  Epstein}{2011}]%
        {parker2011}
\bibfield{author}{\bibinfo{person}{Jon Parker} {and} \bibinfo{person}{Joshua~M
  Epstein}.} \bibinfo{year}{2011}\natexlab{}.
\newblock \showarticletitle{{A Distributed Platform for Global-Scale
  Agent-Based Models of Disease Transmission}}.
\newblock \bibinfo{journal}{\emph{ACM Trans Model Comput Simul}}
  \bibinfo{volume}{22}, \bibinfo{number}{1}, Article \bibinfo{articleno}{2}
  (\bibinfo{date}{12} \bibinfo{year}{2011}), \bibinfo{numpages}{25}~pages.
\newblock
\showISSN{1049-3301}


\bibitem[\protect\citeauthoryear{Pei, Kandula, Yang, and Shaman}{Pei
  et~al\mbox{.}}{2018}]%
        {pei2018forecasting}
\bibfield{author}{\bibinfo{person}{Sen Pei}, \bibinfo{person}{Sasikiran
  Kandula}, \bibinfo{person}{Wan Yang}, {and} \bibinfo{person}{Jeffrey
  Shaman}.} \bibinfo{year}{2018}\natexlab{}.
\newblock \showarticletitle{Forecasting the spatial transmission of influenza
  in the \mbox{United States}}.
\newblock \bibinfo{journal}{\emph{Proceedings of the National Academy of
  Sciences}} \bibinfo{volume}{115}, \bibinfo{number}{11}
  (\bibinfo{year}{2018}), \bibinfo{pages}{2752--2757}.
\newblock


\bibitem[\protect\citeauthoryear{Perez and Wang}{Perez and Wang}{2017}]%
        {perez2017effectiveness}
\bibfield{author}{\bibinfo{person}{Luis Perez} {and} \bibinfo{person}{Jason
  Wang}.} \bibinfo{year}{2017}\natexlab{}.
\newblock \showarticletitle{The effectiveness of data augmentation in image
  classification using deep learning}.
\newblock \bibinfo{journal}{\emph{arXiv preprint arXiv:1712.04621}}
  (\bibinfo{year}{2017}).
\newblock


\bibitem[\protect\citeauthoryear{Reich, Brooks, Fox, Kandula, McGowan, Moore,
  Osthus, Ray, Tushar, Yamana, Biggerstaff, Johansson, Rosenfeld, and
  Shaman}{Reich et~al\mbox{.}}{2019}]%
        {reich2019}
\bibfield{author}{\bibinfo{person}{Nicholas~G. Reich},
  \bibinfo{person}{Logan~C. Brooks}, \bibinfo{person}{Spencer~J. Fox},
  \bibinfo{person}{Sasikiran Kandula}, \bibinfo{person}{Craig~J. McGowan},
  \bibinfo{person}{Evan Moore}, \bibinfo{person}{Dave Osthus},
  \bibinfo{person}{Evan~L. Ray}, \bibinfo{person}{Abhinav Tushar},
  \bibinfo{person}{Teresa~K. Yamana}, \bibinfo{person}{Matthew Biggerstaff},
  \bibinfo{person}{Michael~A. Johansson}, \bibinfo{person}{Roni Rosenfeld},
  {and} \bibinfo{person}{Jeffrey Shaman}.} \bibinfo{year}{2019}\natexlab{}.
\newblock \showarticletitle{A collaborative multiyear, multimodel assessment of
  seasonal influenza forecasting in the United States}.
\newblock \bibinfo{journal}{\emph{Proceedings of the National Academy of
  Sciences}} \bibinfo{volume}{116}, \bibinfo{number}{8} (\bibinfo{year}{2019}),
  \bibinfo{pages}{3146--3154}.
\newblock
\showISSN{0027-8424}
\urldef\tempurl%
\url{https://doi.org/10.1073/pnas.1812594116}
\showDOI{\tempurl}
\showeprint{https://www.pnas.org/content/116/8/3146.full.pdf}


\bibitem[\protect\citeauthoryear{Rizk, Shokry, and Youssef}{Rizk
  et~al\mbox{.}}{2019}]%
        {rizk2019effectiveness}
\bibfield{author}{\bibinfo{person}{Hamada Rizk}, \bibinfo{person}{Ahmed
  Shokry}, {and} \bibinfo{person}{Moustafa Youssef}.}
  \bibinfo{year}{2019}\natexlab{}.
\newblock \showarticletitle{Effectiveness of Data Augmentation in
  Cellular-based Localization Using Deep Learning}.
\newblock \bibinfo{journal}{\emph{arXiv preprint arXiv:1906.08171}}
  (\bibinfo{year}{2019}).
\newblock


\bibitem[\protect\citeauthoryear{Schl{\"u}ter and Grill}{Schl{\"u}ter and
  Grill}{2015}]%
        {schluter2015exploring}
\bibfield{author}{\bibinfo{person}{Jan Schl{\"u}ter} {and}
  \bibinfo{person}{Thomas Grill}.} \bibinfo{year}{2015}\natexlab{}.
\newblock \showarticletitle{Exploring Data Augmentation for Improved Singing
  Voice Detection with Neural Networks.}. In \bibinfo{booktitle}{\emph{ISMIR}}.
  \bibinfo{pages}{121--126}.
\newblock


\bibitem[\protect\citeauthoryear{Shaman and Karspeck}{Shaman and
  Karspeck}{2012}]%
        {shaman2012}
\bibfield{author}{\bibinfo{person}{Jeffrey Shaman} {and}
  \bibinfo{person}{Alicia Karspeck}.} \bibinfo{year}{2012}\natexlab{}.
\newblock \showarticletitle{Forecasting seasonal outbreaks of influenza}.
\newblock \bibinfo{journal}{\emph{Proceedings of the National Academy of
  Sciences}} (\bibinfo{year}{2012}).
\newblock


\bibitem[\protect\citeauthoryear{Shaman, Karspeck, Yang, Tamerius, and
  Lipsitch}{Shaman et~al\mbox{.}}{2013}]%
        {shaman2013realtime}
\bibfield{author}{\bibinfo{person}{Jeffrey Shaman}, \bibinfo{person}{Alicia
  Karspeck}, \bibinfo{person}{Wan Yang}, \bibinfo{person}{James Tamerius},
  {and} \bibinfo{person}{Marc Lipsitch}.} \bibinfo{year}{2013}\natexlab{}.
\newblock \showarticletitle{Real-time influenza forecasts during the 2012-2013
  season}.
\newblock \bibinfo{journal}{\emph{Nature Communications}} \bibinfo{volume}{4},
  \bibinfo{number}{2837} (\bibinfo{year}{2013}).
\newblock


\bibitem[\protect\citeauthoryear{Thinglink}{Thinglink}{2019}]%
        {njmap}
\bibfield{author}{\bibinfo{person}{Thinglink}.}
  \bibinfo{year}{2019}\natexlab{}.
\newblock \bibinfo{title}{{New Jersey Regions Map}}.
\newblock
  \bibinfo{howpublished}{\url{https://www.thinglink.com/scene/788830737167024130}}.
\newblock


\bibitem[\protect\citeauthoryear{Tuite, Greer, Whelan, Winter, Lee, Yan, Wu,
  Moghadas, Buckeridge, Pourbohloul, et~al\mbox{.}}{Tuite
  et~al\mbox{.}}{2010}]%
        {tuite2010}
\bibfield{author}{\bibinfo{person}{Ashleigh~R Tuite}, \bibinfo{person}{Amy~L
  Greer}, \bibinfo{person}{Michael Whelan}, \bibinfo{person}{Anne-Luise
  Winter}, \bibinfo{person}{Brenda Lee}, \bibinfo{person}{Ping Yan},
  \bibinfo{person}{Jianhong Wu}, \bibinfo{person}{Seyed Moghadas},
  \bibinfo{person}{David Buckeridge}, \bibinfo{person}{Babak Pourbohloul},
  {et~al\mbox{.}}} \bibinfo{year}{2010}\natexlab{}.
\newblock \showarticletitle{Estimated epidemiologic parameters and morbidity
  associated with pandemic H1N1 influenza}.
\newblock \bibinfo{journal}{\emph{Cmaj}} \bibinfo{volume}{182},
  \bibinfo{number}{2} (\bibinfo{year}{2010}), \bibinfo{pages}{131--136}.
\newblock


\bibitem[\protect\citeauthoryear{Um, Pfister, Pichler, Endo, Lang, Hirche,
  Fietzek, and Kuli{\'c}}{Um et~al\mbox{.}}{2017}]%
        {um2017data}
\bibfield{author}{\bibinfo{person}{Terry~Taewoong Um}, \bibinfo{person}{Franz
  Michael~Josef Pfister}, \bibinfo{person}{Daniel Pichler},
  \bibinfo{person}{Satoshi Endo}, \bibinfo{person}{Muriel Lang},
  \bibinfo{person}{Sandra Hirche}, \bibinfo{person}{Urban Fietzek}, {and}
  \bibinfo{person}{Dana Kuli{\'c}}.} \bibinfo{year}{2017}\natexlab{}.
\newblock \showarticletitle{Data augmentation of wearable sensor data for
  parkinson's disease monitoring using convolutional neural networks}.
\newblock \bibinfo{journal}{\emph{arXiv preprint arXiv:1706.00527}}
  (\bibinfo{year}{2017}).
\newblock


\bibitem[\protect\citeauthoryear{Vasconcelos and Vasconcelos}{Vasconcelos and
  Vasconcelos}{2017}]%
        {vasconcelos2017increasing}
\bibfield{author}{\bibinfo{person}{Cristina~Nader Vasconcelos} {and}
  \bibinfo{person}{B{\'a}rbara~Nader Vasconcelos}.}
  \bibinfo{year}{2017}\natexlab{}.
\newblock \showarticletitle{Increasing deep learning melanoma classification by
  classical and expert knowledge based image transforms}.
\newblock \bibinfo{journal}{\emph{CoRR, abs/1702.07025}}  \bibinfo{volume}{1}
  (\bibinfo{year}{2017}).
\newblock


\bibitem[\protect\citeauthoryear{Venna, Tavanaei, Gottumukkala, Raghavan,
  Maida, and Nichols}{Venna et~al\mbox{.}}{2019}]%
        {venna2019novel}
\bibfield{author}{\bibinfo{person}{Siva~R Venna}, \bibinfo{person}{Amirhossein
  Tavanaei}, \bibinfo{person}{Raju~N Gottumukkala}, \bibinfo{person}{Vijay~V
  Raghavan}, \bibinfo{person}{Anthony~S Maida}, {and} \bibinfo{person}{Stephen
  Nichols}.} \bibinfo{year}{2019}\natexlab{}.
\newblock \showarticletitle{A novel data-driven model for real-time influenza
  forecasting}.
\newblock \bibinfo{journal}{\emph{IEEE Access}}  \bibinfo{volume}{7}
  (\bibinfo{year}{2019}), \bibinfo{pages}{7691--7701}.
\newblock


\bibitem[\protect\citeauthoryear{Viboud, Bo{\"e}lle, Carrat, Valleron, and
  Flahault}{Viboud et~al\mbox{.}}{2003}]%
        {viboud2003prediction}
\bibfield{author}{\bibinfo{person}{C{\'e}cile Viboud},
  \bibinfo{person}{Pierre-Yves Bo{\"e}lle}, \bibinfo{person}{Fabrice Carrat},
  \bibinfo{person}{Alain-Jacques Valleron}, {and} \bibinfo{person}{Antoine
  Flahault}.} \bibinfo{year}{2003}\natexlab{}.
\newblock \showarticletitle{Prediction of the spread of influenza epidemics by
  the method of analogues}.
\newblock \bibinfo{journal}{\emph{American Journal of Epidemiology}}
  \bibinfo{volume}{158}, \bibinfo{number}{10} (\bibinfo{year}{2003}),
  \bibinfo{pages}{996--1006}.
\newblock


\bibitem[\protect\citeauthoryear{Viboud, Sun, Gaffey, Ajelli, Fumanelli,
  Merler, Zhang, Chowell, Simonsen, Vespignani, et~al\mbox{.}}{Viboud
  et~al\mbox{.}}{2018}]%
        {viboud2018rapidd}
\bibfield{author}{\bibinfo{person}{C{\'e}cile Viboud}, \bibinfo{person}{Kaiyuan
  Sun}, \bibinfo{person}{Robert Gaffey}, \bibinfo{person}{Marco Ajelli},
  \bibinfo{person}{Laura Fumanelli}, \bibinfo{person}{Stefano Merler},
  \bibinfo{person}{Qian Zhang}, \bibinfo{person}{Gerardo Chowell},
  \bibinfo{person}{Lone Simonsen}, \bibinfo{person}{Alessandro Vespignani},
  {et~al\mbox{.}}} \bibinfo{year}{2018}\natexlab{}.
\newblock \showarticletitle{The {RAPIDD Ebola Forecasting Challenge}: Synthesis
  and lessons learnt}.
\newblock \bibinfo{journal}{\emph{Epidemics}}  \bibinfo{volume}{22}
  (\bibinfo{year}{2018}), \bibinfo{pages}{13--21}.
\newblock


\bibitem[\protect\citeauthoryear{Volkova, Ayton, Porterfield, and
  Corley}{Volkova et~al\mbox{.}}{2017}]%
        {volkova2017forecasting}
\bibfield{author}{\bibinfo{person}{Svitlana Volkova}, \bibinfo{person}{Ellyn
  Ayton}, \bibinfo{person}{Katherine Porterfield}, {and}
  \bibinfo{person}{Courtney~D Corley}.} \bibinfo{year}{2017}\natexlab{}.
\newblock \showarticletitle{Forecasting influenza-like illness dynamics for
  military populations using neural networks and social media}.
\newblock \bibinfo{journal}{\emph{PloS one}} \bibinfo{volume}{12},
  \bibinfo{number}{12} (\bibinfo{year}{2017}), \bibinfo{pages}{e0188941}.
\newblock


\bibitem[\protect\citeauthoryear{Wang, Chen, and Marathe}{Wang
  et~al\mbox{.}}{2018}]%
        {wang2018framework}
\bibfield{author}{\bibinfo{person}{Lijing Wang}, \bibinfo{person}{Jiangzhuo
  Chen}, {and} \bibinfo{person}{Achla Marathe}.}
  \bibinfo{year}{2018}\natexlab{}.
\newblock \showarticletitle{A framework for discovering health disparities
  among cohorts in an influenza epidemic}.
\newblock \bibinfo{journal}{\emph{World Wide Web}} (\bibinfo{year}{2018}),
  \bibinfo{pages}{1--24}.
\newblock


\bibitem[\protect\citeauthoryear{Wang, Chen, and Marathe}{Wang
  et~al\mbox{.}}{2019a}]%
        {wang2019defsi}
\bibfield{author}{\bibinfo{person}{Lijing Wang}, \bibinfo{person}{Jiangzhuo
  Chen}, {and} \bibinfo{person}{Madhav Marathe}.}
  \bibinfo{year}{2019}\natexlab{a}.
\newblock \showarticletitle{DEFSI: Deep learning based epidemic forecasting
  with synthetic information}. In \bibinfo{booktitle}{\emph{Proceedings of the
  AAAI Conference on Artificial Intelligence}}, Vol.~\bibinfo{volume}{33}.
  \bibinfo{pages}{9607--9612}.
\newblock


\bibitem[\protect\citeauthoryear{Wang, Chen, and Marathe}{Wang
  et~al\mbox{.}}{2019b}]%
        {gif2019}
\bibfield{author}{\bibinfo{person}{Lijing Wang}, \bibinfo{person}{Jiangzhuo
  Chen}, {and} \bibinfo{person}{Madhav Marathe}.}
  \bibinfo{year}{2019}\natexlab{b}.
\newblock \bibinfo{title}{{TDEFSI: Theory Guided Deep Learning Based Epidemic
  Forecasting with Synthetic Information (Supplement).}}
\newblock
  \bibinfo{howpublished}{\url{https://github.com/christa60/defsi/blob/master/animation.gif}}.
\newblock


\bibitem[\protect\citeauthoryear{Wang, Chakraborty, Mekaru, Brownstein, Ye, and
  Ramakrishnan}{Wang et~al\mbox{.}}{2015}]%
        {wang2015dynamic}
\bibfield{author}{\bibinfo{person}{Zheng Wang}, \bibinfo{person}{Prithwish
  Chakraborty}, \bibinfo{person}{Sumiko~R Mekaru}, \bibinfo{person}{John~S
  Brownstein}, \bibinfo{person}{Jieping Ye}, {and} \bibinfo{person}{Naren
  Ramakrishnan}.} \bibinfo{year}{2015}\natexlab{}.
\newblock \showarticletitle{Dynamic poisson autoregression for
  influenza-like-illness case count prediction}. In
  \bibinfo{booktitle}{\emph{Proceedings of the 21th ACM SIGKDD International
  Conference on Knowledge Discovery and Data Mining}}. ACM,
  \bibinfo{pages}{1285--1294}.
\newblock


\bibitem[\protect\citeauthoryear{WHO}{WHO}{2019}]%
        {whoseasonalinfluenza}
\bibfield{author}{\bibinfo{person}{WHO}.} \bibinfo{year}{2019}\natexlab{}.
\newblock \bibinfo{title}{Seasonal Influenza.}
\newblock
  \bibinfo{howpublished}{\url{http://www.who.int/news-room/fact-sheets/detail/influenza-(seasonal)}}.
\newblock
\newblock
\shownote{Accessed April 01, 2019.}


\bibitem[\protect\citeauthoryear{Wong, Wang, and Shi}{Wong
  et~al\mbox{.}}{2009}]%
        {wong2009}
\bibfield{author}{\bibinfo{person}{Ken~CL Wong}, \bibinfo{person}{Linwei Wang},
  {and} \bibinfo{person}{Pengcheng Shi}.} \bibinfo{year}{2009}\natexlab{}.
\newblock \showarticletitle{Active model with orthotropic hyperelastic material
  for cardiac image analysis}. In \bibinfo{booktitle}{\emph{International
  Conference on Functional Imaging and Modeling of the Heart}}. Springer,
  \bibinfo{pages}{229--238}.
\newblock


\bibitem[\protect\citeauthoryear{Wong, Gatt, Stamatescu, and McDonnell}{Wong
  et~al\mbox{.}}{2016}]%
        {wong2016understanding}
\bibfield{author}{\bibinfo{person}{Sebastien~C Wong}, \bibinfo{person}{Adam
  Gatt}, \bibinfo{person}{Victor Stamatescu}, {and} \bibinfo{person}{Mark~D
  McDonnell}.} \bibinfo{year}{2016}\natexlab{}.
\newblock \showarticletitle{Understanding data augmentation for classification:
  when to warp?}. In \bibinfo{booktitle}{\emph{2016 international conference on
  digital image computing: techniques and applications (DICTA)}}. IEEE,
  \bibinfo{pages}{1--6}.
\newblock


\bibitem[\protect\citeauthoryear{Wu, Yang, Nishiura, and Saitoh}{Wu
  et~al\mbox{.}}{2018}]%
        {wu2018deep}
\bibfield{author}{\bibinfo{person}{Yuexin Wu}, \bibinfo{person}{Yiming Yang},
  \bibinfo{person}{Hiroshi Nishiura}, {and} \bibinfo{person}{Masaya Saitoh}.}
  \bibinfo{year}{2018}\natexlab{}.
\newblock \showarticletitle{Deep learning for epidemiological predictions}. In
  \bibinfo{booktitle}{\emph{The 41st International ACM SIGIR Conference on
  Research \& Development in Information Retrieval}}. ACM,
  \bibinfo{pages}{1085--1088}.
\newblock


\bibitem[\protect\citeauthoryear{Xu, Sapp, Dehaghani, Gao, Horacek, and
  Wang}{Xu et~al\mbox{.}}{2015}]%
        {xu2015}
\bibfield{author}{\bibinfo{person}{Jingjia Xu}, \bibinfo{person}{John~L Sapp},
  \bibinfo{person}{Azar~Rahimi Dehaghani}, \bibinfo{person}{Fei Gao},
  \bibinfo{person}{Milan Horacek}, {and} \bibinfo{person}{Linwei Wang}.}
  \bibinfo{year}{2015}\natexlab{}.
\newblock \showarticletitle{Robust Transmural Electrophysiological Imaging:
  Integrating Sparse and Dynamic Physiological Models into ECG-Based
  Inference}. In \bibinfo{booktitle}{\emph{Medical Image Computing and
  Computer-Assisted Intervention -- MICCAI 2015}},
  \bibfield{editor}{\bibinfo{person}{Nassir Navab}, \bibinfo{person}{Joachim
  Hornegger}, \bibinfo{person}{William~M Wells}, {and}
  \bibinfo{person}{Alejandro Frangi}} (Eds.). \bibinfo{publisher}{Springer
  International Publishing}, \bibinfo{address}{Cham},
  \bibinfo{pages}{519--527}.
\newblock
\showISBNx{978-3-319-24571-3}


\bibitem[\protect\citeauthoryear{Xu, Gel, Ramirez, Nezafati, Zhang, and
  Tsui}{Xu et~al\mbox{.}}{2017}]%
        {xu2017forecasting}
\bibfield{author}{\bibinfo{person}{Qinneng Xu}, \bibinfo{person}{Yulia~R Gel},
  \bibinfo{person}{L~Leticia~Ramirez Ramirez}, \bibinfo{person}{Kusha
  Nezafati}, \bibinfo{person}{Qingpeng Zhang}, {and}
  \bibinfo{person}{Kwok-Leung Tsui}.} \bibinfo{year}{2017}\natexlab{}.
\newblock \showarticletitle{Forecasting influenza in Hong Kong with Google
  search queries and statistical model fusion}.
\newblock \bibinfo{journal}{\emph{PloS one}} \bibinfo{volume}{12},
  \bibinfo{number}{5} (\bibinfo{year}{2017}), \bibinfo{pages}{e0176690}.
\newblock


\bibitem[\protect\citeauthoryear{Yang, Santillana, Brownstein, Gray,
  Richardson, and Kou}{Yang et~al\mbox{.}}{2017}]%
        {yang2017}
\bibfield{author}{\bibinfo{person}{Shihao Yang}, \bibinfo{person}{Mauricio
  Santillana}, \bibinfo{person}{John~S Brownstein}, \bibinfo{person}{Josh
  Gray}, \bibinfo{person}{Stewart Richardson}, {and} \bibinfo{person}{SC Kou}.}
  \bibinfo{year}{2017}\natexlab{}.
\newblock \showarticletitle{Using electronic health records and Internet search
  information for accurate influenza forecasting}.
\newblock \bibinfo{journal}{\emph{BMC infectious diseases}}
  \bibinfo{volume}{17}, \bibinfo{number}{1} (\bibinfo{year}{2017}),
  \bibinfo{pages}{332}.
\newblock


\bibitem[\protect\citeauthoryear{Yang, Santillana, and Kou}{Yang
  et~al\mbox{.}}{2015b}]%
        {yang2015accurate}
\bibfield{author}{\bibinfo{person}{Shihao Yang}, \bibinfo{person}{Mauricio
  Santillana}, {and} \bibinfo{person}{Samuel~C Kou}.}
  \bibinfo{year}{2015}\natexlab{b}.
\newblock \showarticletitle{Accurate estimation of influenza epidemics using
  Google search data via ARGO}.
\newblock \bibinfo{journal}{\emph{Proceedings of the National Academy of
  Sciences}} \bibinfo{volume}{112}, \bibinfo{number}{47}
  (\bibinfo{year}{2015}), \bibinfo{pages}{14473--14478}.
\newblock


\bibitem[\protect\citeauthoryear{Yang, Karspeck, and Shaman}{Yang
  et~al\mbox{.}}{2014}]%
        {yang2014comparison}
\bibfield{author}{\bibinfo{person}{Wan Yang}, \bibinfo{person}{Alicia
  Karspeck}, {and} \bibinfo{person}{Jeffrey Shaman}.}
  \bibinfo{year}{2014}\natexlab{}.
\newblock \showarticletitle{Comparison of filtering methods for the modeling
  and retrospective forecasting of influenza epidemics}.
\newblock \bibinfo{journal}{\emph{PLoS computational biology}}
  \bibinfo{volume}{10}, \bibinfo{number}{4} (\bibinfo{year}{2014}),
  \bibinfo{pages}{e1003583}.
\newblock


\bibitem[\protect\citeauthoryear{Yang, Lipsitch, and Shaman}{Yang
  et~al\mbox{.}}{2015a}]%
        {yang2015inference}
\bibfield{author}{\bibinfo{person}{Wan Yang}, \bibinfo{person}{Marc Lipsitch},
  {and} \bibinfo{person}{Jeffrey Shaman}.} \bibinfo{year}{2015}\natexlab{a}.
\newblock \showarticletitle{Inference of seasonal and pandemic influenza
  transmission dynamics}.
\newblock \bibinfo{journal}{\emph{Proceedings of the National Academy of
  Sciences}} \bibinfo{volume}{112}, \bibinfo{number}{9} (\bibinfo{year}{2015}),
  \bibinfo{pages}{2723--2728}.
\newblock


\bibitem[\protect\citeauthoryear{Yang, Olson, and Shaman}{Yang
  et~al\mbox{.}}{2016}]%
        {yang2016forecasting}
\bibfield{author}{\bibinfo{person}{Wan Yang}, \bibinfo{person}{Donald~R Olson},
  {and} \bibinfo{person}{Jeffrey Shaman}.} \bibinfo{year}{2016}\natexlab{}.
\newblock \showarticletitle{Forecasting influenza outbreaks in boroughs and
  neighborhoods of New York City}.
\newblock \bibinfo{journal}{\emph{PLoS computational biology}}
  \bibinfo{volume}{12}, \bibinfo{number}{11} (\bibinfo{year}{2016}),
  \bibinfo{pages}{e1005201}.
\newblock


\bibitem[\protect\citeauthoryear{Zhang, Bengio, Hardt, Recht, and
  Vinyals}{Zhang et~al\mbox{.}}{2016}]%
        {zhang2016understanding}
\bibfield{author}{\bibinfo{person}{Chiyuan Zhang}, \bibinfo{person}{Samy
  Bengio}, \bibinfo{person}{Moritz Hardt}, \bibinfo{person}{Benjamin Recht},
  {and} \bibinfo{person}{Oriol Vinyals}.} \bibinfo{year}{2016}\natexlab{}.
\newblock \showarticletitle{Understanding deep learning requires rethinking
  generalization}.
\newblock \bibinfo{journal}{\emph{arXiv preprint arXiv:1611.03530}}
  (\bibinfo{year}{2016}).
\newblock


\bibitem[\protect\citeauthoryear{Zhao, Chen, Chen, Wang, Lu, and
  Ramakrishnan}{Zhao et~al\mbox{.}}{2015}]%
        {zhao2015simnest}
\bibfield{author}{\bibinfo{person}{Liang Zhao}, \bibinfo{person}{Jiangzhuo
  Chen}, \bibinfo{person}{Feng Chen}, \bibinfo{person}{Wei Wang},
  \bibinfo{person}{Chang-Tien Lu}, {and} \bibinfo{person}{Naren Ramakrishnan}.}
  \bibinfo{year}{2015}\natexlab{}.
\newblock \showarticletitle{Simnest: Social media nested epidemic simulation
  via online semi-supervised deep learning}. In \bibinfo{booktitle}{\emph{2015
  IEEE International Conference on Data Mining}}. IEEE,
  \bibinfo{pages}{639--648}.
\newblock


\bibitem[\protect\citeauthoryear{Zhu, Park, Isola, and Efros}{Zhu
  et~al\mbox{.}}{2017}]%
        {zhu2017unpaired}
\bibfield{author}{\bibinfo{person}{Jun-Yan Zhu}, \bibinfo{person}{Taesung
  Park}, \bibinfo{person}{Phillip Isola}, {and} \bibinfo{person}{Alexei~A
  Efros}.} \bibinfo{year}{2017}\natexlab{}.
\newblock \showarticletitle{Unpaired image-to-image translation using
  cycle-consistent adversarial networks}. In
  \bibinfo{booktitle}{\emph{Proceedings of the IEEE international conference on
  computer vision}}. \bibinfo{pages}{2223--2232}.
\newblock


\end{thebibliography}
